\documentclass[twocolumn]{aastex631}

\usepackage{CJK}
\shorttitle{FRB Host Galaxies}
\shortauthors{Gordon et al.}
\graphicspath{{./}{figures/}}

\begin{document}
\begin{CJK*}{UTF8}{gbsn}

\title{The Demographics, Stellar Populations, and Star Formation Histories of Fast Radio Burst Host Galaxies: Implications for the Progenitors}

\correspondingauthor{Alexa C. Gordon}
\email{alexagordon2026@u.northwestern.edu}

\newcommand{\NU}{\affiliation{Center for Interdisciplinary Exploration and Research in Astrophysics (CIERA) and Department of Physics and Astronomy, Northwestern University, Evanston, IL 60208, USA}}

\newcommand{\Purdue}{\affiliation{Purdue University, 
Department of Physics and Astronomy, 525 Northwestern Avenue, West Lafayette, IN 47907, USA}}

\newcommand{\CfA}{\affiliation{Center for Astrophysics\:$|$\:Harvard \& Smithsonian, 60 Garden St. Cambridge, MA 02138, USA}}

\newcommand{\UCSC}{\affiliation{Department of Astronomy and Astrophysics, University of California, Santa Cruz, CA 95064, USA}}

\newcommand{\IS}{\affiliation{Centre for Astrophysics and Cosmology, Science Institute, University of Iceland, Dunhagi 5, 107 Reykjav\'ik, Iceland}}

\newcommand{\DAWN}{\affiliation{Cosmic Dawn Center (DAWN), Niels Bohr Institute, University of Copenhagen, Jagtvej 128, 2100 Copenhagen \O, Denmark}}

\newcommand{\PUCV}{\affiliation{Instituto de F\'isica, Pontificia Universidad Cat\'olica de Valpara\'iso, Casilla 4059, Valpara\'iso, Chile}}

\newcommand{\IPMU}{\affiliation{Kavli Institute for the Physics and Mathematics of the Universe (Kavli IPMU), 5-1-5 Kashiwanoha, Kashiwa, 277-8583, Japan}}

\newcommand{\PSU}{\affiliation{Department of Astronomy \& Astrophysics, The Pennsylvania State University, University Park, PA 16802, USA}}

\newcommand{\ICDS}{\affiliation{Institute for Computational \& Data Sciences, The Pennsylvania State University, University Park, PA, USA}}

\newcommand{\IGC}{\affiliation{Institute for Gravitation and the Cosmos, The Pennsylvania State University, University Park, PA 16802, USA}}

\newcommand{\Swin}{\affiliation{ Centre for Astrophysics and Supercomputing, Swinburne University of Technology, Hawthorn, VIC, 3122, Australia}}

\newcommand{\Curtin}{\affiliation{ International Centre for Radio Astronomy Research, Curtin University, Bentley, WA 6102, Australia}}

\newcommand{\MQ}{\affiliation{School of Mathematical and Physical Sciences, Macquarie University, NSW 2109, Australia}}

\newcommand{\MQAAAstro}{\affiliation{Macquarie University Research Centre for Astronomy, Astrophysics \& Astrophotonics, Sydney, NSW 2109, Australia}}

\newcommand{\CSIRO}{\affiliation{CSIRO, Space and Astronomy, PO Box 76, Epping NSW 1710 Australia}}

\newcommand{\KICP}{\affiliation{Kavli Institute for Cosmological Physics, The University of Chicago, 5640 South Ellis Avenue, Chicago, IL 60637, USA}}

\newcommand{\UA}{\affiliation{University of Arizona, Steward Observatory, 933~N.~Cherry~Ave., Tucson, AZ 85721, USA}}

\newcommand{\EFI}{\affiliation{Enrico Fermi Institute, The University of Chicago, 933 East 56th Street, Chicago, IL 60637, USA}}

\newcommand{\mpia}{\affiliation{Max-Planck-Institut f\"{u}r Astronomie (MPIA), K\"{o}nigstuhl 17, 69117 Heidelberg, Germany}}

\newcommand{\GWU}{\affiliation{Department of Physics, The George Washington University, Washington, DC 20052, USA}}

\newcommand{\UCB}{\affiliation{Department of Astronomy, University of California, Berkeley, CA 94720-3411, USA}}

\newcommand{\RU}{\affiliation{Department of Astrophysics/IMAPP, Radboud University, PO Box 9010,
6500 GL, The Netherlands}}

\newcommand{\LJMU}{\affiliation{Astrophysics Research Institute, Liverpool John Moores University, IC2, Liverpool Science Park, 146 Brownlow Hill, Liverpool L3 5RF, UK}}

\newcommand{\LU}{\affiliation{School of Physics and Astronomy, University of Leicester, University Road, Leicester. LE1 7RH, UK}}

\newcommand{\ARC}{\affiliation{ARC Centre of Excellence for All-Sky Astrophysics in 3 Dimensions (ASTRO 3D), Australia}}

\newcommand{\ASTRON}{\affiliation{ASTRON, Netherlands Institute for Radio Astronomy, Oude Hoogeveensedijk 4, 7991 PD Dwingeloo, The Netherlands}}

\newcommand{\VLBI}{\affiliation{Joint institute for VLBI ERIC, Oude Hoogeveensedijk 4, 7991 PD Dwingeloo, The Netherlands
}}

\newcommand{\Anton}{\affiliation{Anton Pannekoek Institute for Astronomy, University of Amsterdam, Science Park 904, 1098 XH, Amsterdam, The Netherlands}}

\newcommand{\Manchester}{\affiliation{Department of Physics and Astronomy, University of Manchester, Oxford Road, M13 9PL, UK}}

\newcommand{\McGill}{\affiliation{Department of Physics, McGill University, Montreal, Quebec H3A 2T8, Canada
}}

\newcommand{\ICRAR}{\affiliation{International Centre for Radio Astronomy Research (ICRAR), Curtin University, Bentley, WA 6102, Australia}}

\newcommand{\SIA}{\affiliation{Sydney Institute for Astronomy, School of Physics A28, University of Sydney, NSW 2006, Australia}}

\newcommand{\NAOJ}{\affiliation{Division of Science, National Astronomical Observatory of Japan,2-21-1 Osawa, Mitaka, Tokyo 181-8588, Japan}}

\newcommand{\MPIA}{\affiliation{Max-Planck-Institut f\"ur Astronomie, K\"onigstuhl 17, 69117 Heidelberg, Germany}}
\author[0000-0002-5025-4645]{Alexa C. Gordon}
\NU

\author[0000-0002-7374-935X]{Wen-fai Fong}
\NU

\author[0000-0002-5740-7747]{Charles D. Kilpatrick}
\NU

\author[0000-0003-0307-9984]{Tarraneh Eftekhari}\thanks{NHFP Einstein Fellow}
\NU

\author[0000-0001-6755-1315]{Joel Leja}
\PSU
\ICDS
\IGC

\author[0000-0002-7738-6875]{J. Xavier Prochaska}
\UCSC
\IPMU
\NAOJ

\author[0000-0002-2028-9329]{Anya E. Nugent}
\NU

\author[0000-0003-3460-506X]{Shivani Bhandari}
\ASTRON
\VLBI
\Anton
\CSIRO

\author[0000-0003-0526-2248]{Peter K. Blanchard}
\NU

\author[0000-0002-4079-4648]{Manisha Caleb}
\SIA
\ARC

\author[0000-0002-8101-3027]{Cherie K. Day}
\McGill

\author[0000-0001-9434-3837]{Adam T. Deller}
\Swin

\author[0000-0002-9363-8606]{Yuxin Dong (董雨欣)} 
\NU

\author[0000-0002-5067-8894]{Marcin Glowacki}
\ICRAR

\author[0000-0002-0152-1129]{Kelly Gourdji}
\Swin

\author{Alexandra G. Mannings}
\UCSC

\author[0000-0002-5053-2828]{Elizabeth K. Mahoney}
\CSIRO

\author[0000-0003-1483-0147]{Lachlan Marnoch}
\CSIRO
\MQ
\MQAAAstro
\ARC

\author[0000-0001-9515-478X]{Adam A. Miller}
\NU

\author[0000-0001-8340-3486]{Kerry Paterson}
\MPIA

\author[0000-0002-9267-6213]{Jillian C. Rastinejad}
\NU

\author[0000-0003-4501-8100]{Stuart D. Ryder}
\MQ
\MQAAAstro

\author[0000-0002-1136-2555]{Elaine M. Sadler}
\CSIRO
\SIA

\author[0000-0002-6895-4156]{Danica R. Scott}
\ICRAR

\author[0000-0001-8023-4912]{Huei Sears}
\NU

\author[0000-0002-7285-6348]{Ryan~M.~Shannon}
\Swin

\author[0000-0003-3801-1496]{Sunil Simha}
\UCSC

\author[0000-0001-9242-7041]{Benjamin W. Stappers}
\Manchester

\author[0000-0002-1883-4252]{Nicolas Tejos}
\PUCV



\begin{abstract}

We present a comprehensive catalog of observations and stellar population properties for 23 highly secure host galaxies of fast radio bursts (FRBs). Our sample comprises six repeating FRBs and 17 apparent non-repeaters. We present 82 new photometric and eight new spectroscopic observations of these hosts. Using stellar population synthesis modeling and employing non-parametric star formation histories (SFHs), we find that FRB hosts have a median stellar mass of $\approx 10^{9.9}\,M_{\odot}$, mass-weighted age $\approx 5.1$~Gyr, and ongoing star formation rate $\approx 1.3\,M_{\odot}$~yr$^{-1}$ but span wide ranges in all properties. Classifying the hosts by degree of star formation, we find that 87\% (20/23 hosts) are star-forming, two are transitioning, and one is quiescent. The majority trace the star-forming main sequence of galaxies, but at least three FRBs in our sample originate in less active environments (two non-repeaters and one repeater). Across all modeled properties, we find no statistically significant distinction between the hosts of repeaters and non-repeaters. However, the hosts of repeating FRBs generally extend to lower stellar masses, and the hosts of non-repeaters arise in more optically luminous galaxies. While four of the galaxies with the most clear and prolonged rises in their SFHs all host repeating FRBs, demonstrating heightened star formation activity in the last $\lesssim 100$~Myr, one non-repeating host shows this SFH as well. Our results support progenitor models with short delay channels (i.e., magnetars formed via core-collapse supernova) for most FRBs, but the presence of some FRBs in less active environments suggests a fraction form through more delayed channels. 

\end{abstract}



\section{Introduction} \label{sec:intro}

Fast radio bursts (FRBs) are extremely bright ($\approx 10$~mJy--$100$~Jy), (sub-)millisecond pulses of MHz--GHz emission. First discovered over a decade ago in 2007 \citep{Lorimer07}, it was not until 2017 that an FRB was precisely localized and traced back to its host galaxy at $z=0.19$, providing the first direct evidence of their cosmological origins \citep{Chatterjee+17,Marcote+17,Tendulkar+17}. While this first localization was facilitated by targeted observations made possible by the repeating nature of the source, an increasing number of FRBs are now being localized upon detection by fast transient searches using radio interferometers \citep[e.g.,][]{Bannister+19,Ravi+19}. Over the last decade, the commissioning of sensitive and wide-field fast transient detection instrumentation has led to an enormous increase in the detected FRB population \citep[e.g.,][]{CHIME+18,Macquart+10}. It is now established that some fraction (currently $\sim$4\% observed; \citealt{Petroff+22}) of the FRB population produce repeat bursts from the same cosmic source \citep[repeating FRBs,][]{Spitler+16,chime-repeater} while the majority of discovered FRBs have not been observed to repeat to date \citep[apparent non-repeating FRBs,][]{CHIME-catalog}. Despite over 600 published FRBs to date \citep{Petroff+22}, their origins and the nature of their repetition remains uncertain.

Due to their short durations and inferred high energies, many of the leading FRB progenitor models invoke a magnetically-powered neutron star, known as a magnetar \citep{Platts+19}. The connection between FRBs and magnetars was bolstered with the detection of a bright radio burst from the Galactic magnetar SGR\,1935+2154, which accompanied emission at higher energies \citep{Bochenek20, CHIME-magnetar,Mereghetti+20}. While the radio burst energies from that source fall a few orders of magnitude short of the typical extragalactic FRBs detected to date \citep{Kirsten-SGR}, they begin to bridge the gap between Galactic and extragalactic coherent radio sources (e.g., \citealt{Margalit20-Galacticmagnetar,Nimmo+22}).

The FRB signal properties, namely their dispersion measure (DM), rotation measure (RM), duration, spectro-temporal morphology, and polarization, provide key information on their central engines, and serve as important probes of the intervening ionized matter \citep[e.g.,][]{Michilli18, Hessels+19,Hilmarsson+21,Cook+2022,Mannings+2022,Ryder22,Wu_McQuinn2022,Ocker+2023}. However, these signals alone provide only a highly model-dependent, low-precision estimate of the distance to the FRB, making it difficult or impossible to infer the parent stellar populations and FRB energetics directly. However, if a host galaxy can be identified, a wealth of information can be gleaned from their local and global environments -- precise redshifts, energy scales, and properties of the environment on local and galactic scales. Concurrent to the findings that repeater and non-repeater burst morphologies may be statistically distinct \citep{Pleunis21}, host properties may provide additional distinguishing power on the physical distinction between these classes. Aligned with this goal, advancements in FRB experiments have enabled routine (sub-)arcsecond localizations, making robust host galaxy associations feasible.
 
 For FRBs, the first host associations were in seemingly distinct environments. The first repeating FRB 20121102A was found in a star-forming, metal-poor dwarf galaxy \citep{Chatterjee+17,Tendulkar+17}, while the first well-localized apparently non-repeating FRB\,20180924B resided in an older and more massive star-forming galaxy \citep{Bannister+19}, providing early signs that they might arise from distinct populations.  The next well-localized repeater FRB\,20180916B was pinpointed to a massive spiral galaxy with a moderate star formation rate \citep{Marcote_etal_2020}, complicating the picture that their host properties alone could be discerning. One interesting feature that has only concerned two repeating FRBs thus far (FRBs\,20121102A and 20190520B) are their co-location with persistent radio sources (PRSs; \citealt{Chatterjee+17,Marcote+17,Niu+22}): compact radio emission that cannot be attributed to star formation. Based on the size and brightness temperature ($\lesssim$ 0.7 pc and $T_{b}\gtrsim 5\times10^{7}$ K, respectively), \citet{Marcote+17} claim the PRS associated with FRB\,20121102A is compact and directly linked to the FRB event, while the PRS associated with FRB\,20190520B was too luminous to be explained by star formation \citep{Niu+22}. Notably both bursts reside in dwarf host galaxies \citep{Tendulkar+17,Niu+22}, with their PRSs coincident with or close to a star-forming knot in the galaxy. While no other FRBs have known PRSs to date, it is worth characterizing their host galaxies to connect to their multi-wavelength properties.
 
 As a population, FRBs are generally found in star-forming galaxies across a wide range of masses (e.g., \citealt{Tendulkar+17,Bannister+19,Prochaska+19,Marcote_etal_2020,Fong+21,Bhandari_210117,Ravi22b,Niu+22}). Based on early studies of a limited number of hosts, no statistically significant distinction was found between the stellar population properties of repeating and apparent non-repeating FRBs \citep{Heintz+20,Bhandari+20}. In some cases, milliarcsecond-scale localizations pinpointed FRBs to very different sub-galactic environments, e.g., the discoveries of some repeating FRBs in or proximal to knots of star formation \citep{Tendulkar+17,Bassa+17,Tendulkar+21,Piro+21,Niu+22} or in non-star-forming environments altogether, such as the old $\sim 9$~Gyr globular cluster environment \citep{Bhardwaj21,Kirsten21}. From a modest sample of sub-arcsecond localized FRBs, one finds the majority occur within or near the spiral arms of their hosts \citep{Marcote_etal_2020,Chittidi21,Mannings+21,Tendulkar+21}, although there are exceptions \citep{Xu22}.

 While such sub-galactic details can be afforded by the fairly local $z\lesssim 0.3$ FRB population, the bulk of newly-discovered FRBs will be found at higher redshift by more sensitive searches, and only integrated galaxy properties will be available in almost all cases.  As spectral energy distribution (SED) modeling techniques diversify in their specific assumptions and methodologies \citep[see ][for a comparison of 14 SED fitting codes]{Pacifici+2022}, it is necessary to compare the FRB host population to similarly-modeled field galaxies and to derive their individual star formation histories (SFHs). Thus, it is timely and complementary to more local studies to perform a uniform analysis of the population of FRB host galaxies, their SFHs, and derive implications for their progenitors.

In this work, we compile a sample of 23 highly secure FRB host galaxies, including the hosts of six repeating and 17 apparent non-repeating FRBs, discovered across a range of facilities, and perform uniform modeling of the stellar population properties and SFHs of their host galaxies. The FRBs in this sample were discovered over roughly a decade, November 2012--January 2022. In Section~\ref{sec:sample} we discuss the sample selection. In Section~\ref{sec:observations} we detail the data acquisition and processing. We detail our modeling assumptions and methodology  using the \texttt{Prospector} SED modeling code \citep{Johnson+21} in Section~\ref{sec:prospector}. We present our results of the stellar population properties and SFHs and compare to the general galaxy population in Section~\ref{sec:results} and discuss the implications of these results in Section~\ref{sec:discussion}. Finally, we summarize and conclude in Section~\ref{sec:conclusion}.

\section{Sample} \label{sec:sample}

Our sample is comprised of known FRB hosts from the literature as well as new FRBs and host identifications. We start with all FRBs localized before the end of January 2022  (FRB\,20121102A to FRB\,20220105A). As part of the Fast and Fortunate for FRB Follow-up\footnote{http://frb-f4.org/} (F$^{4}$) collaboration, we receive the positional information of new FRBs from the Commensal Real-Time ASKAP Fast-Transients ({\tt CRAFT}; \citealt{mbb+10}) survey on the Australian Square Kilometre Array Pathfinder \citep{ASKAP} and the More TRansients and Pulsars ({\tt MeerTRAP}; \citealt{Rajwade+22}) project on the MeerKAT radio telescope \citep{MeerKAT}.  We next search imaging archives such as the Sloan Digital Sky Survey (SDSS; \citealt{SDSS}), Pan-STARRS1 (PS1; \citealt{PS1}), and the Dark Enegy Camera Legacy Survey (DECaLS; \citealt{DECALS}) for any plausible host galaxies at or near the FRB position. If no host galaxy candidates are visible in these images, we obtain deep $r$- or $I$-band imaging with 4-meter to 10-meter class telescopes to aid with host galaxy identification (see Section \ref{sec: imaging}). We also use known FRB host identifications from the literature, with the FRBs discovered by Arecibo \citep{Spitler+14,Spitler+16,Scholz+16}, Parkes \citep{Price+19}, the Canadian Hydrogen Intensity Mapping Experiment (CHIME; \citealt{CHIME+19}), and the Five-hundred-meter Aperture Spherical radio Telescope (FAST; \citealt{Niu+22}).

\subsection{Sample Selection} \label{sec: host criteria}

We begin with a parent sample of 27 events that have been localized to $\lesssim 1-2$\arcsec, which is necessary for unambiguous host associations with luminous $L^*$ galaxies across the redshift range $z \sim 0.1 -1$ \citep{Eftekhari2017}. We then apply the following criteria for inclusion in this work.

\begin{enumerate}
\item We require a PATH (Probabilistic Association of Transients to their Hosts; \citealt{path}) posterior probability $\geq90\%$, following the same convention used in \citet{bha+22}. PATH employs a Bayesian framework to calculate the likelihood a transient is associated with a galaxy given the transient's localization, the galaxy's position on the sky and angular size, and its apparent magnitude. Combined with prior assumptions on the probability that the host is undetected (i.e., fainter than the flux limit of the imaging observations) and the offset of FRBs from the centers of their hosts, PATH reports the posterior probability of association to every galaxy in the field. Higher probabilities correspond to higher likelihood of association. By requiring the posterior probability to be $\geq90\%$, we construct a sample of high probability host galaxy associations. We apply PATH to four FRBs -- 20190520B (using CFHT $r^\prime$ archival data; \citealt{Niu+22} and \citealt{Gwyn08} for details on  the $r^\prime$ transmission curve), 20211203C, 20210410D, and 20220105A -- for which there are no published PATH probabilities. We otherwise use published PATH probabilities from \citet{James+22, path}, and \citet{bha+22}. This criterion removes the hosts of FRB\,20190614D and FRB\,20190523A.

\item There is no bright ($\lesssim$ 10~AB~mag) foreground star within 5\arcmin~of the FRB position. We employ this criterion to ensure that the observations are not contaminated by scattered light from a nearby bright star; in particular, this can inhibit accurate photometry of the host galaxy. 

\item There are detections of the host galaxy in at least three photometric bands in the optical/IR or otherwise overlapping the observed spectrum's wavelength coverage. For our methodology used to model the FRB hosts, described in Section~\ref{sec:prospector}, the absolute flux calibration of the spectrum depends on the photometry. We have found three photometric bands to be the minimum required in order to obtain a trustworthy model. 

\item We exclude FRBs with burst spectral energies below 10$^{27}$~erg~Hz$^{-1}$, to exclude low-energy bursts that would be undetectable over the majority of the redshift range spanned by our sample. Assuming an ASKAP detection limit of 4.4 Jy~ms \citep{James+22}, this energy cut corresponds to excluding FRBs that can be seen to a maximum redshift of $z\sim 0.003$ by the telescope that contributed the bulk of our sample. This criterion excludes two very nearby FRBs/FRB-like signals: FRB\,20200120E and the Galactic source SGR\,J1935+2154. These sources are significant outliers in luminosity space, as is clear from fluence-redshift diagrams such as the one in \citet{Ryder22}. We note that while all bursts seen to date from both of these sources are excluded by this cut, the highest energy bursts from the FRB-like SGR\,J1935+2154 are more energetic than the lowest energy bursts of FRB\,20200120E \citep{Nimmo+22}, making them more comparable to each other than to the higher-z FRB population. 

\item A spectrum of the host galaxy is available and contains detectable spectral features (i.e., emission and absorption lines) for redshift determination and a signal-to-noise (S/N) $\gtrsim$ 3/\AA\ in the continuum. We note that spectroscopy is essential for breaking the known degeneracy in age, dust, and metallicity in SED modeling (e.g., \citealt{Bell+01,Leja+17}). Additionally, the thousands of data points inherent in a spectrum provide more information to fit than photometry alone. We note that even though there is relative difficulty in detecting high S/N features for quiescent versus star-forming galaxies, no hosts are excluded on this criterion alone.

\end{enumerate} 

While these sample criteria are enforced for the majority of the FRBs, we made two exceptions (requiring additional care in our data processing) to ensure that the small sample of repeating FRBs was not further reduced. The repeating FRB\,20190520B \citep{Niu+22} fails the bright star criterion due to its $\approx 1$\arcmin\ proximity to the bright, $r$$\sim$8 AB mag, variable star V1042 Sco B. In addition to being a repeating FRB, this host contains the second known PRS, making it especially noteworthy. We took appropriate steps during the data collection and reduction phases to ensure the photometry and spectroscopy were not contaminated by extra flux from the star (see Sections~\ref{sec: imaging} and \ref{sec:spectra} for details). As mentioned above, we use archival CFHT $r^\prime$ \citep[which is close to Sloan $r^\prime$;][]{Gwyn08} data for calculating the PATH probability of this host galaxy. Our analysis assigns the host identified in \citet{Niu+22} a posterior probability of nearly 100\%, robustly confirming the host association. The host galaxy of the repeating FRB\,20190711A fails the spectroscopic continuum signal requirement; at a declination of $-80^{\circ}$, the source is always at high airmass for most available follow-up resources and is also quite faint, making observation challenging. While we use the existing Gemini South/Gemini Multi-Object Spectrograph (GMOS) spectrum (first reported in \citealt{mpm+20}) to determine the redshift, this spectrum does not have a high enough S/N for our subsequent modeling. In this case, we still include this host, and take advantage of the seven photometric data points for modeling. 

When applying these criteria to the existing literature and new FRBs obtained from the ASKAP/{\tt CRAFT} and MeerKAT/{\tt MeerTRAP} collaborations, we obtain a sample of 23 FRB host galaxies (six repeating FRBs and 17 apparent non-repeaters). We list the properties of the FRBs, discovery surveys, and optical host magnitudes in Table~\ref{tab: FRB properties}. This compilation is the largest sample of highly secure localized FRB host galaxies to date and allows for a systematic study of their host properties and SFHs.

\startlongtable
\begin{deluxetable*}{l|cccccccc}
\tabletypesize{\footnotesize}
\tablewidth{0pc}
\tablecaption{Basic FRB Properties
\label{tab: FRB properties}}
\tablehead{
\colhead{FRB} &
\colhead{Survey/Facility$^{a}$}	 &
\colhead{RA (FRB)} &
\colhead{Decl. (FRB)} & 
\colhead{$z$} &
\colhead{Repeater} &
\colhead{Host Magnitude} &
\colhead{Filter} & 
\colhead{Reference} \\
\colhead{} &
\colhead{} &
\colhead{[J2000]} &
\colhead{[J2000]} &
\colhead{} &
\colhead{} &
\colhead{[AB]} &
\colhead{} &
\colhead{}
}
\startdata
20121102A & Arecibo/VLA/EVN & 05:31:58.70 & 33:08:52.6 & 0.1927 & Y & 23.73 & GMOS-N $r$ & 1,2 \\
20180301A & Parkes/VLA & 06:12:54.44 & 04:40:15.8 & 0.3304 & Y & 21.21 & NOT $r$ & 3 \\
20180916B & CHIME/EVN & 01:58:00.75 & 65:43:00.3 & 0.0337 & Y & 16.17 & SDSS $r$ & 4,5 \\
20180924B & CRAFT & 21:44:25.26 & $-$40:54:00.1 & 0.3212 & N & 20.33 & DECaLS $r$ & 6, This Work \\
20181112A & CRAFT & 21:49:23.63 & $-$52:58:15.4 & 0.4755 & N & 21.68 & DES $r$ & 7 \\
20190102C & CRAFT & 21:29:39.76 & $-$79:28:32.5 & 0.2912 & N & 20.77 & VLT/FORS2 $I$ & 8, 10 \\
20190520B & FAST/VLA & 16:02:04.27 & $-$11:17:17.3 & 0.2418 & Y & 22.16 & SOAR $r$ & 9, This Work \\
20190608B & CRAFT & 22:16:04.77 & $-$07:53:53.7 & 0.1178 & N & 17.41 & DECaLS $r$ & 8, This Work \\
20190611B & CRAFT & 21:22:58.94 & $-$79:23:51.3 & 0.3778 & N & 22.15 & GMOS-S $r$ & 4,8 \\
20190711A & CRAFT & 21:57:40.62 & $-$80:21:28.8 & 0.522 & Y & 23.54 & GMOS-S $r$ & 4,8 \\
20190714A & CRAFT & 12:15:55.13 & $-$13:01:15.6 & 0.2365 & N & 20.34 & Pan-STARRS $r$ & 4 \\
20191001A & CRAFT & 21:33:24.31 & $-$54:44:51.9 & 0.234 & N & 18.36 & DECaLS $r$ & 10 \\
20200430A & CRAFT & 15:18:49.54 & 12:22:36.3 & 0.1608 & N & 21.05 & DECaLS $r$ & 4 \\
20200906A & CRAFT & 03:33:59.08 & $-$14:04:59.5 & 0.3688 & N & 19.95 & DES $r$ & 3 \\
20201124A & CHIME/ASKAP & 05:08:03.51 & 26:03:38.5 & 0.0979 & Y & 17.86 & Pan-STARRS $r$ & 11 \\
20210117A & CRAFT & 22:39:55.07 & $-$16:09:05.4 & 0.2145 & N & 22.97 & DEIMOS $R$ & 12, 13, This Work \\
20210320C & CRAFT & 13:37:50.60 & $-$16:07:21.7 & 0.2797 & N & 19.47 & SOAR $r$ & 13, 14, This Work \\
20210410D & MeerTRAP & 21:44:20.70 & $-$79:19:05.5 & 0.1415 & N & 20.65 & SOAR $r$ & 15, This Work \\
20210807D & CRAFT & 19:56:53.14 & $-$00:45:44.5 & 0.1293 & N & 17.17 & Pan-STARRS $r$ & 13, 16, This Work \\
20211127I & CRAFT & 13:19:13.97 & $-$18:50:16.1 & 0.0469 & N & 14.97 & SOAR $r$ & 13, 16, 17, This Work \\
20211203C & CRAFT & 13:38:15.00 & $-$31:22:48.2 & 0.3439 & N & 19.64 & SOAR $r$ & 13, 14, This Work \\
20211212A & CRAFT & 10:29:24.22 & 01:21:39.4 & 0.0707 & N & 16.44 & SOAR $r$ & 13, 16, This Work \\
20220105A & CRAFT & 13:55:12.94 & 22:27:59.4 & 0.2785 & N & 21.19 & Pan-STARRS $r$ & 13, 14, This Work 
\enddata
\tablecomments{Properties of FRBs included in this work. The localization uncertainties are on the order of $1\arcsec$, with the majority $<1\arcsec$. All photometry is corrected for Galactic extinction following the \cite{Fitzpatrick:2007} extinction law. Redshift values are pulled from the literature or derived from \texttt{PypeIt}. \\
\footnotesize{\textit{a}: This column denotes the discovery telescope or collaboration and the localization facility. For those with one entry, the FRB was discovered and localized by the same group. We note that FRB\,20201124A was further localized with the EVN \citep{Piro+21}. The coordinates provided here correspond to the ASKAP localization.} \\
References:
1. \citet{Bassa+17},
2. \citet{Chatterjee+17}, 
3. \citet{bha+22}, 
4. \citet{Heintz+20},
5. \citet{Marcote_etal_2020},
6. \citet{Bannister+19},
7. \citet{Prochaska+19},
8. \citet{mpm+20},
9. \citet{Niu+22},
10. \citet{Bhandari+20},
11. \citet{Fong+21},
12.\citet{Bhandari_210117},
13. \citet{James+22},
14. R.~Shannon+23 in prep.,
15. \citet{Caleb+23},
16. A.~Deller+23 in prep.,
17. \citet{Glowacki+23}}
\end{deluxetable*}

\section{Observations and data reduction} \label{sec:observations}

\subsection{Imaging} \label{sec: imaging}

Once the host for a given FRB is identified, we first search for archival photometry from optical and near-infrared (NIR) surveys: the Sloan Digital Sky Survey (SDSS; \citealt{SDSS}), the Dark Energy Camera Legacy Survey (DECaLS; \citealt{DECALS}), the Dark Energy Survey (DES; \citealt{DES}), the Pan-STARRS 3$\pi$ survey (PS1; \citealt{PS1}), the Two Micron All Sky Survey (2MASS; \citealt{2MASS}), the VISTA Hemisphere Survey (VHS; \citealt{VISTA}), and the Wide-field Infrared Survey Explorer (WISE; \citealt{WISE}). 
For the VHS data, we use a custom script to stack the available, reduced frames. This script implements \texttt{SWarp} \citep{swarp} to coadd the frames. We then set a default zero-point of 27.5 AB mag using the flux calibration from the European Southern Observatory (ESO) archive data reductions. For the WISE data, for which there is a choice of aperture, we select photometry values from the `aperture 2' instrumental aperture (8.25\arcsec\ for W1, W2, and W3, and 16.5\arcsec\ for W4); these apertures were chosen as a balance between encompassing the entire host galaxy, allowing for the change in point spread function with wavelength, and avoiding flux from unassociated sources. We then convert all values to the AB magnitude system if needed.

For nine observations, particularly for FRBs in crowded regions or for hosts of large angular size which are not adequately encompassed by the default apertures, we perform manual photometry. We created a custom script that implements the \texttt{aperture\char`_photometry} module of \texttt{photutils} \citep{photutils}\footnote{https://github.com/charliekilpatrick/photometry}. We determined the aperture and annuli sizes by loading the images into \texttt{SAOImageDS9} \citep{DS9} and adjusting the region sizes to ensure the host and background were modeled accurately. In most cases, the zero-points were taken from the header of the images after verifying the values were consistent with nearby photometric standards, but in cases where no zero-point was provided, we performed point spread function (PSF) photometry on all point sources in the image using {\tt DoPhot} \citep{Schechter93} and compared their instrumental magnitudes to SkyMapper DR2 standard stars \citep{Onken19} for southern fields ($\delta<-30^{\circ}$) or Pan-STARRS DR2 standard stars \citep{Flewelling16} for more northern fields ($\delta>-30^{\circ}$). We then converted to the AB magnitude system if needed for consistency with the rest of the data. Finally, we corrected the values for Galactic extinction using the \cite{Fitzpatrick:2007} extinction law. We report the data source, filters, photometry, uncertainties and references in Appendix \ref{sec: phot}.

To complement the existing archival data and fill out the host SEDs, we observed the fields of 10 FRBs with the Goodman High Throughput Spectrograph on the 4-m Southern Astrophysical Research Telescope (SOAR; PIs Fong, Gordon; \citealt{SOAR}); GMOS on the 8-m Gemini South Telescope (PI Tejos; \citealt{Gemini}); the Low Resolution Imaging Spectrometer (LRIS) on the 10-m Keck I Telescope (PI Nugent; \citealt{LRIS}); the MMT and Magellan Infrared Spectrograph (MMIRS) on the 6.5-m MMT (PI Nugent; \citealt{MMIRS}); and, the FOcal Reducer and low dispersion Spectrograph 2 (FORS2; \citealt{FORS2}) and High Acuity Wide-field K-band Imager (HAWK-I; \citealt{HAWKI}) on the 8-m Very Large Telescope (VLT; PIs Macquart, Shannon).  Details of these observations, including the observation dates, filter, and corresponding magnitude, are reported in Table~\ref{tab:imaging}. For six FRBs, we duplicate filters used in archival observations to increase the S/N of the detected host, or perform deeper searches if they were previously undetected. 

We reduced the new Gemini, Keck, and MMT imaging data with the \texttt{POTPyRI}\footnote{https://github.com/CIERA-Transients/POTPyRI} pipeline. \texttt{POTPyRI} creates master bias, master dark, and master flat frames (depending on the types of calibrations available). These calibrations are applied to the science frames before alignment and stacking. A World Coordinate System (WCS) is calculated then applied by calibrating to the Gaia DR3 catalog \citep{Gaia,GaiaDR3}. In the case where automatic WCS alignment fails (i.e., the RMS of the astrometric fit is $\geq$ 0.5\arcsec), we perform manual WCS alignment by interactively matching sources detected in the science image with known counterparts in either the Gaia DR3 (for optical images) or 2MASS (for NIR images) catalogs. Once the RMS of the astrometric fit is on order 0.2\arcsec, we calculate the zero-point and proceed with PSF photometry as described above. 

For the SOAR data, we use the \texttt{photpipe} pipeline \citep{Rest+05} for reduction. This pipeline performs bias correction and flat-fielding and aligns the WCS against the Gaia DR3 catalog. The science frames are then sky subtracted, stacked, and regridded to a common pixel scale and field center using \texttt{SWarp} \citep{swarp}.  PSF photometry is performed on all point sources in the stacked image using a custom version of {\tt DoPhot}.  Finally, the pipeline calculates the zero-point of the final calibrated science image by comparing the instrumental PSF magnitudes to SkyMapper DR2 and Pan-STARRS DR2 standard stars using the same methods described above. We then applied these zero-points to aperture magnitudes obtained using {\tt photutils}. The VLT data were reduced and further processed using the procedure detailed in \citet{Bhandari_210117}. 

\begin{figure*}
    \centering
\includegraphics[width=0.245\textwidth]{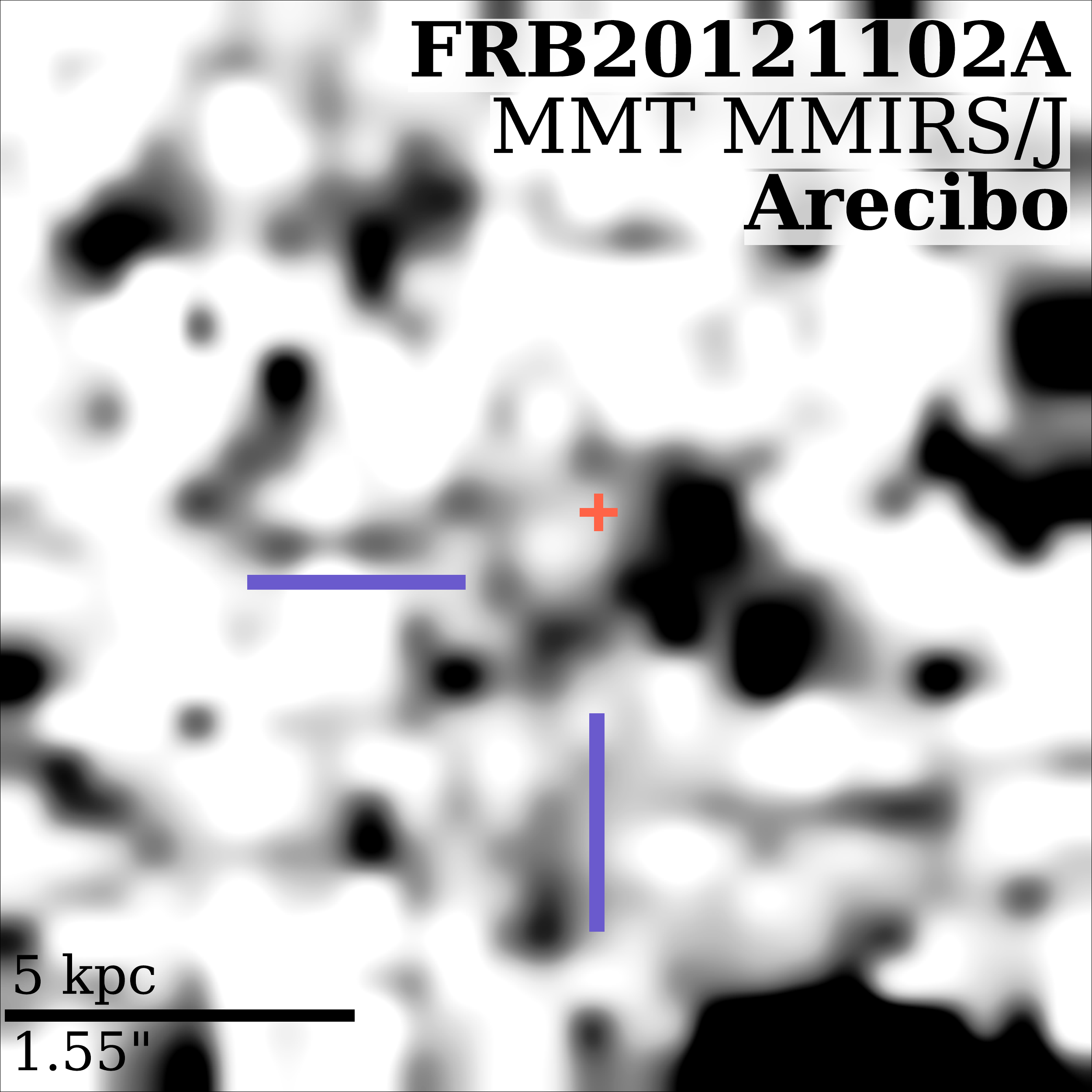}
\includegraphics[width=0.245\textwidth]{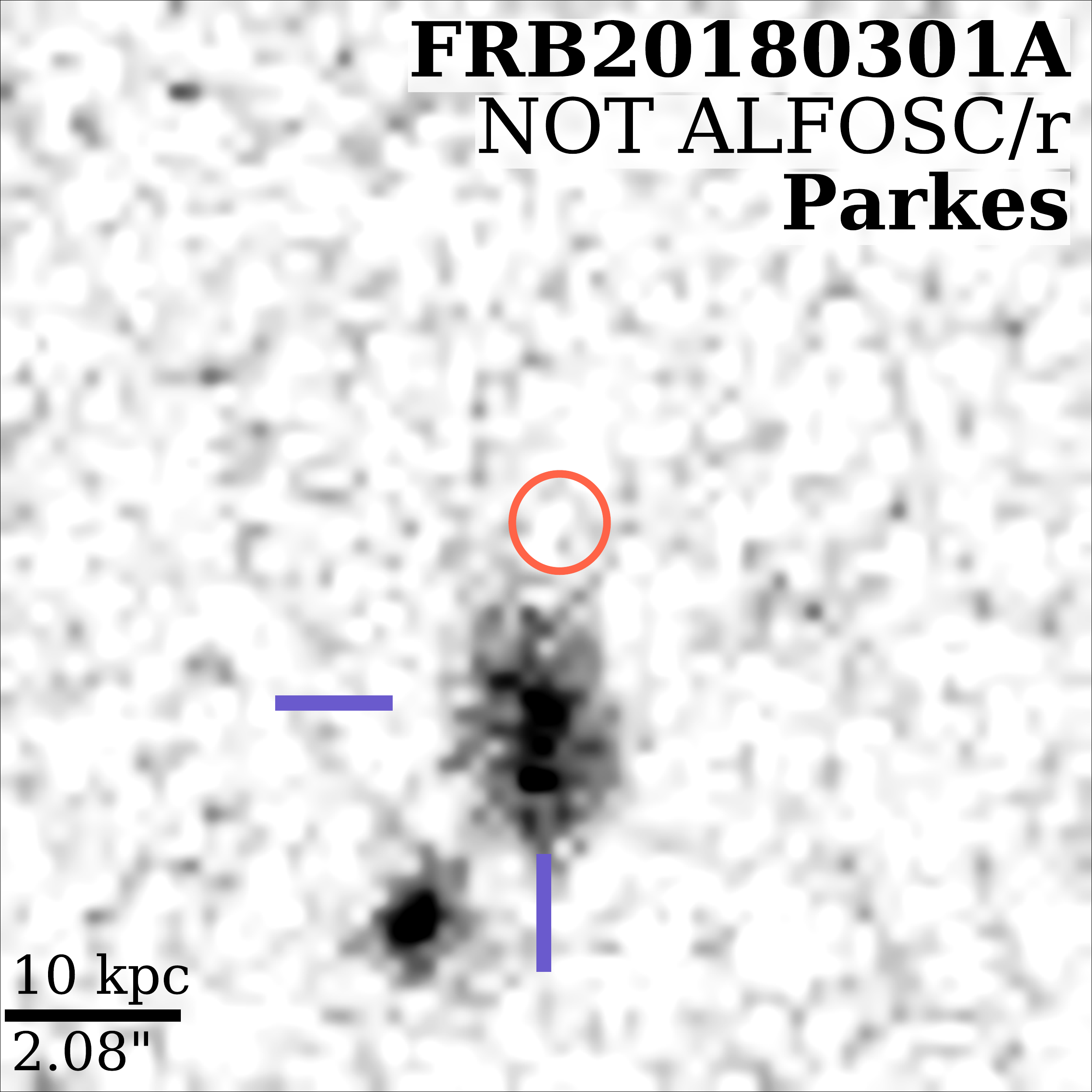}
\includegraphics[width=0.245\textwidth]{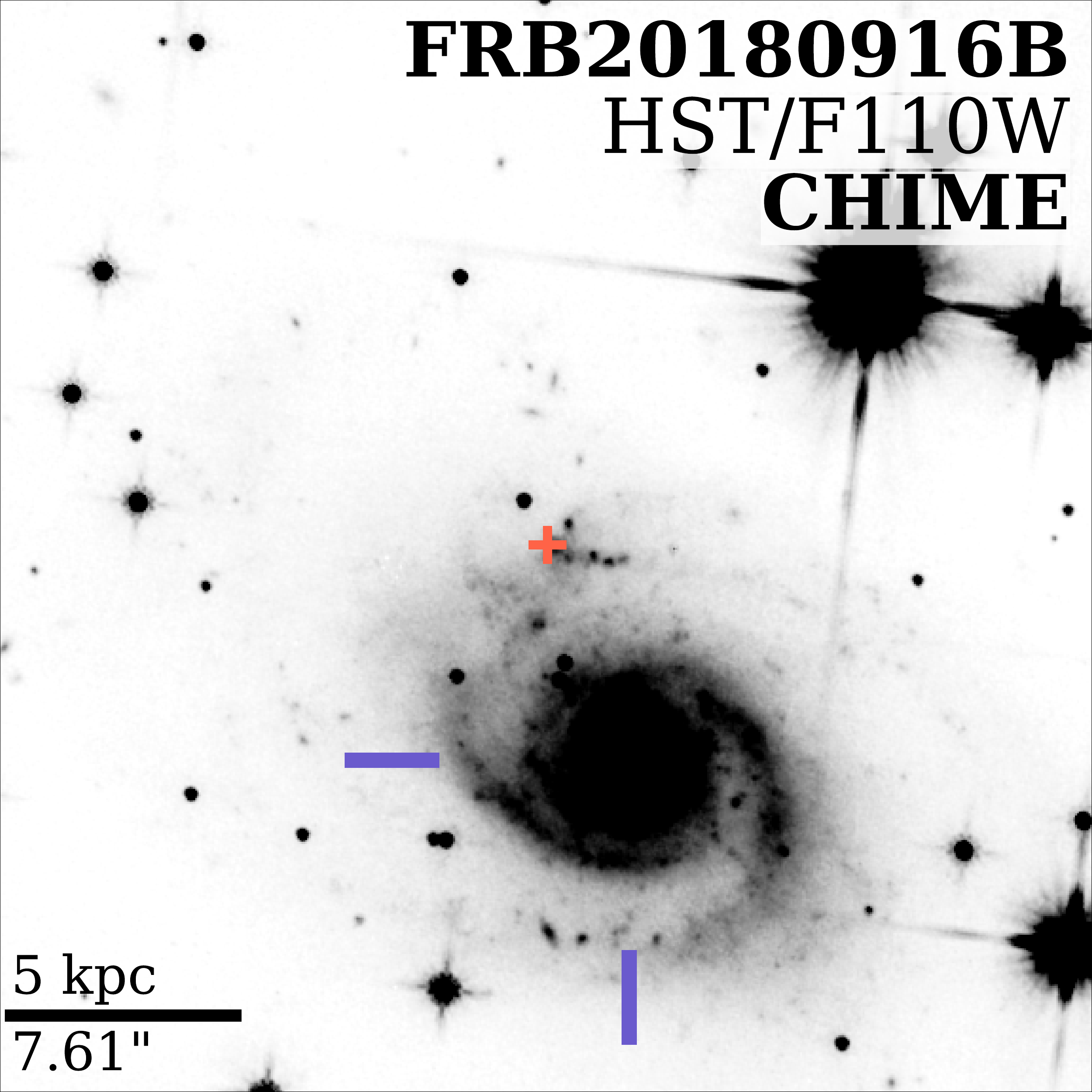}
\includegraphics[width=0.245\textwidth]{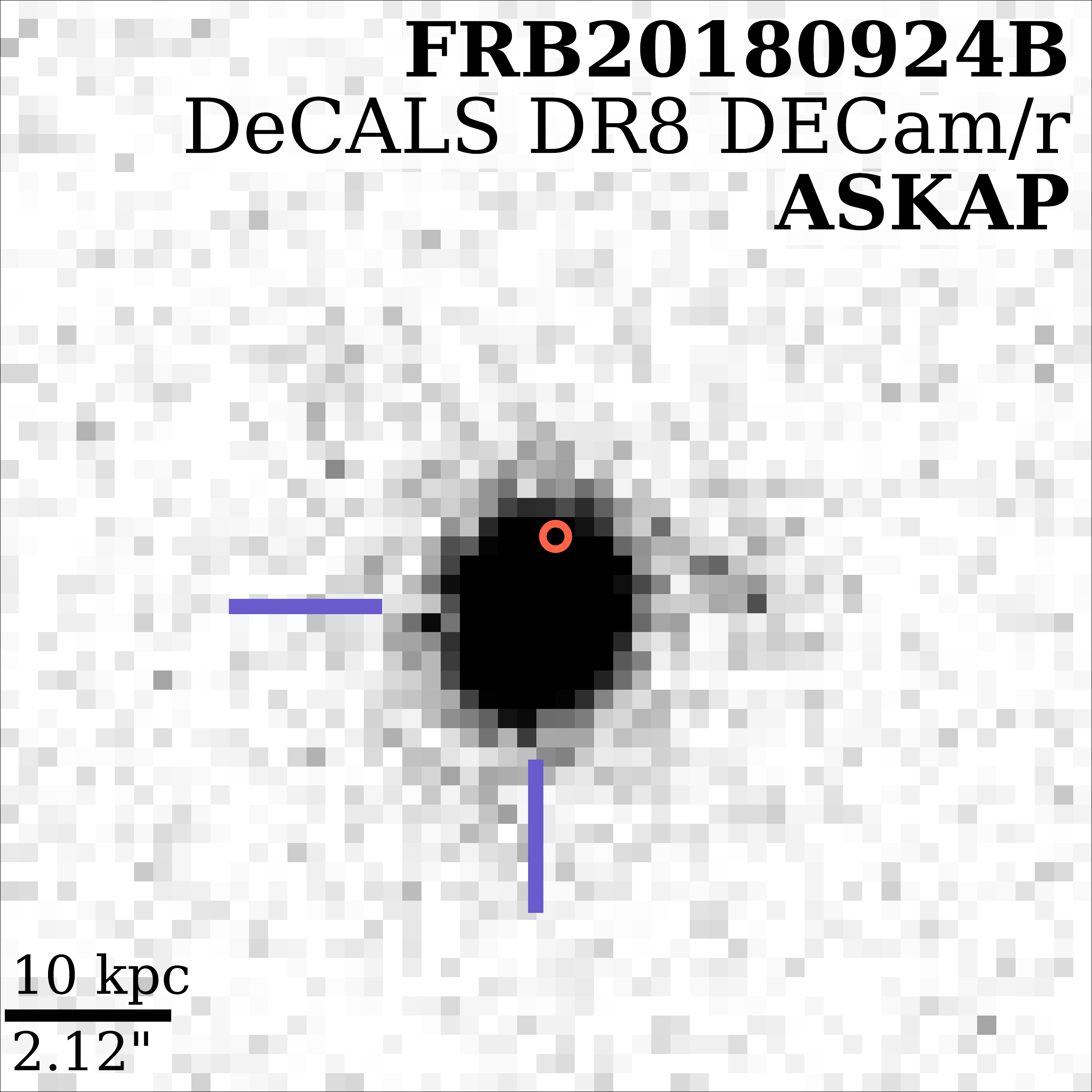}
\includegraphics[width=0.245\textwidth]{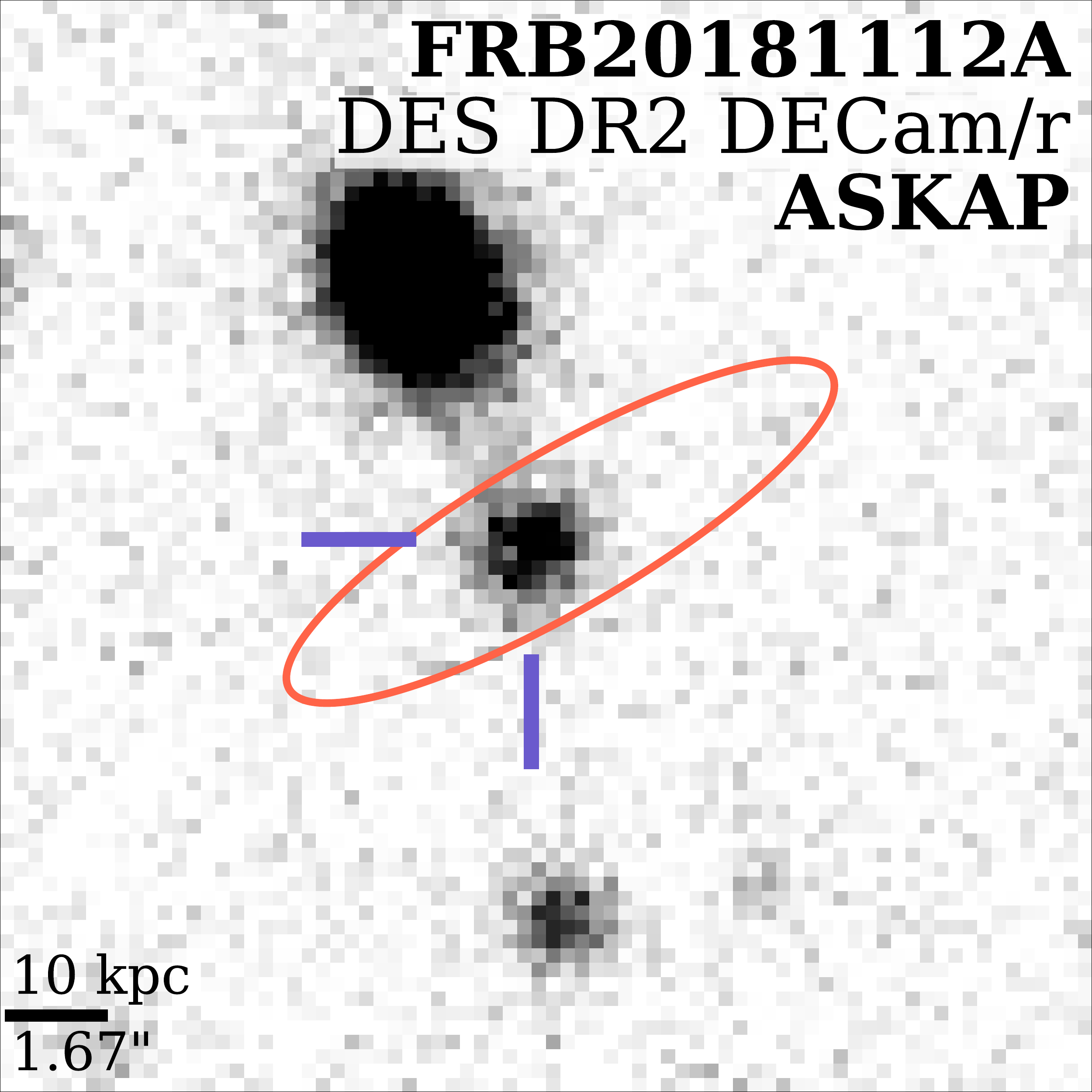}
\includegraphics[width=0.245\textwidth]{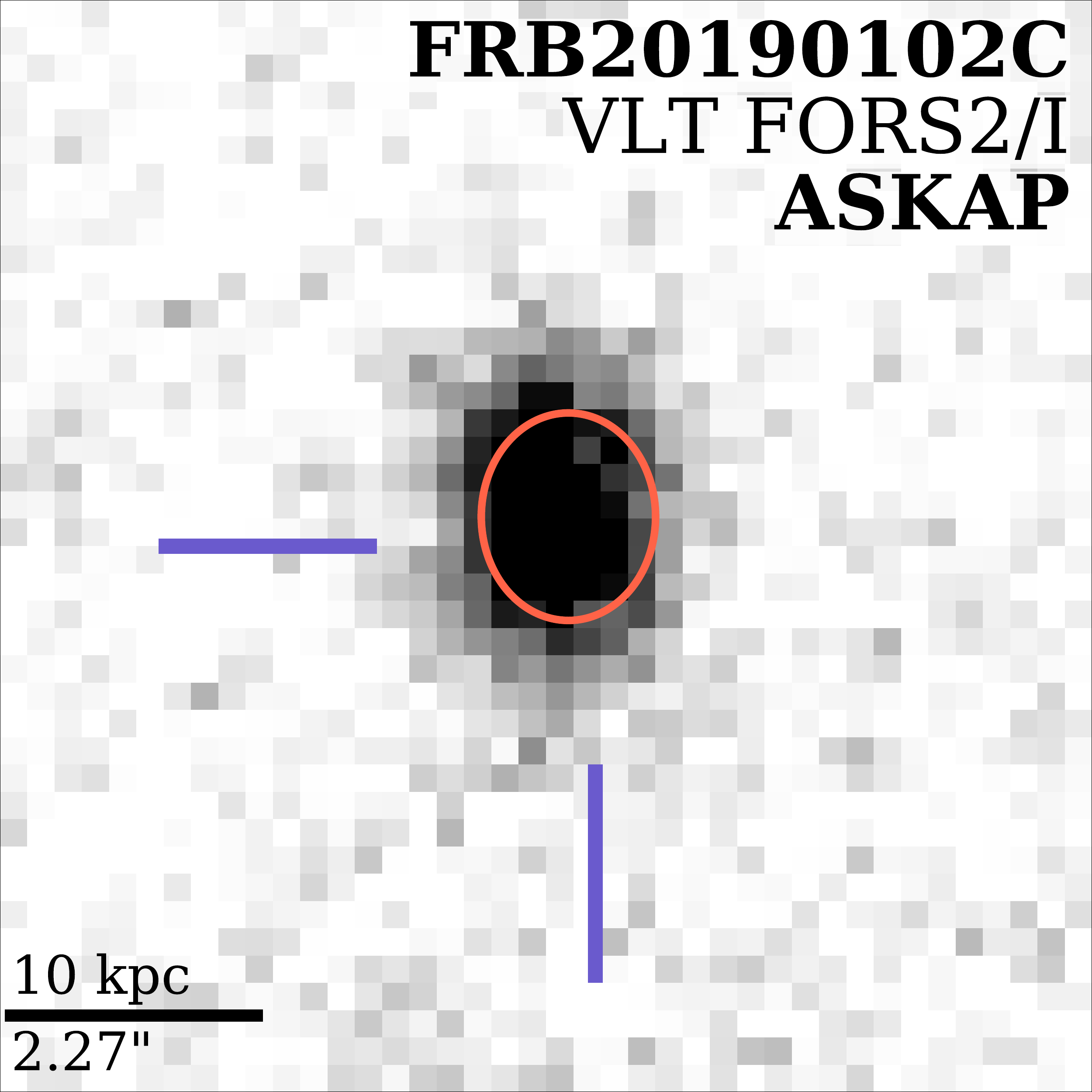}
\includegraphics[width=0.245\textwidth]{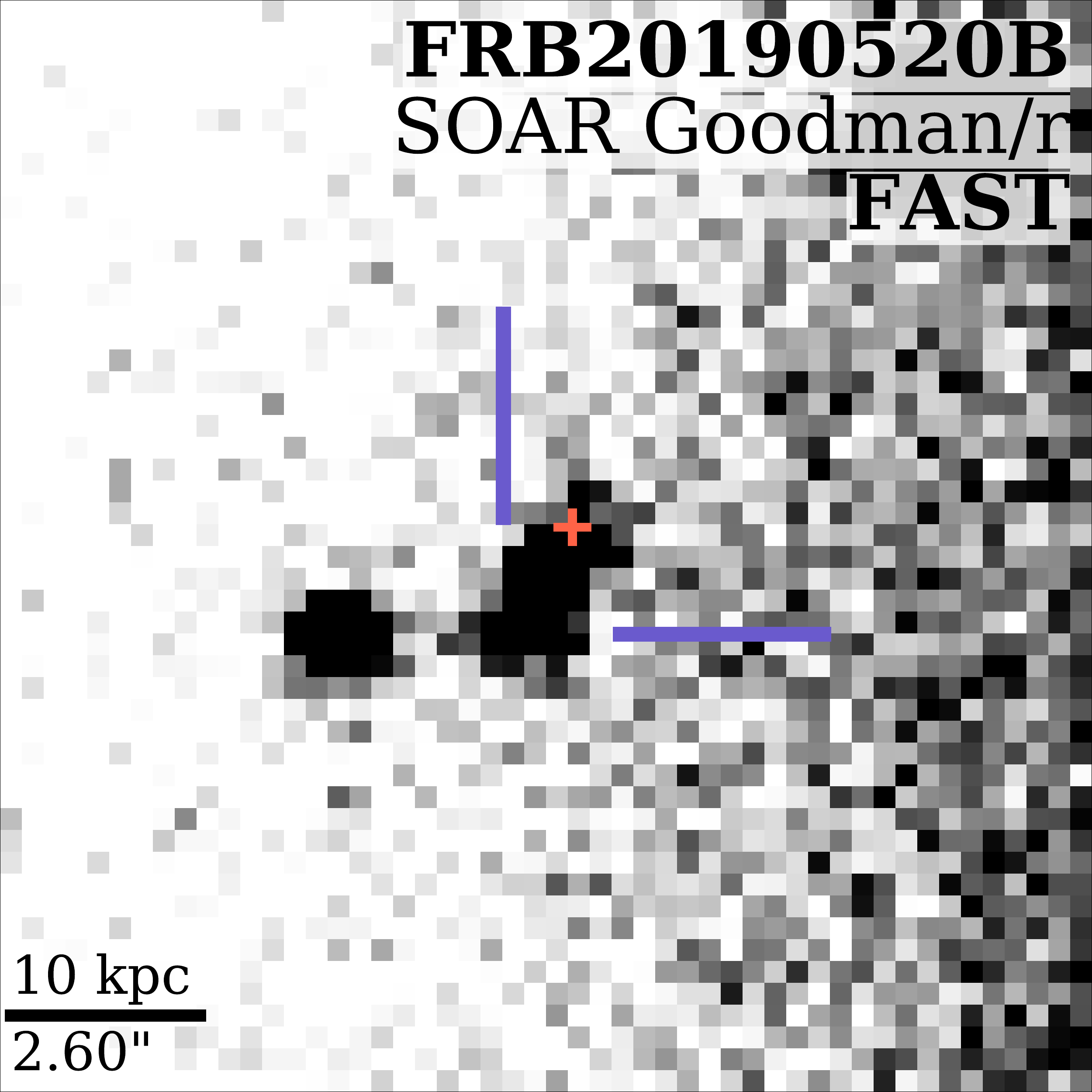}
\includegraphics[width=0.245\textwidth]{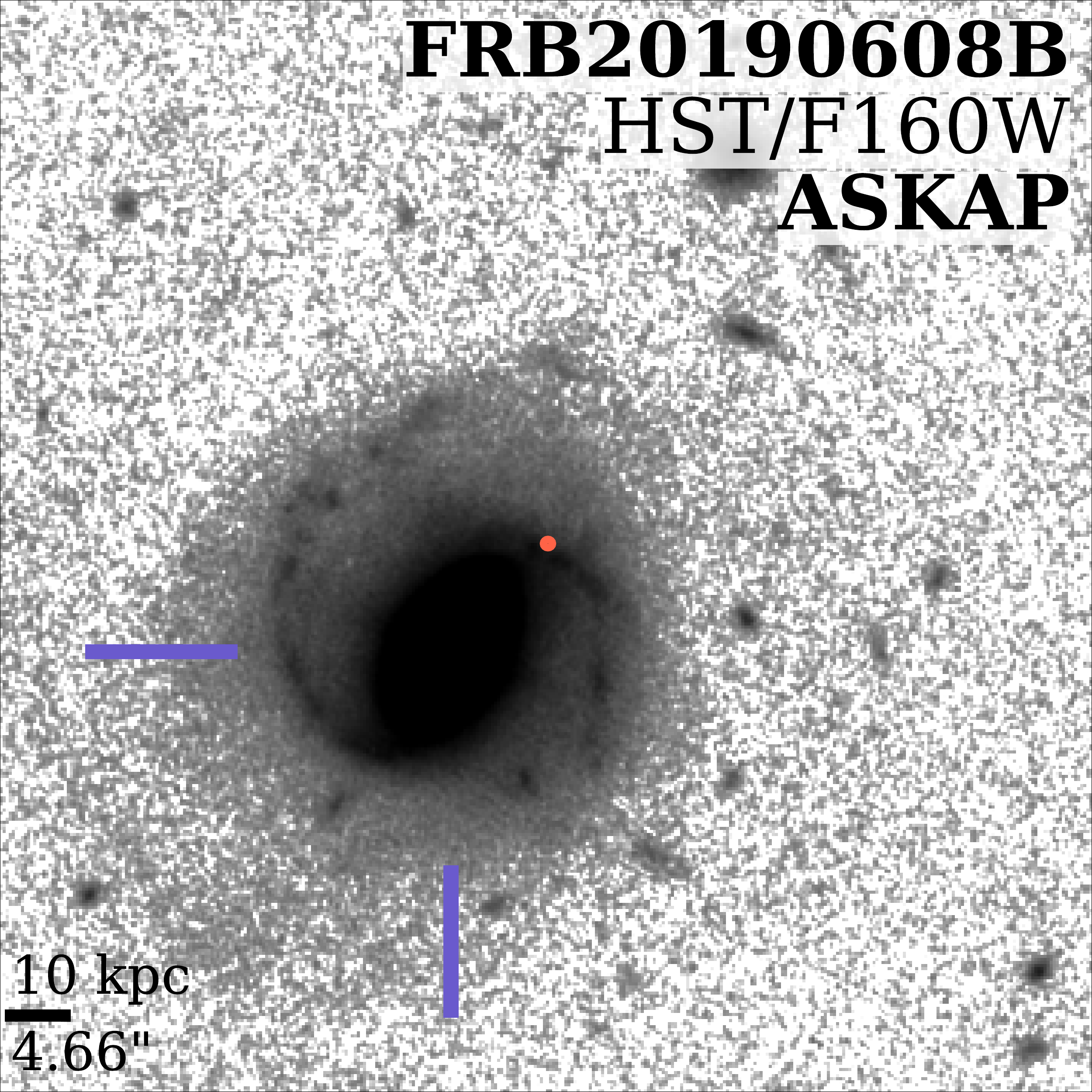}
\includegraphics[width=0.245\textwidth]{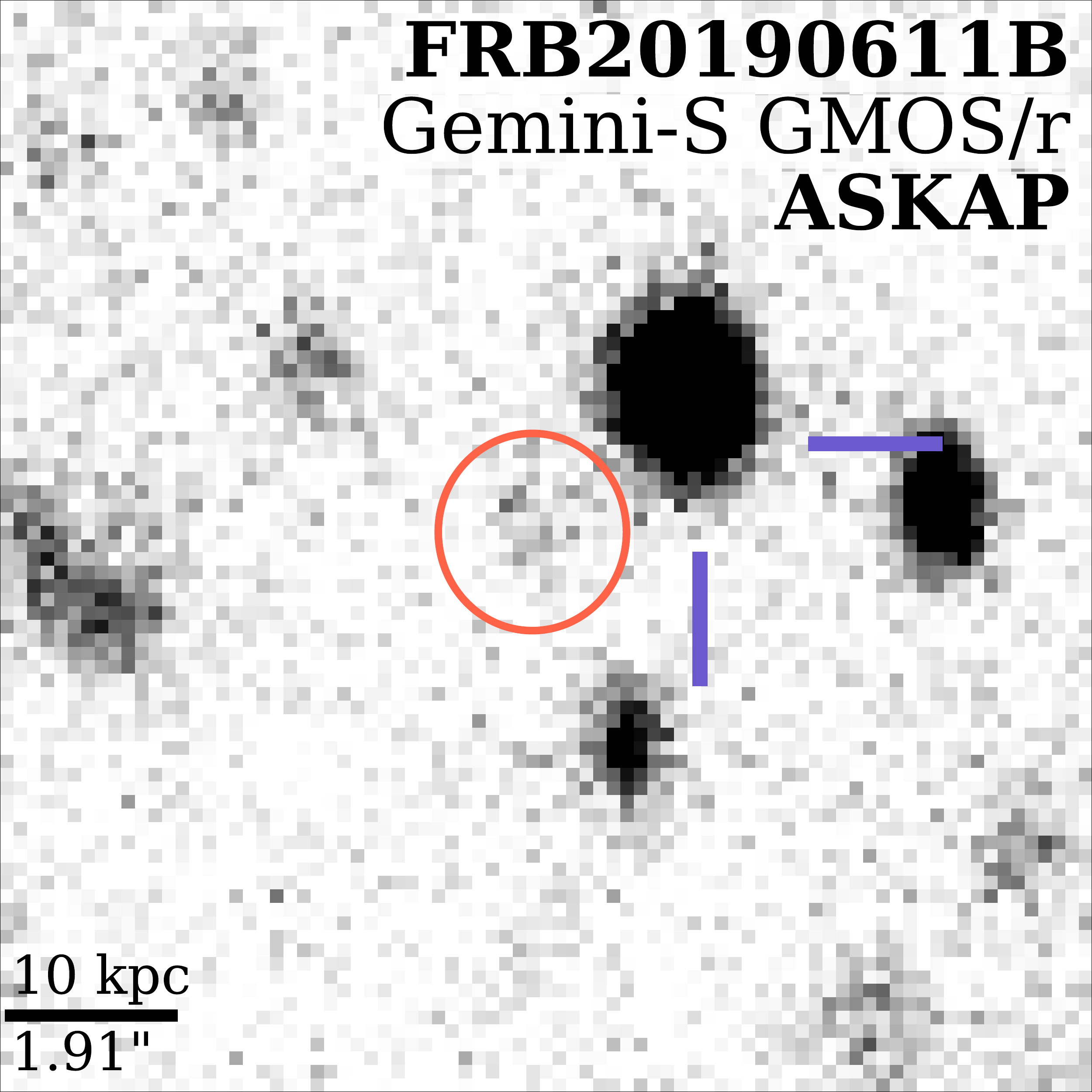}
\includegraphics[width=0.245\textwidth]{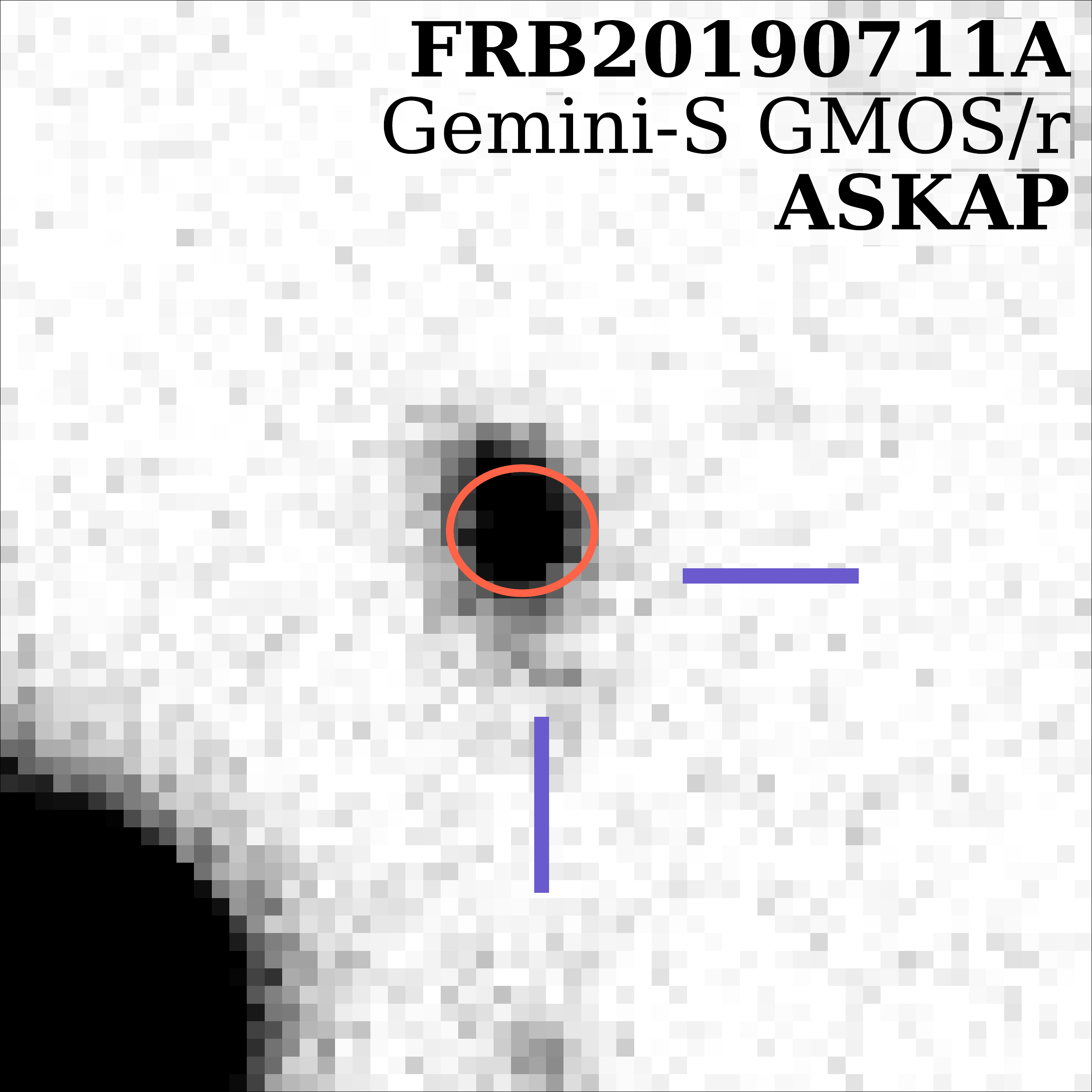}
\includegraphics[width=0.245\textwidth]{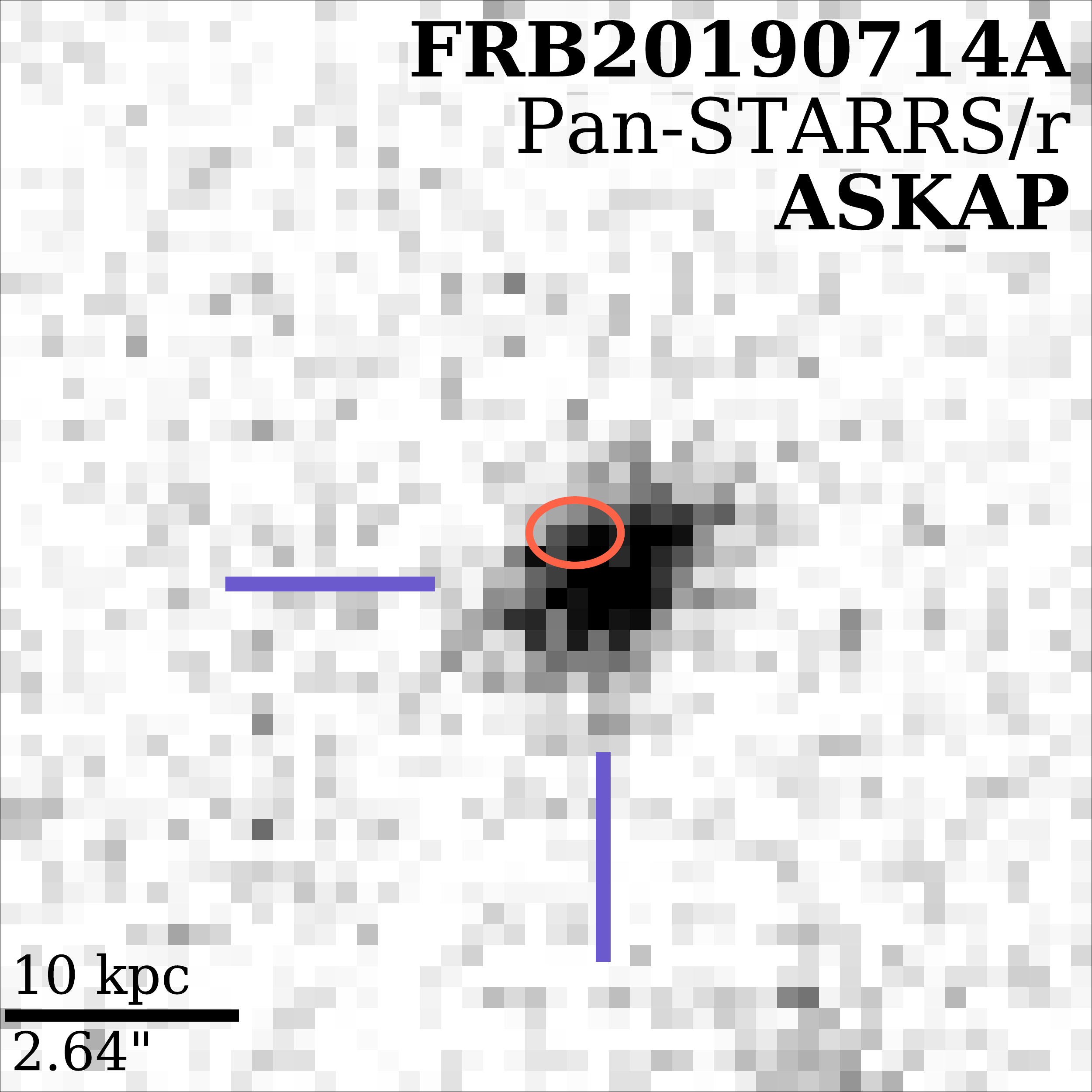}
\includegraphics[width=0.245\textwidth]{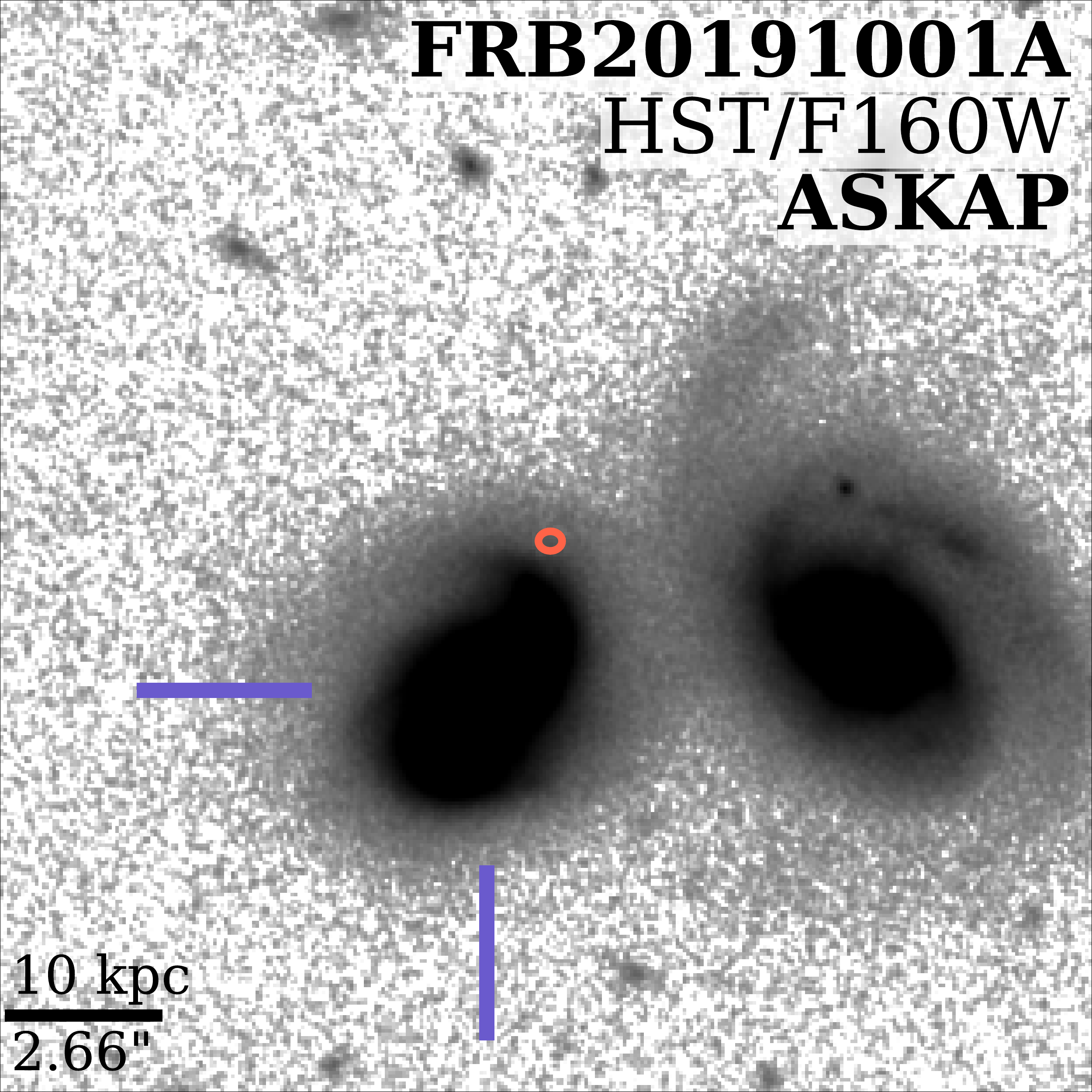}
\includegraphics[width=0.245\textwidth]{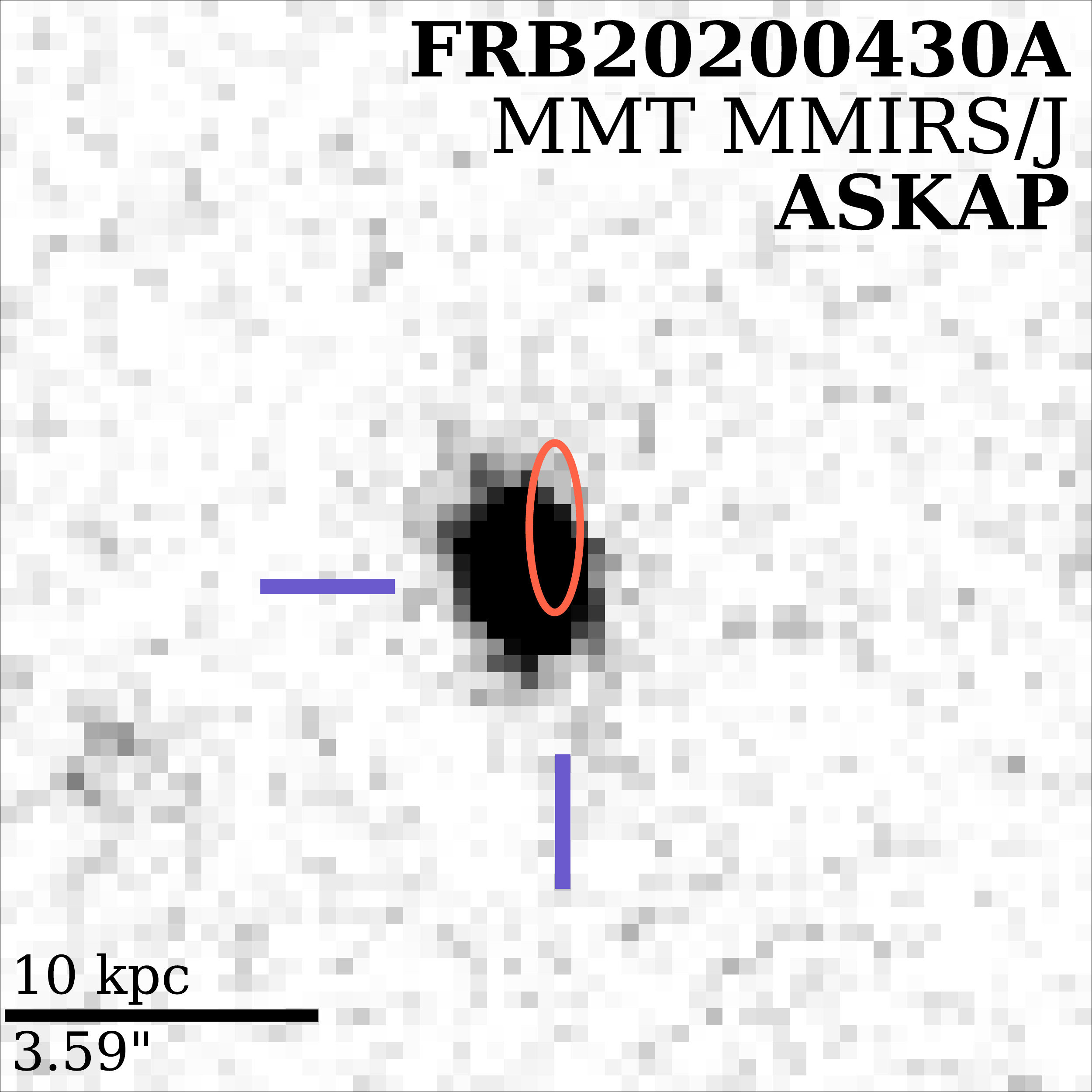}
\includegraphics[width=0.245\textwidth]{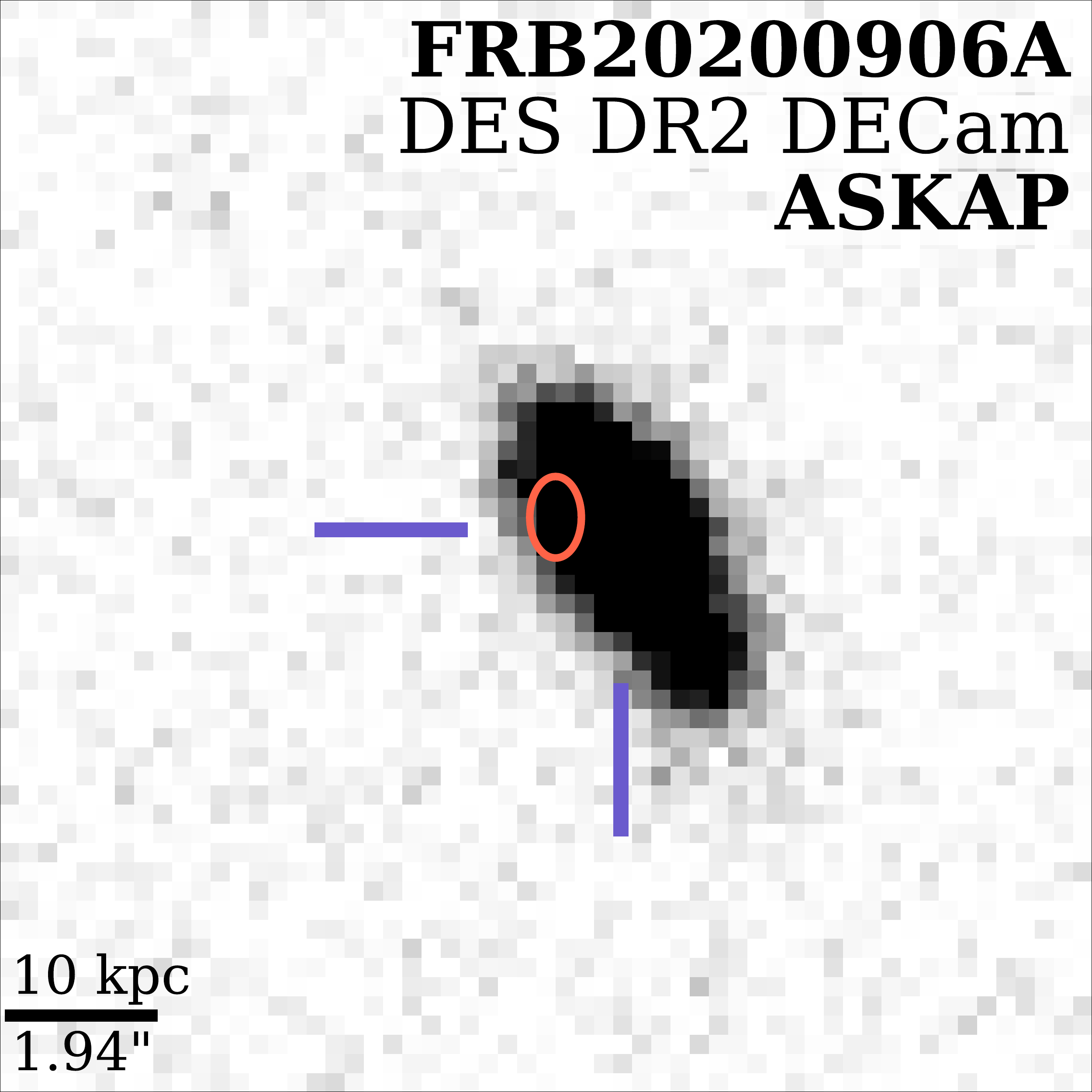}
\includegraphics[width=0.245\textwidth]{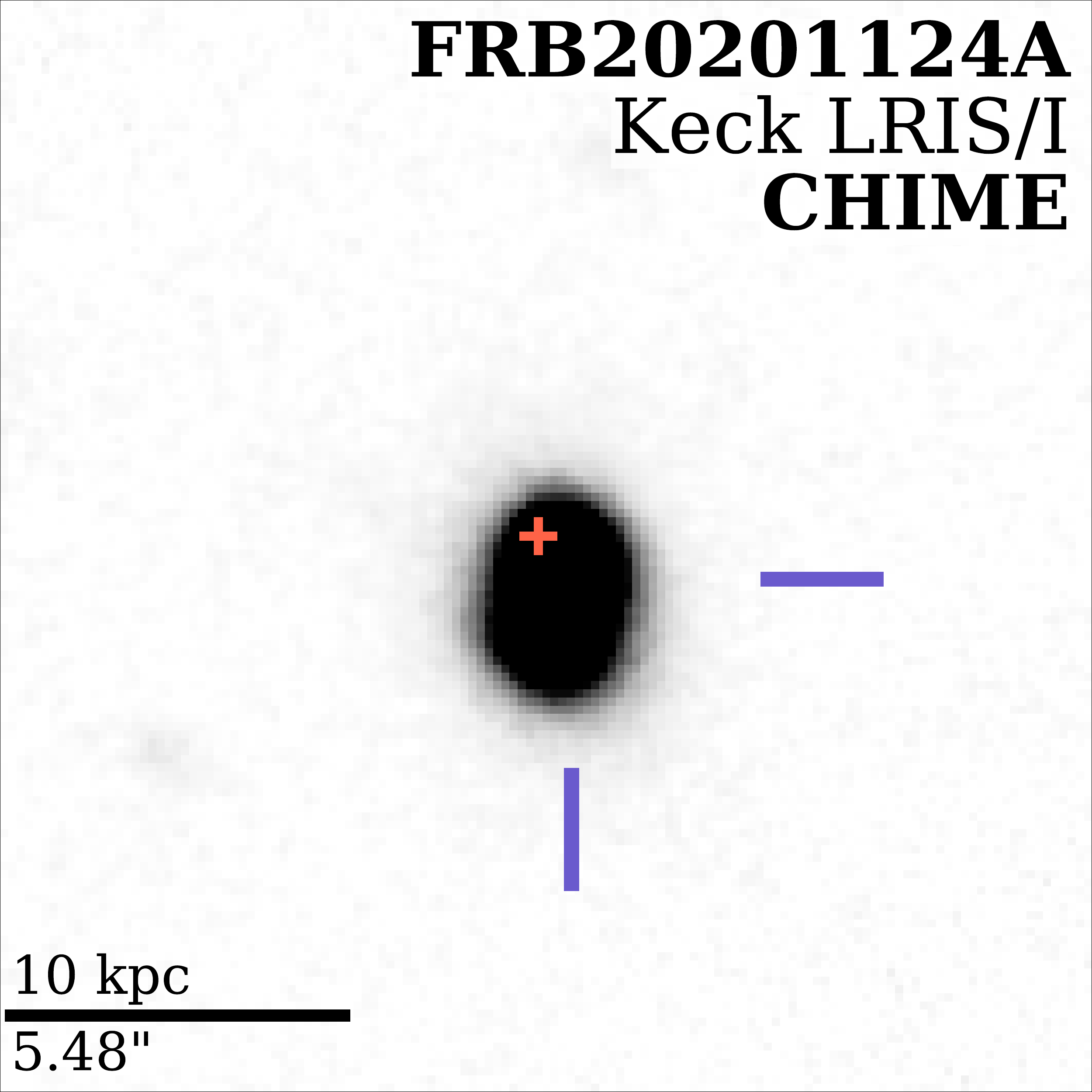}
\includegraphics[width=0.245\textwidth]{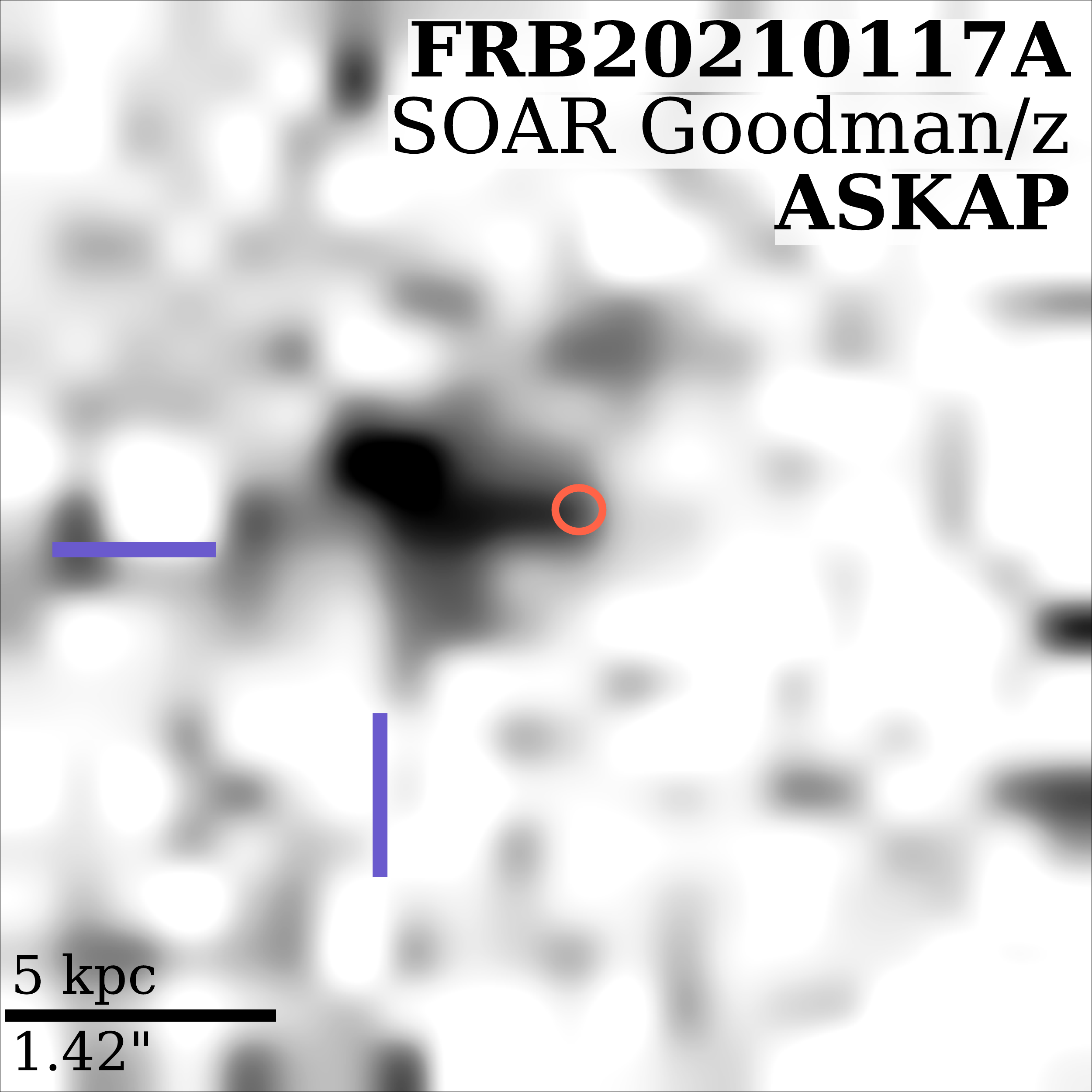}
    \caption{Imaging gallery of the 23 FRB hosts included in our sample, with images oriented North up and East to the left. The host galaxies are marked by the purple crosshairs and the 68\% confidence interval FRB localizations are denoted in red by an ellipse. For the three FRBs with milliarcsecond-scale localizations (FRB\,20121102A, \citealt{Marcote+17}; FRB\,20180916B, \citealt{Marcote_etal_2020}; and FRB\,20201124A, \citealt{Nimmo+22}) and one with $\sim 0.1\arcsec$ localization (FRB\,20190520B; \citealt{Niu+22})}, the position is indicated by a plus sign. The facility or survey which discovered the FRB is also listed.
    \label{fig:imaging gallery}
\end{figure*}

We show representative images of all FRB hosts in the sample, including new observations and images from the literature \citep{Heintz+20,mpm+20,Mannings+21,Tendulkar+21,Bhandari_210117,bha+22,Caleb+23}, in Figure~\ref{fig:imaging gallery}. In total, we present 29 new photometric measurements from our own imaging and 53 measurements from archival imaging. Finally, we collect 114 published photometric values from the literature and archives for 17 FRBs. These values and their references are listed in Table \ref{tab:imaging}.

For FRB\,20190520B, we aligned the SOAR/Goodman field of view so that the bright, nearby star would not land on the detector. We then reduced the data with \texttt{photpipe} following the procedure described above. However, even with the care taken to avoid excess flux from the neighboring star, the background surrounding the host was still significantly greater than the rest of the field. In order to accurately calculate the photometry of the host, we applied a more complex background model to our photometry code\footnote{https://github.com/charliekilpatrick/photometry}, assuming a spatially varying one-dimensional background that varies with the $x$, $y$, and $x*y$ pixel value from the center of the host galaxy. We then placed a 2.0\arcsec\ aperture around the host galaxy and all point-like sources of emission close to FRB\,20190520B. The resulting values are within 0.1-0.3 AB magnitudes of the values derived from the original photometry script, but we proceed with the complex-background subtracted values in our modeling as these are more representative of the true brightness of the host.

\renewcommand{\thefigure}{\arabic{figure} (Cont.)}
\addtocounter{figure}{-1}

\begin{figure*}[!ht]
    \centering
\includegraphics[width=0.245\textwidth]{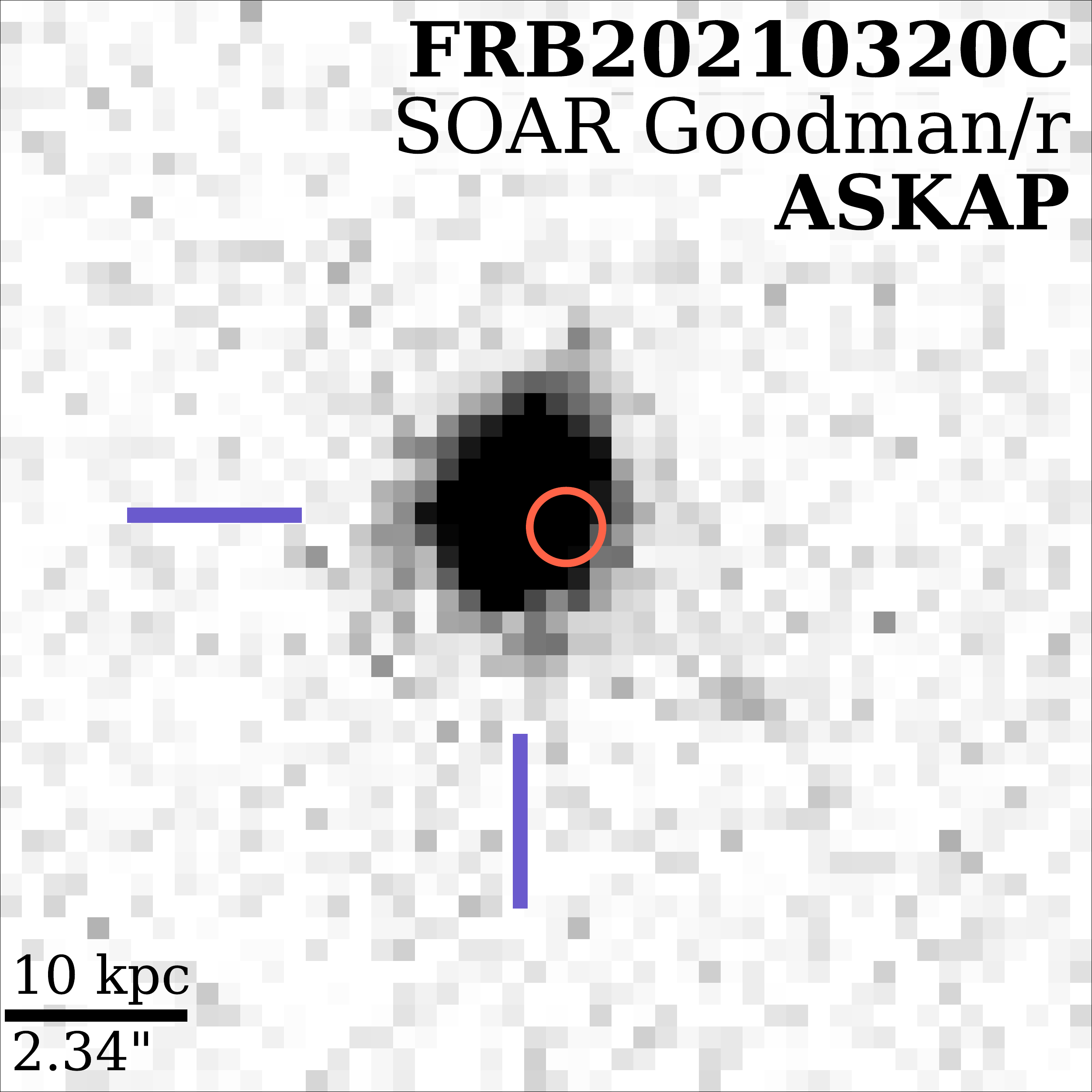}
\includegraphics[width=0.245\textwidth]{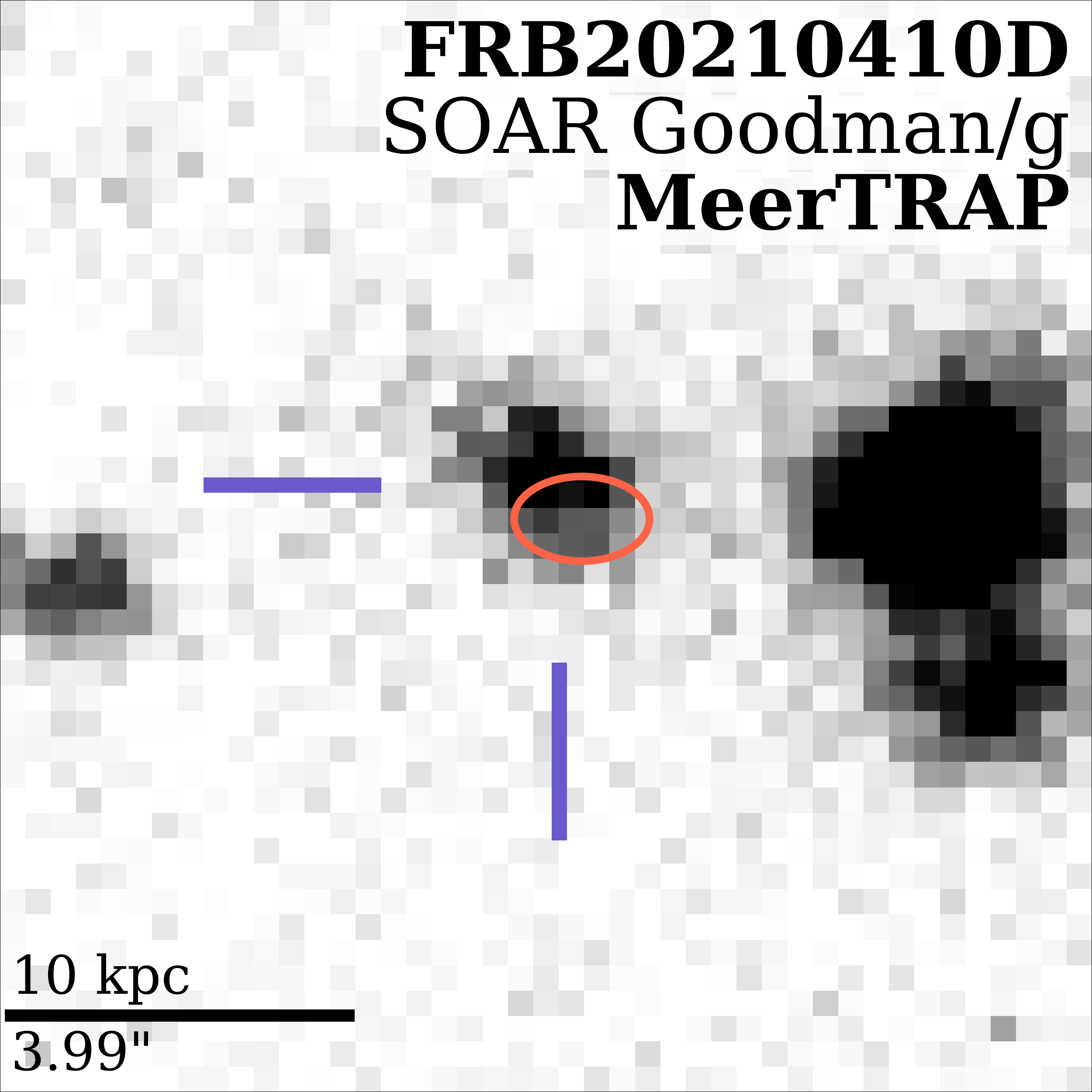}
\includegraphics[width=0.245\textwidth]{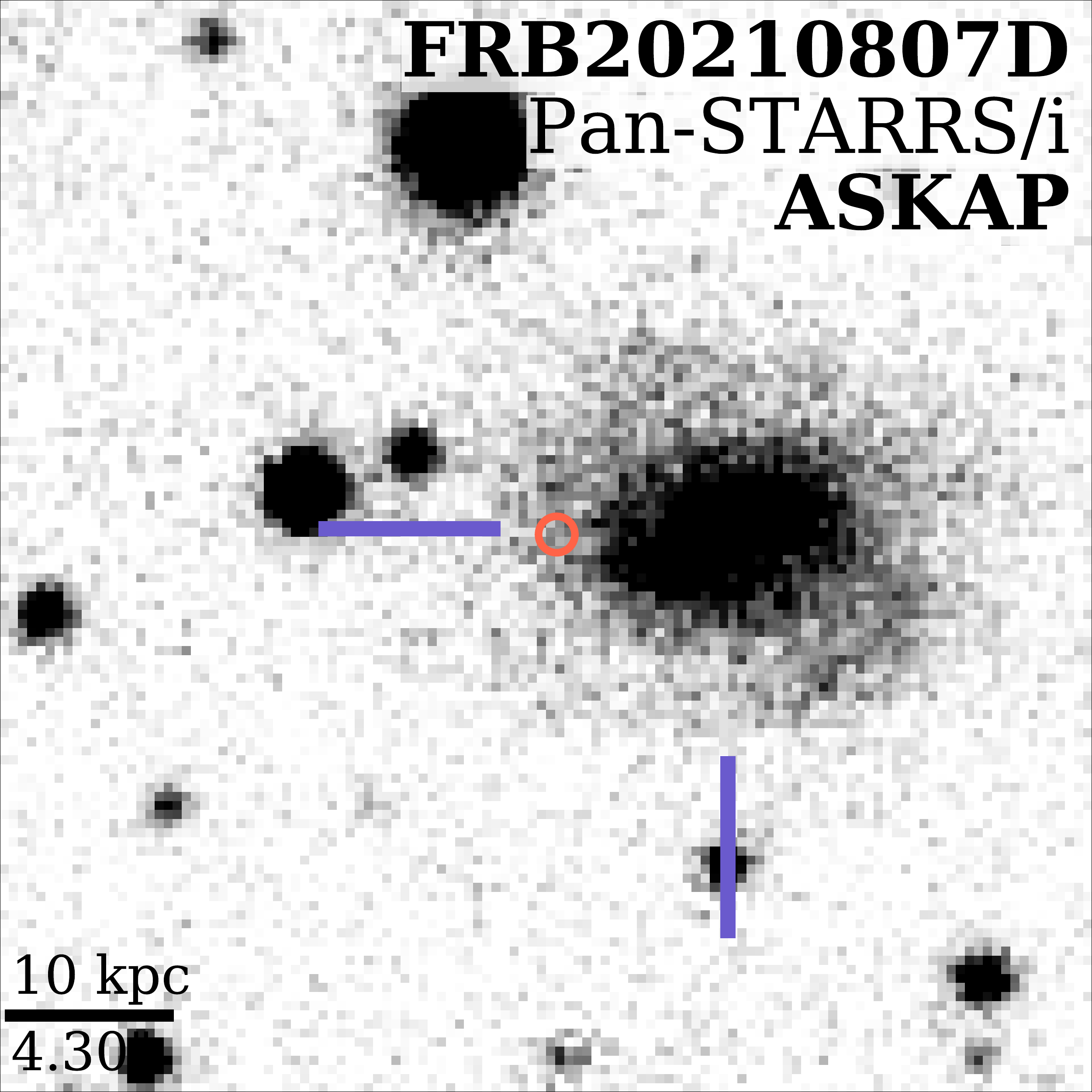}
\includegraphics[width=0.245\textwidth]{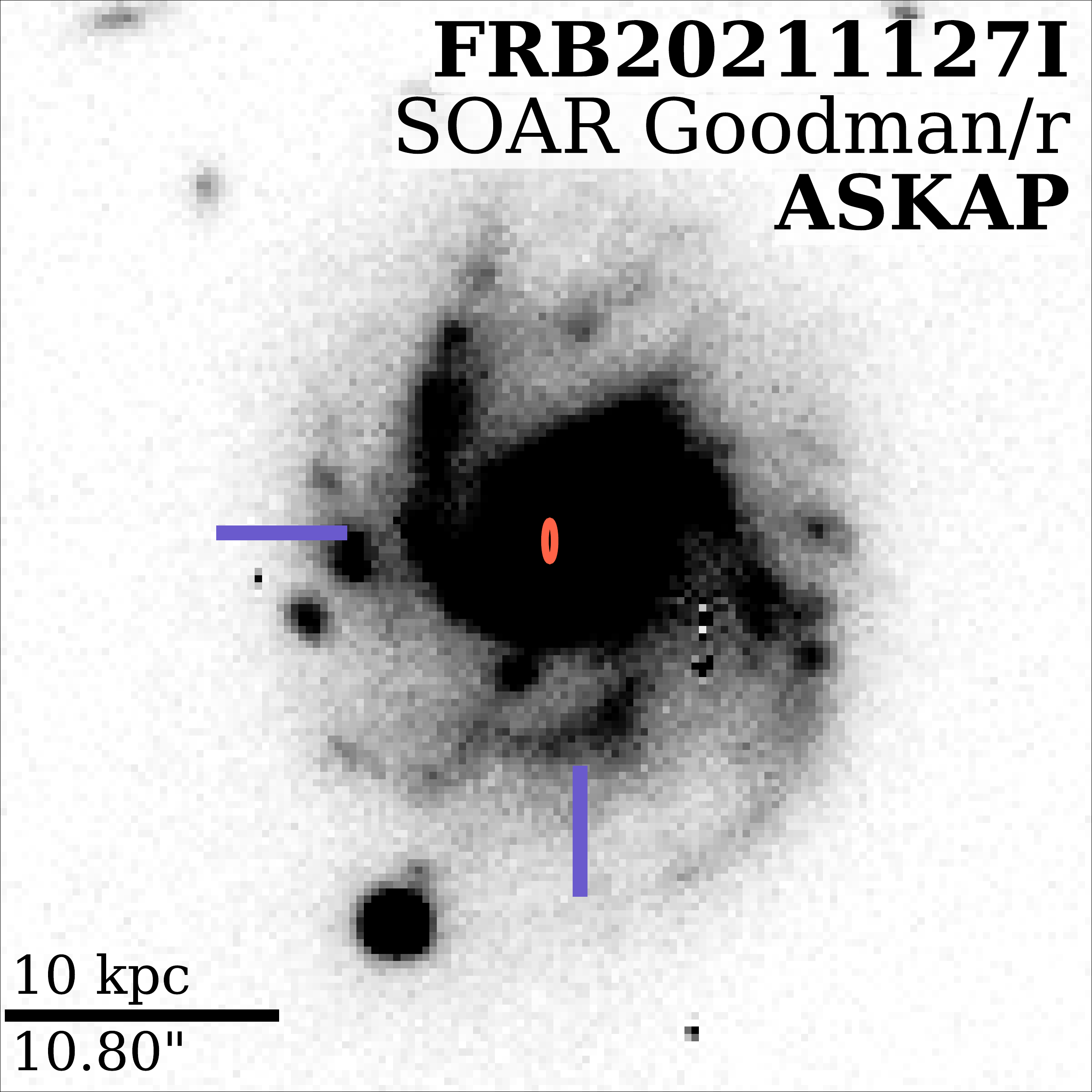}
\includegraphics[width=0.245\textwidth]{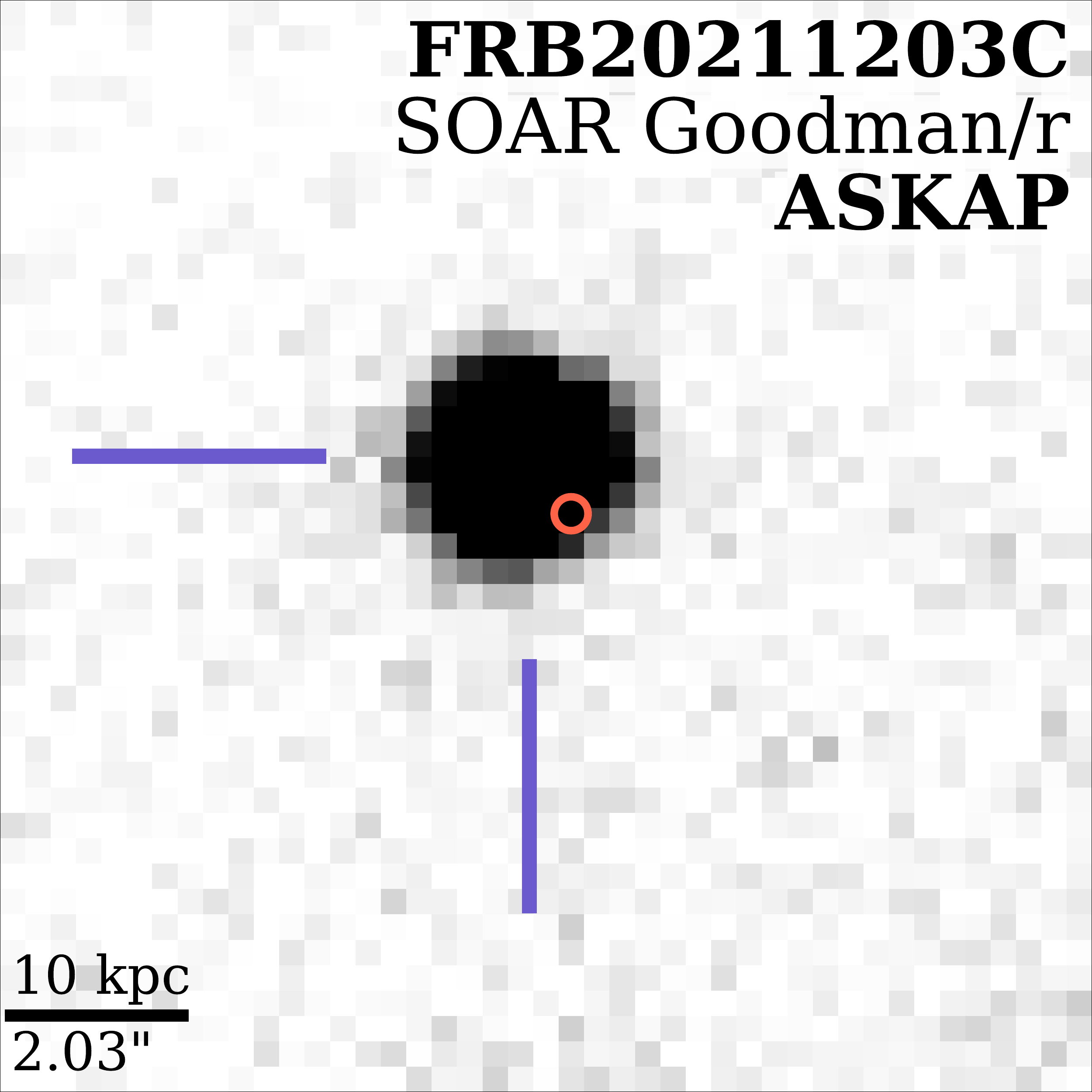}
\includegraphics[width=0.245\textwidth]{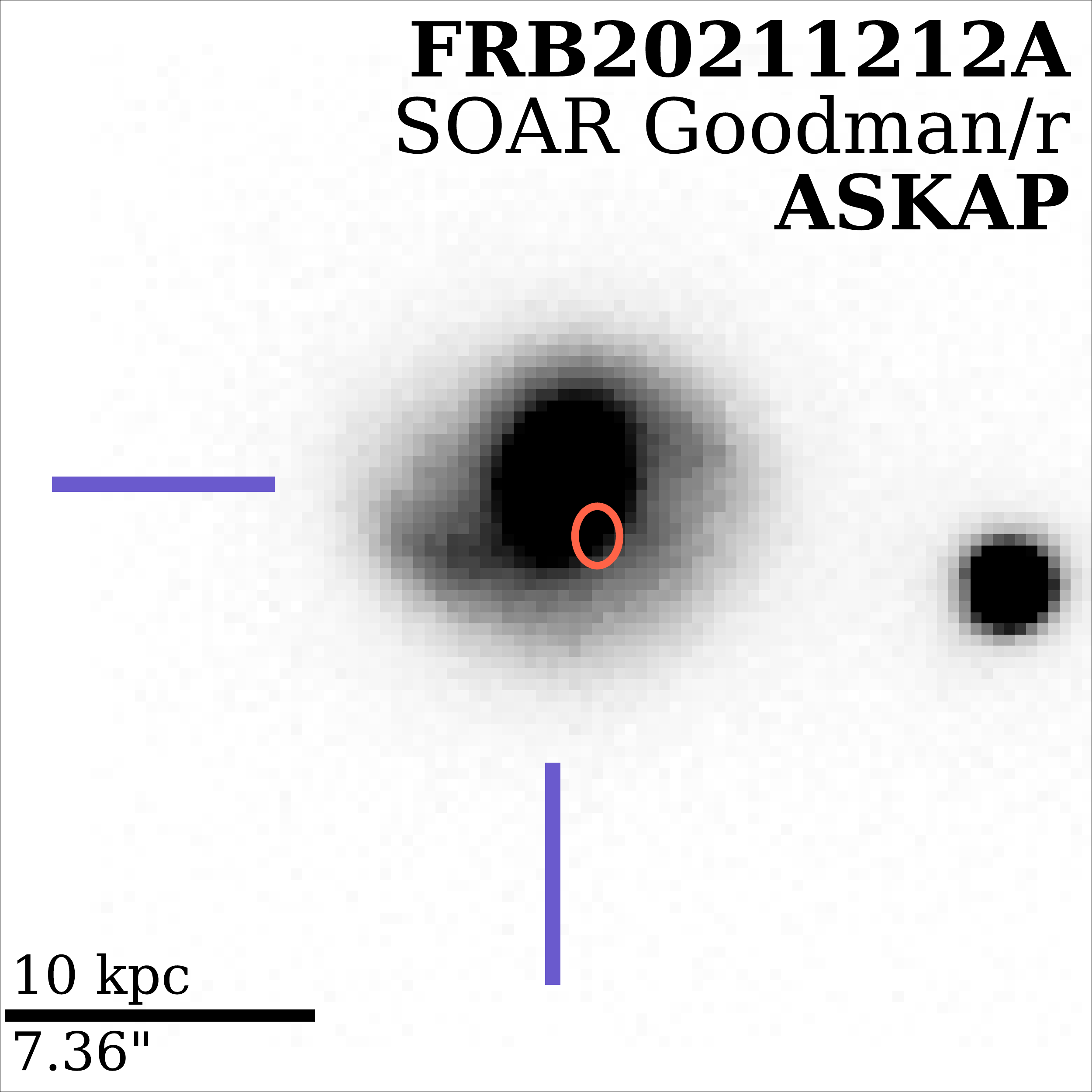}
\includegraphics[width=0.245\textwidth]{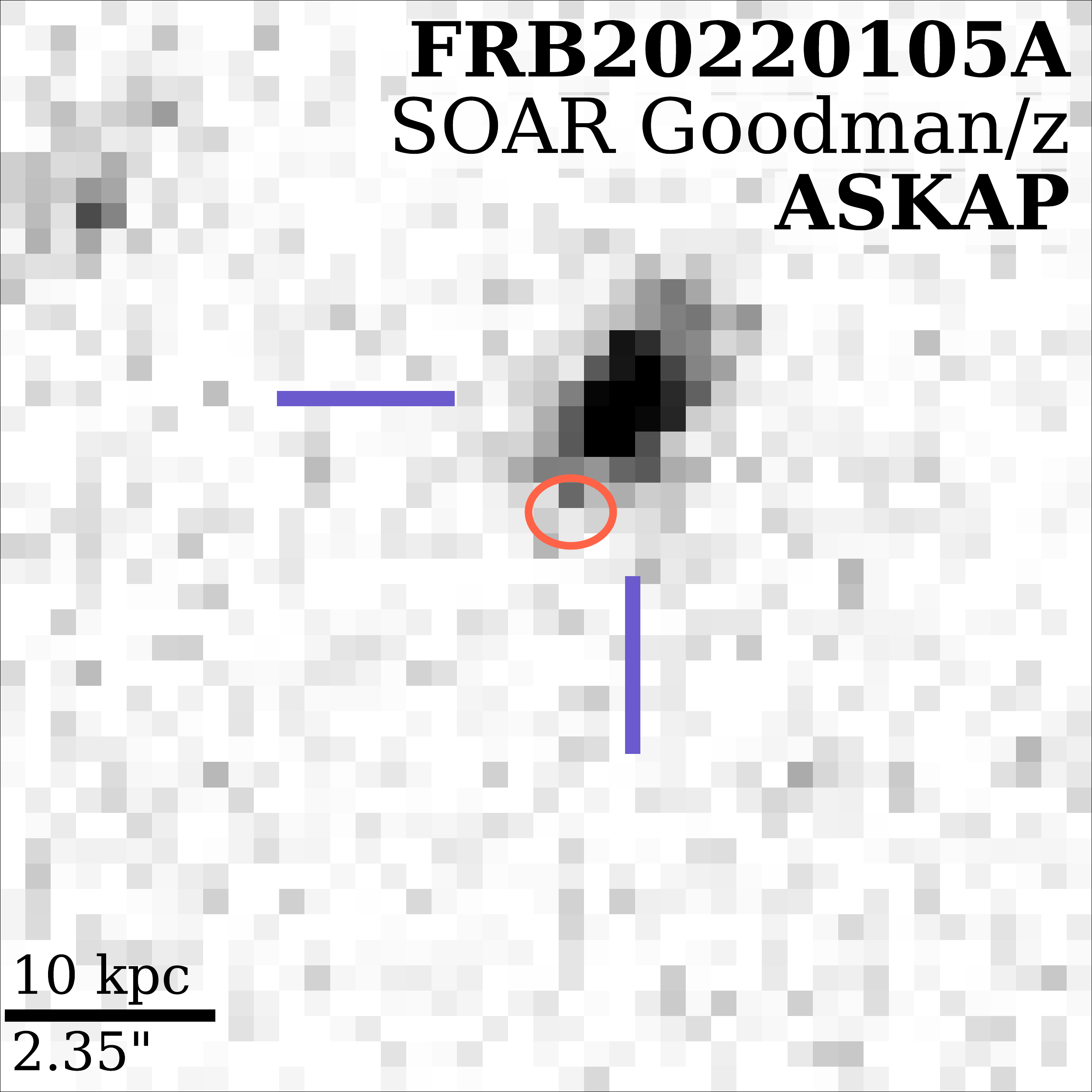}
    \caption{}
\end{figure*}

\renewcommand{\thefigure}{\arabic{figure}}

\subsection{Spectroscopy} \label{sec:spectra}

We obtained long-slit optical spectroscopy for the host galaxies of seven FRBs (20180916B, 20210320C, 20210807D, 20211127I, 20211203C, 20211212A, and 20220105A) and one mask observation for the field of FRB\,20190520B (with one of the slits centered on the host galaxy center, used in this work). The observations were taken with Keck/LRIS (PIs Miller, Blanchard; Programs O304, O314), Keck/DEIMOS (PI Blanchard; Programs O287, O300), and SOAR/Goodman (PI Fong; Programs SOAR2021A-010, SOAR2021B-002). We designed the observations to span wavelengths of $\lambda \approx$ 5000-10000\AA.  We list this new spectroscopy, along with the gratings/grisms, and slit widths for each data set in Table~\ref{tab:spectra}.

For our subsequent host galaxy modeling (Section~\ref{sec:prospector}), we require a S/N $\gtrsim$ 3/\AA\ in the continuum across most of the wavelength range. The slits were oriented to cover the center of the host galaxy and the center of the FRB position to search for possible anomalous emission at the FRB location. We manually inspected all of the spectra and found no additional continuum or line-emission detected at the FRB positions. 

For the data taken with Keck, we obtained calibrations including arc lamps (Hg, Ne, Ar, Cd, Zn, Kr, Xe, or a subset of these), flat fields, bias frames, and spectrophotometric standard star spectra taken on the same night as the science data. For the SOAR data, we also obtained flat-field and arc lamp spectra (Hg, Ar, Ne) on the same night as the science data and at a similar airmass. In order to flux calibrate the SOAR data, we used spectrophotometric standards from the SOAR archive\footnote{\url{https://archive.lco.global}} that used the same spectral set-up and were observed as close to the date of the science observations as possible. 

We reduced these data using the Python Spectroscopic Data Reduction Pipeline (\texttt{PypeIt}; \citealt{pypeit:joss_pub, pypeit:zenodo}). \texttt{PypeIt} performs bias-subtraction, flat-fielding, cosmic ray masking, and wavelength-calibration of the raw frames. After the initial processing to generate a calibrated 2D spectral image, the pipeline extracts 1D spectra using the standard Horne algorithm \citep{Horne1986}. At this step, we manually inspected the quality of the calibrated 1D and 2D spectra; in some cases, we implemented minor changes to default parameters to improve the extraction. 

For instance, for SOAR/Goodman spectra, we include the flexure parameter \texttt{spec\char`_method = boxcar} in the wavelength calibration module to account for 
instrumental flexure. In two cases, the host galaxy was too faint to be detected in the individual frames with the default S/N threshold settings and required us to lower \texttt{snr\char`_thresh}\footnote{We note this was called \texttt{sig\char`_thresh} in earlier versions.} in the object finding module to successfully identify the host trace. Finally, for two spectra, very strong emission lines in some of the host galaxies were misinterpreted as cosmic rays and masked in the extraction. If this was found to be the case during the manual inspection, we reran the pipeline with the profile masking turned off
(\texttt{use\_2dmodel\_mask = False}). 

After extraction, we apply relative flux calibration using spectrophotometric standard spectra. If multiple exposures were taken, we reduced each exposure separately and co-added the wavelength-calibrated 1D spectra. The final error spectrum generated by the pipeline is a combination of uncertainties propagated from each calibration step and shot noise in the electron counts. We then perform telluric correction on the co-added spectra using the corresponding atmospheric grids for each observatory site and apply correction for Galactic extinction according to the \citet{Fitzpatrick:2007} extinction law.

To measure a preliminary redshift, we use the XSpecGUI viewer from \texttt{linetools} \citep{linetools} included in \texttt{PypeIt} to determine the redshift of the host galaxies. This tool overlays the locations of common lines onto the observed spectrum after specifying a fiducial redshift. We adjust the redshift estimate until major features are matched. In particular, we base our initial redshift determination on the Balmer lines (H$\alpha$, H$\beta$), [O\,{\sc iii}]$\lambda 4959, 5007$ emission features, and Ca II H\&K absorption features when available. These serve as input redshifts for our full galaxy modeling (Section~\ref{sec:prospector}).

We supplement our new spectra with those of 13 FRB hosts previously published in the literature. These include FRBs\,20180301A, 20200430A, 20200906A \citep{bha+22}, 20180924B \citep{Bannister+19}, 20181112A \citep{Prochaska+19}, 20190102C, 20190608B \citep{Bhandari+20}, 20190611B, 20190714A, 20191001A \citep{Heintz+20},  20201124A \citep{Fong+21}, 20210117A \citep{Bhandari_210117}, and 20210410D \citep{Caleb+23}. These fully-reduced datasets are publicly available and accessible via the F4 Github repository (\texttt{FRBs/FRB}; \citealt{F4_repo}). In addition, we retrieved archival Gemini North/GMOS spectroscopy for the host of FRB\,20121102A (first published in \citealt{Tendulkar+17}, PI Tendulkar, Program GN-2016B-DD-2) from the Gemini Observatory Archive\footnote{\url{https://archive.gemini.edu}} and reduced the data with \texttt{PypeIt} as described above. As this host is faint, we reduced the S/N threshold so the host trace would be detectable in the individual frames and turned off the profile masking module.

\begin{figure*}
    \centering   
    \includegraphics[width=\textwidth]{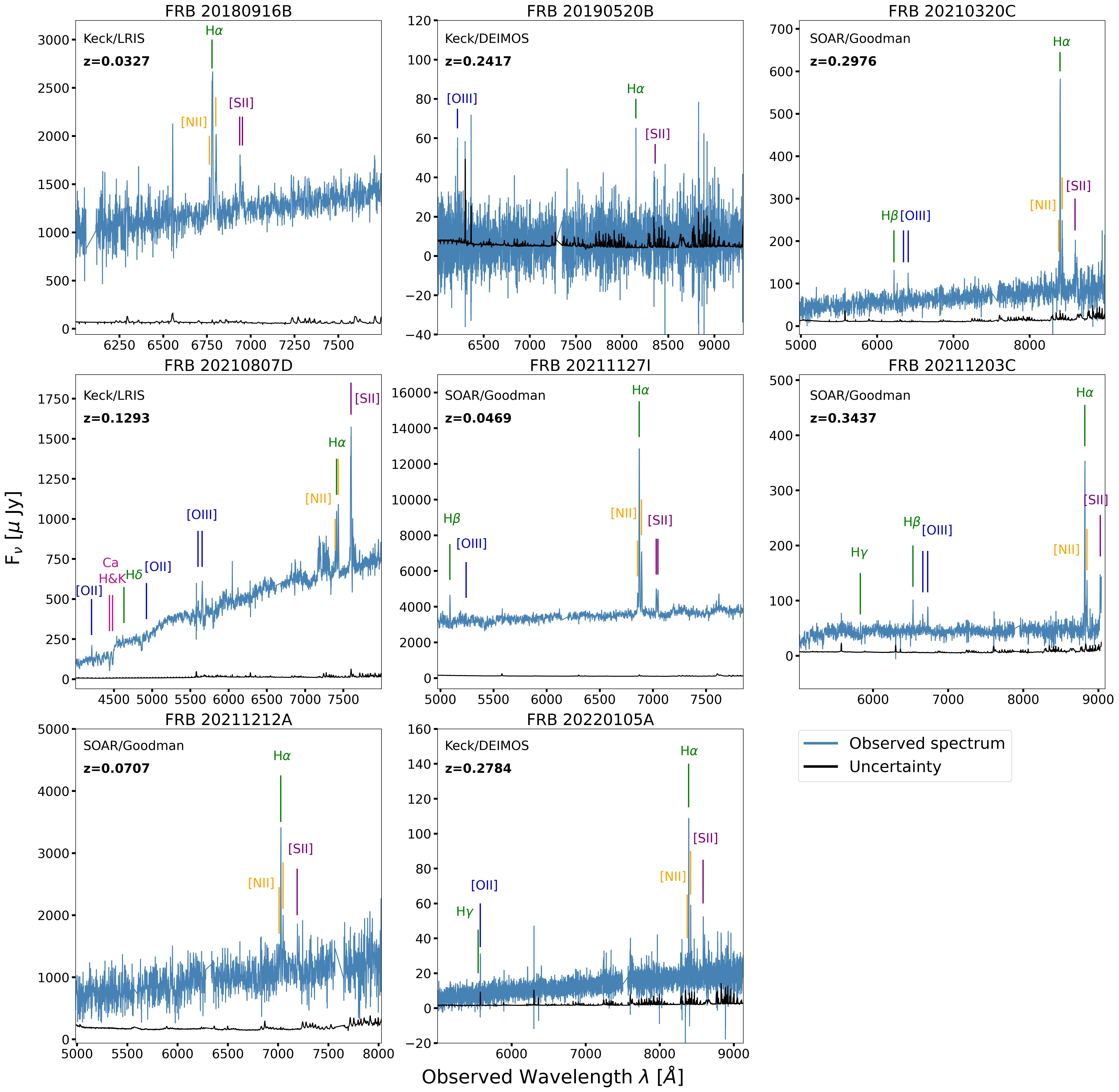}
    \caption{New spectroscopic observations included in this work from SOAR/Goodman, Keck/DEIMOS, and Keck/LRIS. Major emission and absorption lines are denoted by colored lines: Balmer lines in green, oxygen lines in blue, nitrogen lines in orange, sulphur lines in purple, and calcium lines in pink. The spectra are normalized to the photometry in \texttt{Prospector}. The \texttt{Prospector}-derived redshifts are listed in each panel along with the facility/instrument of observation.}
    \label{fig:new spectra}
\end{figure*}

In total, our sample comprises 22 FRB hosts with usable spectroscopy. Details of these observations and those taken from the literature are presented in Table \ref{tab:spectra}. We use all of these spectra in the modeling (Section~\ref{sec:prospector}), with the exception of FRB\,20190711A, which has insufficient S/N to include in the fitting. In Figure \ref{fig:new spectra}, we show the spectra of all new observations presented in this work with major lines denoted and organized by chemical species (see Section~\ref{sec:prospector} for more details on the normalization process).

\begin{deluxetable*}{l|ccccccc}
\tablewidth{0pc}
\tablecaption{FRB Host Galaxy Spectroscopy Details
\label{tab:spectra}}
\tablehead{
\colhead{FRB}	 &
\colhead{Facility} &
\colhead{Instrument} &
\colhead{Observation Date} &
\colhead{Grating/Grism} &
\colhead{Slit Width} &
\colhead{Program ID} &
\colhead{Reference} \\
\colhead{} &
\colhead{} &
\colhead{} &
\colhead{} &
\colhead{} &
\colhead{[$\arcsec$]} &
\colhead{} &
\colhead{} 
}
\startdata
20121102A & Gemini North & GMOS & 2016 Nov 09 UT & R400 & 1.0 & GN-2016B-DD-2 & 1 \\
20180301A & Keck & DEIMOS & 2020 Sept 17 UT & 600ZD & 1.0 & O298 & 2 \\
20180916B & Keck & LRIS & 2020 Aug 18 UT & R400/8500, B400/3400 & 1.0 & O304 & This Work \\
20180924B & VLT & MUSE & 2018 Nov 05 UT & VPHG & IFU & 2101.A-5005 & 3 \\
20181112A & VLT & FORS2 & 2018 Dec 05 UT & GRIS\_300I & 1.0 & 0102.A-0450(A) & 4 \\
20190102C & VLT & FORS2 & 2019 Mar 25 UT & GRIS\_300I & 1.3 & 0102.A-0450(A) & 5 \\
20190520B & Keck & DEIMOS & 2022 Aug 28 UT & 600ZD & 1.0 & O287 & This Work \\
20190608B & SDSS 2.5-M & SDSS & 2001 Oct 21 UT & - & - & - & 5 \\
20190611B & VLT & FORS2 & 2019 July 12 UT & GRIS\_300I & 1.3 & 0103.A-0101(A) & 6 \\
20190714A & Keck & LRIS & 2020 Jan 28 UT & R600/7500 & 1.0 & U180 & 6 \\
20191001A & Gemini South & GMOS & 2019 Oct 04 UT & R400 & 1.0 & GS-2019B-Q-132 & 6 \\
20200430A & Keck & DEIMOS & 2020 June 07 UT & 600ZD & 1.0 & E353 & 2 \\
20200906A & Keck & DEIMOS & 2020 Sept 17 UT & 600ZD & 0.7 & O298 & 2 \\
20201124A & MMT & Binospec & 2021 Apr 03 UT & 270l & 1.0 & UAO-G195-21A & 7 \\
20210117A & VLT & FORS2 & 2021 Sept 06 UT & GRIS\_300I & 1.0 & 105.204W.003 & 8 \\
20210320C & SOAR & Goodman & 2021 Apr 05 UT & R400, M2 & 1.0 & SOAR2021A-010 & This Work \\
20210410D & Gemini South & GMOS & 2021 Oct 14 UT & R400 & 1.0 & GS-2021B-Q-138 & 9 \\
20210807D & Keck & LRIS & 2021 Aug 11 UT & R400/8500, B400/3400 & 1.0 & O314 & This Work \\
20211127I & SOAR & Goodman & 2022 Jan 01 UT & R400, M2 & 1.0 & SOAR2021B-002 & This Work \\
20211203C & SOAR & Goodman & 2022 Feb 01 UT & R400, M2 & 1.0 & SOAR2021B-002 & This Work \\
20211212A & SOAR & Goodman & 2021 Dec 08 UT & R400, M2 & 1.0 & SOAR2021B-002 & This Work \\
20220105A & Keck & DEIMOS & 2022 Mar 31 UT & 600ZD & 1.0 & O300 &  This Work 
\enddata
\tablecomments{Details of the spectroscopic observations included in this work. \\
References:
1. \citet{Tendulkar+17},
2. \citet{bha+22},
3. \citet{Bannister+19},
4. \citet{Prochaska+19},
5. \citet{Bhandari+20},
6. \citet{Heintz+20},
7. \citet{Fong+21},
8. \citet{Bhandari_210117},
9. \citet{Caleb+23}}
\end{deluxetable*}

\section{Host Galaxy Modeling} \label{sec:prospector}

To determine the host stellar population properties, we use the Bayesian modeling code \texttt{Prospector} \citep{Johnson+21}. \texttt{Prospector} is a stellar population synthesis code that derives the posterior probability distributions of stellar population properties for a given observational data set. We use the stellar population synthesis library \texttt{python-fsps} to generate the model SEDs and \texttt{Prospector} to jointly fit the photometry and spectroscopy \citep{Conroy2009, Conroy2010}. The posteriors are sampled using the dynamic nested sampling routine \texttt{dynesty} \citep{Speagle2020}. We initiate our fits with a number of assumptions. First, we employ a \citet{Kroupa01} initial mass function (IMF) and  \citet{KriekandConroy13} dust attenuation curve. We also require that the fits roughly adhere to the \citet{Gallazzi2005} mass-metallicity ($M-Z$) relation by assuming a Gaussian scatter around the relationship with a standard deviation twice the measured scatter to allow for potential redshift evolution. 

Unique to this work and the study of FRB host galaxies, we use a non-parametric SFH with a \texttt{continuity} prior (c.f., \citealt{Leja2019}) represented by eight age bins. While more computationally expensive than stellar population modeling with a parametric SFH, non-parametric modeling is more physically realistic due to the lack of strong priors dictating how and when galaxies form their mass \citep{Leja+17, Leja2019}. We provide further information on the implementation of the non-parametric SFH and the continuity prior in Appendix~\ref{app:prospector}. Furthermore, we employ a spectral smoothing model, a model to normalize the spectrum to the photometry, a pixel outlier model to marginalize over poorly modeled noise, and a jitter model to deal with noise in the observed spectrum (see Appendix D of \citealt{Johnson+21} for further details of these procedures). For the spectrum normalization model, we use a 12$^{\rm th}$ order Chebyshev polynomial to fit the model spectrum to the observed spectrum. 

In general, we fit for total mass formed (M$_{\rm F}$), stellar metallicity (Z$_*$), redshift ($z$), the dust attenuation of stellar light (\texttt{dust2}), the fraction of dust attenuation of young stellar light (\texttt{dust1\char`_fraction}), the offset in slope from the \citet{Calzetti00} dust attenuation curve (\texttt{dust\char`_index}), the velocity dispersion of the spectrum ($\sigma_{\rm smooth}$), and the ratios of star formation rate between each of the age bins (\texttt{logsfr\char`_ratios}). If a spectrum is included, which is the case for all of our fits except the host of FRB\,20190711A, we fit for the gas-phase metallicity (Z$_{\rm gas}$) and the gas ionization parameter ($U_{\rm min}$). For galaxies with rest-frame infrared data $\geq$ 2 microns, we fit for the mass fraction of polycyclic aromatic hydrocarbons ($q_{\rm PAH}$) and include a two-component active galactic nucleus (AGN) model in the fit: the fraction of total AGN luminosity relative to the bolometric stellar luminosity (\texttt{fagn}) and the optical depth of the AGN dust torus ($\tau_{\rm AGN}$). We list further details on the exact prior ranges and distributions used in Appendix \ref{sec: priors}.

To initiate the observations for fitting, we use the Galactic extinction-corrected photometry and spectroscopy and apply a mask to all spectra, which limits the rest-frame wavelength coverage to that of the MILES spectral library (masking everything above approximately 7500\AA, rest-frame) and removes the Na\,\textsc{i}~D absorption lines from the fit. We apply additional masking as needed to regions where the error spectrum dominates the observed spectrum or to account for detector chip gaps. While the redshifts are known for each of the hosts in the sample (c.f., Section~\ref{sec:spectra}), we treat redshift as a free but tightly constrained parameter, allowing a $\pm0.01$ deviation from the initial value determined from our manual inspection of the spectral features. This freedom allows for some flexibility due to small uncertainties propagated from the data reduction and redshift determination. We report these redshifts in Table~\ref{tab:SPP_Results} and in Figure~\ref{fig:new spectra}.

For the non-parametric SFH, we use eight age bins for the \texttt{continuity} SFH prior. \citet{Leja2019} found that varying the number of age bins between 4 and 14 bins produced little variation in the results. We choose eight bins to balance resolving features in the SFH (e.g., starburst events) and the computational resources required to run the models (which increase with the model dimensionality). The first two bins are fixed to 0--30 Myr and 30--100 Myr, and the maximum of the last bin is fixed to the age of the Universe at the redshift of the host. The remaining six bins are spaced evenly in logarithmic time. We then use the SFH combined with other parameters to determine a number of key inferred properties: the star formation rate integrated over the past 100~Myr (${\rm SFR}_{\rm 0-100 Myr}$), the mass-weighted age ($t_{\rm m}$), dust attenuation of young and old stellar light ($A_{\rm V,young}$ and $A_{\rm V, old}$, respectively) and stellar mass (M$_*$). In particular, the mass-weighted age is more sensitive to the older stars in the galaxy than light-weighted ages, which tend to be dominated by younger, brighter stars \citep{Conroy2013}. We allow \texttt{dust1\char`_fraction}, to be a free parameter. We convert this to \texttt{dust1} by multiplying it by \texttt{dust2}. We then convert both \texttt{dust1} and \texttt{dust2} to extinction in $V$-band in magnitudes by multiplying by 1.086 to convert from optical depth to magnitudes of dust attenuation for $A_{\rm V,young}$ and $A_{\rm V,old}$, respectively. We use the total mass formed, combined with the SFH, IMF, and metallicity, to calculate stellar mass by multiplying M$_F$ by the surviving mass fraction. For each of the inferred properties, we construct a posterior distribution and report the median, 16th, and 84th quantiles. 

Finally, as part of the fits, we also self-consistently model and measure the strength of the emission lines using a nebular marginalization template. We refer the reader to Appendix~\ref{app:prospector} for further details on the \texttt{Prospector} fitting and conversions from fit to calculated parameters. We report the median of the posterior distributions of the stellar population properties for all 23 FRB hosts modeled, as well as the 68\% credible intervals in Table~\ref{tab:SPP_Results}. As an example of our process, we present the SED for FRB\,20211127I in Figure~\ref{fig:exSED}, while the remaining host SED fits are presented in Appendix~\ref{sec:SEDs}.

\begin{deluxetable*}{l|cccccccccc}
\tabletypesize{\footnotesize}
\tablewidth{0pc}
\tablecaption{Stellar Population Properties
\label{tab:SPP_Results}}
\tablehead{
\colhead{FRB}	 &
\colhead{$z$} &
\colhead{log(M$_{\rm F}$/M$_{\odot}$)} &
\colhead{log(Z$_*$/Z$_{\odot}$)} &
\colhead{A$_{\rm V, young}$} &
\colhead{A$_{\rm V, old}$} &
\colhead{AGN} &
\colhead{log(Z$_{\rm gas}$/Z$_{\odot}$)} &
\colhead{${\rm SFR}_{\rm 0-100 Myr}$} &
\colhead{log(M$_*$/M$_{\odot}$)} &
\colhead{$t_{\rm m}$} \\
\colhead{}	 &
\colhead{}	 &
\colhead{}	 &
\colhead{}	 &
\colhead{[mag]}	 &
\colhead{[mag]} &
\colhead{}	 &
\colhead{}	 &
\colhead{[M$_{\odot}$~yr$^{-1}$]}	 &
\colhead{}	 &
\colhead{[Gyr]}	  
 }
\startdata
20121102A & $0.1931$ & $8.34^{+0.10}_{-0.11}$ & $0.11^{+0.05}_{-0.11}$ & $1.10^{+0.25}_{-0.26}$ & $0.11^{+0.09}_{-0.06}$ & N & $-0.62^{+0.07}_{-0.06}$ & $0.05^{+0.02}_{-0.01}$ & $8.14^{+0.09}_{-0.10}$ & $5.71^{+0.96}_{-1.26}$ \\
20180301A & $0.3305$ & $9.84^{+0.12}_{-0.13}$ & $-0.99^{+0.27}_{-0.21}$ & $1.18^{+0.28}_{-0.27}$ & $0.31^{+0.17}_{-0.12}$ & N & $-0.33^{+0.05}_{-0.04}$ & $1.91^{+0.64}_{-0.55}$ & $9.64^{+0.11}_{-0.11}$ & $4.34^{+1.10}_{-1.29}$ \\
20180916B & $0.0330$ & $10.13^{+0.04}_{-0.05}$ & $-1.80^{+0.12}_{-0.11}$ & $0.94^{+0.27}_{-0.25}$ & $0.35^{+0.07}_{-0.01}$ & N & $0.18^{+0.18}_{-0.18}$ & $0.04^{+0.03}_{-0.02}$ & $9.91^{+0.03}_{-0.05}$ & $7.73^{+0.86}_{-1.22}$ \\
20180924B & $0.3212$ & $10.60^{+0.02}_{-0.03}$ & $-0.14^{+0.04}_{-0.04}$ & $1.10^{+0.30}_{-0.25}$ & $0.11^{+0.03}_{-0.03}$ & N & $0.03^{+0.08}_{-0.08}$ & $0.62^{+0.32}_{-0.24}$ & $10.39^{+0.02}_{-0.02}$ & $5.63^{+0.53}_{-0.75}$ \\
20181112A & $0.4755$ & $10.06^{+0.07}_{-0.08}$ & $-0.19^{+0.27}_{-0.32}$ & $1.16^{+0.26}_{-0.31}$ & $0.13^{+0.13}_{-0.08}$ & N & $-0.17^{+0.12}_{-0.12}$ & $1.54^{+0.99}_{-0.65}$ & $9.87^{+0.07}_{-0.07}$ & $3.82^{+0.84}_{-0.98}$ \\
20190102C & $0.2909$ & $9.90^{+0.09}_{-0.09}$ & $-1.15^{+0.37}_{-0.39}$ & $1.09^{+0.29}_{-0.29}$ & $0.20^{+0.18}_{-0.13}$ & N & $-0.51^{+0.78}_{-0.51}$ & $0.40^{+0.31}_{-0.11}$ & $9.69^{+0.09}_{-0.11}$ & $4.76^{+1.02}_{-1.47}$ \\
20190520B & $0.2417$ & $9.30^{+0.08}_{-0.11}$ & $-1.55^{+0.33}_{-0.29}$ & $1.18^{+0.27}_{-0.33}$ & $0.15^{+0.15}_{-0.10}$ & N & $-0.68^{+0.45}_{-0.55}$ & $0.04^{+0.04}_{-0.02}$ & $9.08^{+0.08}_{-0.09}$ & $5.27^{+1.02}_{-1.32}$ \\
20190608B & $0.1178$ & $10.78^{+0.02}_{-0.02}$ & $-0.03^{+0.04}_{-0.04}$ & $1.09^{+0.22}_{-0.21}$ & $0.08^{+0.02}_{-0.02}$ & Y & $0.02^{+0.05}_{-0.05}$ & $7.03^{+1.43}_{-1.15}$ & $10.56^{+0.02}_{-0.02}$ & $7.13^{+0.70}_{-1.21}$ \\
20190611B & $0.3778$ & $9.77^{+0.13}_{-0.13}$ & $-0.84^{+0.55}_{-0.54}$ & $1.20^{+0.28}_{-0.30}$ & $0.45^{+0.35}_{-0.23}$ & N & $0.00^{+0.11}_{-0.42}$ & $0.53^{+0.77}_{-0.26}$ & $9.57^{+0.12}_{-0.12}$ & $4.45^{+0.98}_{-1.34}$ \\
20190711A & $0.5218$ & $9.29^{+0.17}_{-0.25}$ & $-0.99^{+0.53}_{-0.54}$ & $1.06^{+0.26}_{-0.28}$ & $0.28^{+0.34}_{-0.16}$ & N & - & $0.95^{+0.96}_{-0.50}$ & $9.10^{+0.15}_{-0.23}$ & $3.54^{+0.96}_{-1.36}$ \\
20190714A & $0.2365$ & $10.42^{+0.04}_{-0.05}$ & $-0.09^{+0.22}_{-0.55}$ & $1.05^{+0.28}_{-0.27}$ & $0.69^{+0.32}_{-0.19}$ & Y & $0.12^{+0.22}_{-0.21}$ & $1.89^{+1.22}_{-0.72}$ & $10.22^{+0.04}_{-0.04}$ & $5.48^{+0.75}_{-1.02}$ \\
20191001A & $0.2342$ & $10.92^{+0.08}_{-0.09}$ & $-0.52^{+0.11}_{-0.10}$ & $1.15^{+0.28}_{-0.25}$ & $1.06^{+0.10}_{-0.10}$ & N & $-0.08^{+0.11}_{-0.11}$ & $18.28^{+17.24}_{-8.95}$ & $10.73^{+0.07}_{-0.08}$ & $3.89^{+1.68}_{-1.56}$ \\
20200430A & $0.1607$ & $9.51^{+0.07}_{-0.10}$ & $-0.99^{+0.32}_{-0.35}$ & $1.08^{+0.32}_{-0.33}$ & $0.38^{+0.14}_{-0.15}$ & Y & $-0.12^{+0.06}_{-0.06}$ & $0.11^{+0.06}_{-0.04}$ & $9.30^{+0.07}_{-0.10}$ & $5.99^{+0.96}_{-1.31}$ \\
20200906A & $0.3688$ & $10.57^{+0.05}_{-0.06}$ & $-0.39^{+0.18}_{-0.15}$ & $1.09^{+0.27}_{-0.23}$ & $0.20^{+0.10}_{-0.10}$ & Y & $-0.26^{+0.14}_{-0.13}$ & $4.93^{+3.46}_{-2.34}$ & $10.37^{+0.05}_{-0.05}$ & $4.30^{+0.86}_{-1.11}$ \\
20201124A & $0.0980$ & $10.43^{+0.05}_{-0.05}$ & $-0.58^{+0.11}_{-0.11}$ & $1.25^{+0.27}_{-0.25}$ & $0.73^{+0.10}_{-0.10}$ & Y & $0.18^{+0.19}_{-0.11}$ & $2.72^{+1.65}_{-1.22}$ & $10.22^{+0.05}_{-0.05}$ & $6.13^{+0.95}_{-1.16}$ \\
20210117A & $0.2145$ & $8.80^{+0.05}_{-0.07}$ & $-1.82^{+0.18}_{-0.12}$ & $1.19^{+0.26}_{-0.32}$ & $0.05^{+0.06}_{-0.03}$ & N & $-0.30^{+0.07}_{-0.08}$ & $0.02^{+0.01}_{-0.01}$ & $8.59^{+0.05}_{-0.06}$ & $5.01^{+0.95}_{-1.21}$ \\
20210320C & $0.2796$ & $10.57^{+0.06}_{-0.06}$ & $-0.82^{+0.16}_{-0.17}$ & $1.26^{+0.26}_{-0.26}$ & $0.64^{+0.15}_{-0.17}$ & N & $0.01^{+0.12}_{-0.16}$ & $3.51^{+2.44}_{-1.45}$ & $10.37^{+0.05}_{-0.06}$ & $4.56^{+0.99}_{-1.15}$ \\
20210410D & $0.1415$ & $9.70^{+0.05}_{-0.06}$ & $-1.04^{+0.19}_{-0.27}$ & $1.14^{+0.28}_{-0.30}$ & $0.39^{+0.13}_{-0.11}$ & N & $0.03^{+0.26}_{-0.23}$ & $0.03^{+0.03}_{-0.01}$ & $9.47^{+0.05}_{-0.05}$ & $6.78^{+1.02}_{-1.48}$ \\
20210807D & $0.1293$ & $11.20^{+0.02}_{-0.02}$ & $-0.52^{+0.04}_{-0.05}$ & $1.08^{+0.17}_{-0.15}$ & $0.04^{+0.03}_{-0.03}$ & Y & $-0.26^{+0.07}_{-0.07}$ & $0.63^{+0.18}_{-0.17}$ & $10.97^{+0.02}_{-0.02}$ & $8.36^{+2.25}_{-1.84}$ \\
20211127I & $0.0469$ & $9.58^{+0.08}_{-0.02}$ & $-0.53^{+0.03}_{-0.02}$ & $1.22^{+0.25}_{-0.31}$ & $0.06^{+0.01}_{-0.01}$ & Y & $0.29^{+0.13}_{-0.12}$ & $35.83^{+1.02}_{-1.46}$ & $9.48^{+0.06}_{-0.02}$ & $3.85^{+2.13}_{-3.65}$ \\
20211203C & $0.3437$ & $9.90^{+0.09}_{-0.10}$ & $0.00^{+0.12}_{-0.19}$ & $1.08^{+0.26}_{-0.25}$ & $0.04^{+0.04}_{-0.02}$ & N & $-0.25^{+0.15}_{-0.14}$ & $15.91^{+2.82}_{-2.98}$ & $9.76^{+0.07}_{-0.09}$ & $2.47^{+2.00}_{-1.25}$ \\
20211212A & $0.0707$ & $10.49^{+0.06}_{-0.07}$ & $-0.77^{+0.11}_{-0.12}$ & $1.19^{+0.26}_{-0.27}$ & $0.19^{+0.04}_{-0.03}$ & N & $0.20^{+0.17}_{-0.26}$ & $0.73^{+0.62}_{-0.39}$ & $10.28^{+0.05}_{-0.06}$ & $5.83^{+1.05}_{-1.16}$ \\
20220105A & $0.2784$ & $10.22^{+0.06}_{-0.07}$ & $-0.81^{+0.16}_{-0.14}$ & $1.15^{+0.26}_{-0.28}$ & $0.76^{+0.15}_{-0.17}$ & Y & $-0.14^{+0.13}_{-0.13}$ & $0.42^{+0.31}_{-0.19}$ & $10.01^{+0.05}_{-0.07}$ & $5.67^{+0.73}_{-1.24}$ \\
\enddata
\tablecomments{Median and 68\% confidence intervals of the stellar population properties. $z$ is the \texttt{Prospector}-derived redshift. These values are highly consistent with those reported in Table~\ref{tab: FRB properties} to within 0.1\%. log(M$_{\rm F}$/M$_{\odot}$) represents total mass formed. log(Z$_*$/Z$_{\odot}$) is the stellar metallicity. A$_{\rm V, young}$ and A$_{\rm V, old}$ are the magnitudes of dust extinction for young and old stars, respectively. AGN denotes if the AGN model was used in the fitting process - this does not necessarily imply the presence of a known AGN in the system. log(Z$_{\rm gas}$) is the gas-phase metallicity. ${\rm SFR}_{\rm 0-100 Myr}$ is the integrated 0--100~Myr star formation rate. log(M$_*$/M$_{\odot}$) is the stellar mass. Finally, $t_{\rm m}$ is the mass-weighted age. The values and uncertainties for all derived measurements will be made available via the F4 repository (\texttt{FRBs/FRB}; \citealt{F4_repo}).}
\end{deluxetable*}

\begin{figure*}
    \centering
    \includegraphics[width=\textwidth]{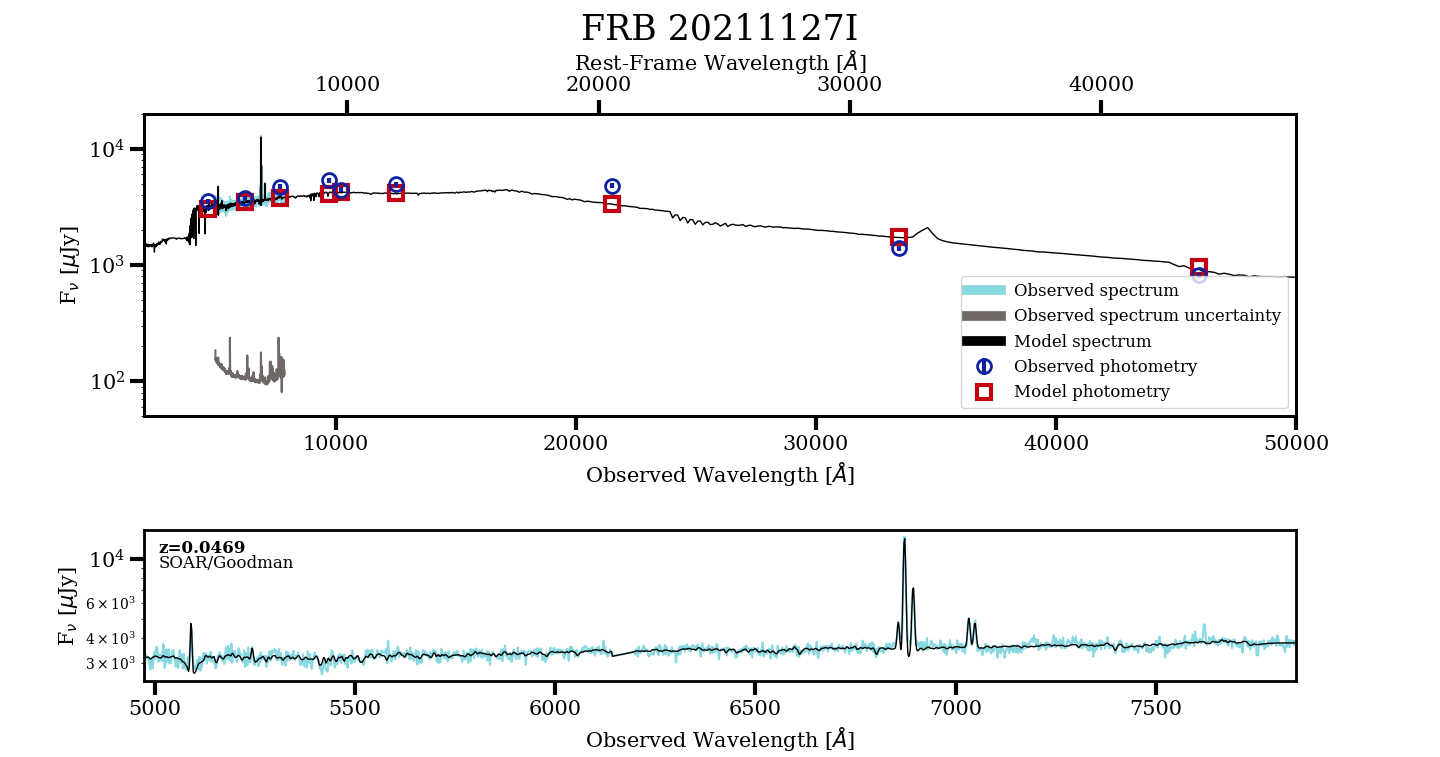}
    \caption{Example SED for the host of FRB\,20211127I. \textbf{Top panel:} The observed spectrum (light blue) and its associated error spectrum (light grey) are modeled jointly with the observed photometry (blue circles). The best-fit {\tt Prospector} model spectrum (black) and corresponding model photometry (red squares) are also displayed. \textbf{Bottom panel:} Zoom-in of the observed and model spectra following the same color scheme as the top panel. As the nebular emission lines are self-consistently modeled, the {\tt Prospector} model provides excellent agreement to the spectroscopic data. The remaining SEDs are in Appendix~\ref{sec:SEDs}.}
    \label{fig:exSED}
\end{figure*}

\section{Results} \label{sec:results}

\subsection{Stellar Population Properties} \label{sec: SPPs}

We now present the inferred stellar population properties of the 23~FRB host galaxies in our sample as a whole. To calculate population medians, we draw 1000 representative samples of log(M$_*$/M$_{\odot}$), log(Z$_*$/Z$_{\odot}$), $t_{\rm m}$, $A_{\rm V,young}$ and $A_{\rm V,old}$, and construct distributions of log(${\rm SFR}_{\rm 0-100 Myr}$) and specific star formation rate (star formation rate divided by stellar mass; log(${\rm sSFR}_{\rm 0-100 Myr}$)) as described above for each host galaxy. We chose 1000 draws as this constitutes a representative sample from which deviations in the median and 68\% confidence intervals are negligible with increasing numbers of draws. We then combine the distributions for each parameter and derive the medians and 68\% credible intervals for the total population. 

\begin{figure*}
    \centering
    \includegraphics[width=\textwidth]{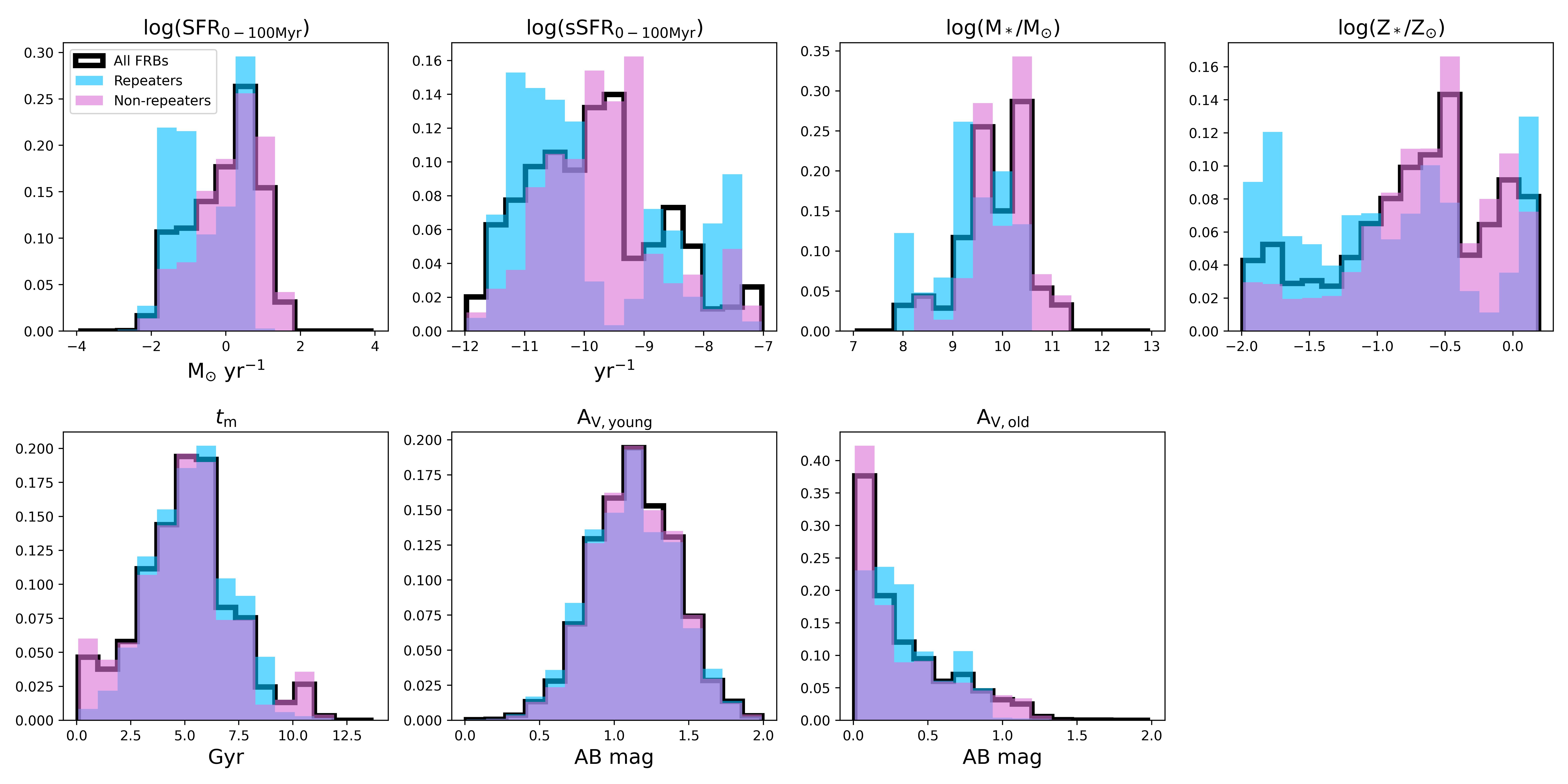}
    \caption{ Posterior distributions of log(${\rm SFR}_{\rm 0-100 Myr}$), log(${\rm sSFR}_{\rm 0-100 Myr}$), log(M$_*$/M$_{\odot}$), log(Z$_*$/Z$_{\odot}$), $t_{\rm m}$, $A_{\rm V, young}$, and $A_{\rm V, old}$ for the full sample. We show the repeater (blue) and non-repeater (pink) populations separately. While the two populations are similar for most properties, repeaters tend to have lower log(M$_*$/M$_{\odot}$) than non-repeaters.}
    \label{fig:prior-posterior_plot}
\end{figure*}

We find that the FRB host population has a median $t_{\rm m} \approx 5.12$~Gyr (interquartile range of $3.72-6.30$~Gyr) and log(M$_*$/M$_{\odot})\approx 9.86$ (interquartile range of $9.46-10.33$). The two hosts with the lowest log(M$_*$/M$_{\odot}$) are those of the repeating FRB\,20121102A and the non-repeating FRB\,20210117A, with $\approx8.1$ and $8.6$, respectively; both fall in the dwarf galaxy class ($\lesssim10^{9}~$M$_{\odot}$; \citealt{dwarf}). For comparison, these are an order of magnitude less massive than the Large Magellanic Cloud with log(M$_*$/M$_{\odot})\approx 9.43$ \citep{LMC} and closer to that of the Small Magellanic Cloud with log(M$_*$/M$_{\odot})\approx 8.49$ \citep{SMC}. The median SFR$_{\rm 0-100 Myr} \approx 1.3\,M_{\odot}$~yr$^{-1}$ (interquartile range of $0.20-4.0\,M_{\odot}$~yr$^{-1}$, while the median log(${\rm sSFR}_{\rm 0-100 Myr})\approx -9.86$~yr$^{-1}$ (interquartile range of $-10.69$ to $-9.17$~yr$^{-1}$). We present the posterior distributions of a selection of stellar population properties and derived properties for the full host sample in Figure~\ref{fig:prior-posterior_plot}. We report these results numerically in Table~\ref{tab:SPP_stats}.

\begin{deluxetable*}{l|cccccccc}
\linespread{1.4}
\tablewidth{0pc}
\tablecaption{FRB Host Galaxy Stellar Population Property Statistics
\label{tab:SPP_stats}}
\tablehead{
\colhead{Population}	 &
\colhead{log(${\rm SFR}_{\rm 0-100 Myr}$)} &
\colhead{log(${\rm sSFR}_{\rm 0-100 Myr}$)} &
\colhead{log(M$_{\rm F}$/M$_{\odot}$)} &
\colhead{log(M$_*$/M$_{\odot}$)} &
\colhead{log(Z$_*$/Z$_{\odot}$)} &
\colhead{$t_{\rm m}$} &
\colhead{A$_{\rm V, young}$} &
\colhead{A$_{\rm V, old}$} \\
\colhead{} & 
\colhead{[M$_{\odot}$~yr$^{-1}$]} & 
\colhead{[yr$^{-1}$]} &
\colhead{} &  
\colhead{} &  
\colhead{} & 
\colhead{[Gyr]} & 
\colhead{[mag]} & 
\colhead{[mag] }
 }
\startdata
Full Sample & $0.11^{+0.81}_{-1.31}$ & $-9.86^{+1.26}_{-1.16}$ & $10.06^{+0.55}_{-0.67}$ & $9.86^{+0.55}_{-0.68}$ & $-0.63^{+0.54}_{-0.75}$ & $5.12^{+1.97}_{-2.15}$ & $1.13^{+0.28}_{-0.28}$ & $0.23^{+0.46}_{-0.17}$ \\
Repeaters & $-0.56^{+1.01}_{-0.95}$ & $-10.37^{+2.39}_{-0.77}$ & $9.59^{+0.75}_{-1.11}$ & $9.40^{+0.73}_{-1.11}$ & $-0.96^{+0.89}_{-0.80}$ & $5.31^{+1.83}_{-1.83}$ & $1.11^{+0.29}_{-0.29}$ & $0.30^{+0.37}_{-0.2}$ \\
Non-repeaters & $0.26^{+0.84}_{-1.03}$ & $-9.77^{+0.86}_{-1.06}$ & $10.23^{+0.50}_{-0.63}$ & $10.01^{+0.54}_{-0.55}$ & $-0.56^{+0.48}_{-0.54}$ & $5.07^{+1.99}_{-2.30}$ & $1.13^{+0.28}_{-0.27}$ & $0.20^{+0.51}_{-0.14}$
\enddata
\tablecomments{Median and 68\% confidence intervals of the stellar population properties for the full sample, repeaters, and non-repeaters. log(${\rm SFR}_{\rm 0-100 Myr}$) is the logarithm of the integrated 0--100~Myr star formation rate. log(${\rm sSFR}_{\rm 0-100 Myr}$) is the logarithm of the specific star formation rate. log(M$_{\rm F}$/M$_{\odot}$) represents total mass formed. log(M$_*$/M$_{\odot}$) is the stellar mass. log(Z$_*$/Z$_{\odot}$) is the stellar metallicity. $t_{\rm m}$ is the mass-weighted age. Finally, A$_{\rm V, young}$ and A$_{\rm V, old}$ are the magnitudes of dust extinction for young and old stars, respectively.}
\end{deluxetable*}

\begin{figure*}
    \centering
\includegraphics[width=\textwidth]{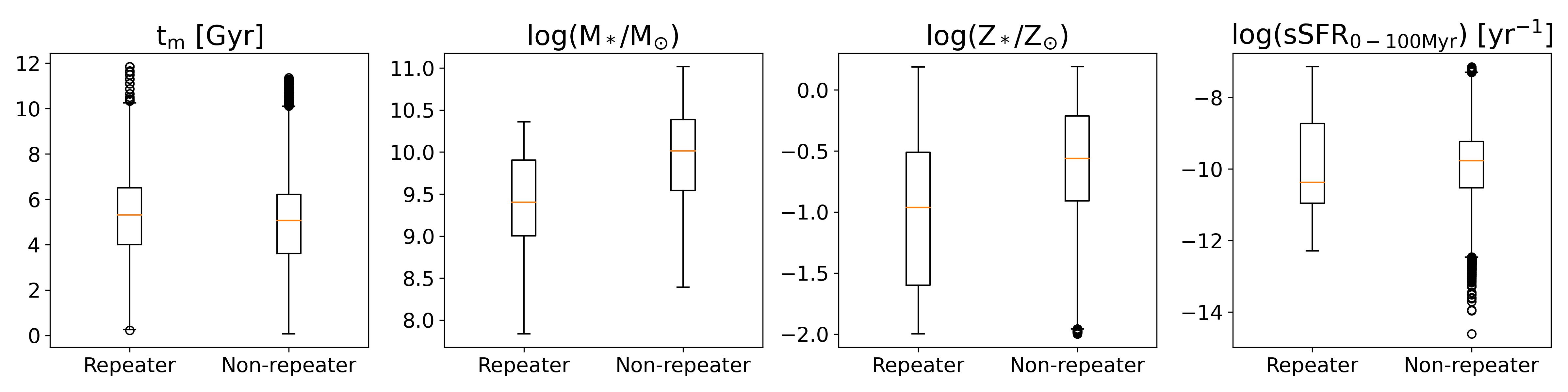}
    \caption{Boxplots of the full sample of FRB hosts split by repeater and non-repeaters derived from 1000 representative samples of the hosts' posterior probability distributions. The box represents the range between the first and third quartiles, with the median denoted by an orange line. The whiskers denote the maxima and minima, and outliers are represented by circles.}
    \label{fig:boxplot_R-NR}
\end{figure*}

We next compare the host galaxy properties of repeating and non-repeating FRBs by presenting their five-number summaries derived from the full posterior distributions as described above. These statistics are represented visually by boxplots in Figure~\ref{fig:boxplot_R-NR}. We find that their distributions of mass-weighted ages, stellar metallicities, and sSFR values are similar, spanning nearly the full range available to galaxies (although in the latter property, non-repeaters span a wider range). We also find that the hosts of non-repeaters tend to have slightly larger stellar masses (population median log(M$_*$/M$_{\odot})\approx 10.01$ versus $9.40$ for the hosts of repeaters). To test whether the stellar population properties of repeaters and non-repeaters could originate from the same underlying distribution, we perform an Anderson-Darling (AD) two sample test with a chosen cut-off p-value of 0.05 (95\% confidence). We find p-values~$>$~0.2 in all properties, with the exception of stellar mass for which we derive a p-value of 0.060 and stellar metallicity for which we derive a p-value of 0.164. Thus, we do not find any evidence that the stellar population properties of repeaters and non-repeaters are statistically distinct.

We also note a few outliers from this analysis. The host of the non-repeating FRB\,20211127I has a ${\rm sSFR}_{\rm 0-100 Myr}\approx 10^{-8.29}$~yr$^{-1}$, over an order of magnitude higher than the next greatest FRB in the same redshift range -- the highly active FRB\,20201124A at $10^{-9.73}$~yr$^{-1}$. At $z=0.0469$, FRB\,20211127I is a relatively young ($\approx 4$~Gyr), nearby spiral galaxy with an age and redshift among the lowest in the sample. This host additionally shows an elevated HI-to-stellar mass ratio and has a slight asymmetry in HI \citep{Glowacki+23}. The other notable outlier is the host age of FRB\,20210807D. This is the oldest and most massive FRB host ($8.36^{+2.25}_{-1.84}$ Gyr and log(M$_*$/M$_{\odot})\approx 10.97$) and is also the only quiescent galaxy in the sample (c.f., Section~\ref{sec:discuss-SFMS}) with a ${\rm sSFR}_{\rm 0-100 Myr}\approx 10^{-11.41}$~yr$^{-1}$, the lowest in the sample.

\citet{Heintz+20} and \citet{bha+22} were the first FRB host population studies based on 12 and 16 hosts, respectively. They found FRB hosts span log(M$_*$/M$_{\odot}) \approx 8-10.8$, $t_{\rm m}\approx 0.06-1.6$~Gyr, and SFRs $\approx 0.05-10$\,M$_{\odot}$~yr$^{-1}$, essentially much of the parameter space expected for galaxies within the redshift ranges of their samples. Their SFRs were primarily derived from H$\alpha$ emission line measurements, which trace more recent SF (timescales of 10-30 Myr). These previous studies performed stellar population modeling using the photometry-only code CIGALE \citep{CIGALE}, employing a parametric delayed-$\tau$ SFH. One known difference between parametric and non-parametric SFHs is that non-parametric SFHs allow for older, more massive galaxies \citep{Leja2019b}, essentially giving the galaxies the freedom to form more mass over a longer time period, and are more physically realistic. In our analysis, the strength of the 4000\AA\ break in the spectrum also drives the older ages, with more considerable breaks implying older stellar populations. Indeed, our new analysis finds significantly older ages by a factor of $\approx$~5 and slightly larger stellar masses compared to previous studies (see Section~\ref{sec:discuss-SPPs} for further discussion). An additional difference between earlier works and this is the IMF used (Chabrier IMF in \citealt{Heintz+20} and \citealt{bha+22} versus Kroupa here). However, the expected differences in mass and SFR attributed to the assumed IMF between these models are very small \citep{Conroy2013}.

\subsection{The Relationship Between FRBs and Current Star Formation} \label{sec:Star Formation}

\begin{figure*}
    \centering
    \includegraphics[width=\textwidth]{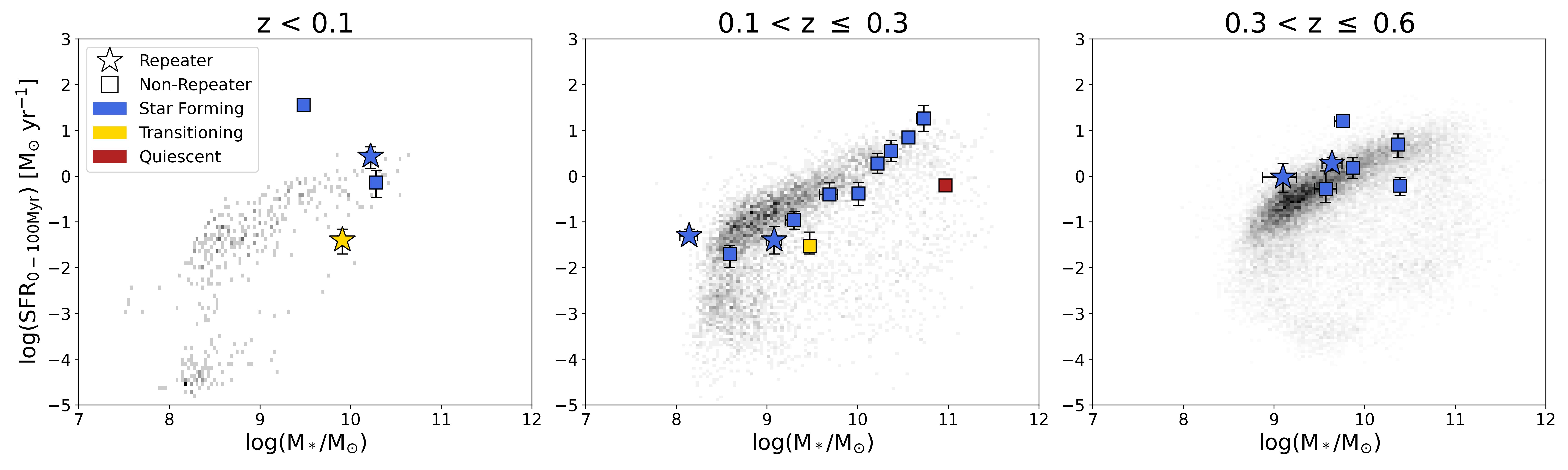}
    \caption{The recent star formation rate versus stellar mass for the full host sample plotted against the COSMOS field galaxies, tracing the star-forming main sequence. We split the sample into three redshift bins (individual panels) to account for evolution in the sequence. Repeaters and non-repeaters are denoted by stars and squares, respectively. The symbols are color-coded by their \citet{Tacchella+22} classification.}
    \label{fig:SFR-Mstar}
\end{figure*}

One of the main context clues for transients and their host galaxies is how they trace active star formation in galaxies. This is particularly important for FRBs given their potential association with magnetars \citep{CHIME-magnetar,Bochenek20}. To systematically classify the degree of star formation in FRB hosts, we use the mass-doubling number criterion from \citet{Tacchella+22} to classify the hosts into star-forming, transitioning (off the main sequence), or quiescent galaxies. From Equation 2 of \citet{Tacchella+22},
\begin{equation}
    \mathcal{D}(z) = \mathrm{sSFR}(z) \times t_{\mathrm{H}}(z)
\end{equation}

\noindent where $t_{\mathrm{H}}(z)$ is the age of the Universe at the redshift of the host galaxy. Following their classification, if $\mathcal{D}(z) > 1/3$, the galaxy is star-forming; if $1/3 < \mathcal{D}(z) < 1/20$, the galaxy is transitioning; and, if $\mathcal{D}(z) < 1/20$, the galaxy is quiescent. To determine the classification, we take the distribution of log(${\rm sSFR}_{\rm 0-100 Myr}$) (generated from 1000 representative draws of ${\rm sSFR}_{\rm 0-100 Myr}$ and log(M$_*$/M$_{\odot}$)) and 1000 draws of $t_{\rm m}$, ensuring the values come from the same models. We then calculate the mass-doubling number for each of the 1000 models and take the mode to determine the most common classification. We find that the large majority of FRB hosts are star-forming (20 hosts), two are transitioning (FRBs\,20180916B and 20210410D), and one is quiescent (FRB\,20210807D). Two of the transitioning and/or quiescent hosts are associated with apparent non-repeating FRBs, although the repeating FRB\,20180916B is classified as transitioning as well. 

We next compare the SFRs and stellar masses of FRB hosts (specifically, the log(${\rm SFR}_{\rm 0-100 Myr}$)---log(M$_*$/M$_{\odot}$) phase space) to field galaxies from the COSMOS sample \citep{Laigle_etal_2016}, as shown in Figure~\ref{fig:SFR-Mstar}. We emphasize that these background galaxies were similarly modeled using \texttt{Prospector} with a non-parametric \texttt{continuity} SFH in \citet{Leja_etal_2020}, allowing for a direct comparison to our derived properties for FRB hosts. We divide the FRB hosts and COSMOS galaxies into three redshift bins, spanning $z<0.1$, $0.1<z\leq0.3$, and $0.3<z\leq0.6$, to enable a proper comparison as there is redshift evolution in this phase space. The well-known star-forming main sequence of galaxies (SFMS; e.g., \citealt{Whitaker12,Speagle14,Leja+22}) is apparent; in general, galaxies below the SFMS are transitioning off or completely quiescent.

We find that the majority of FRB hosts trace the SFMS across all redshifts\footnote{We stress that this is not a selection effect in our data. Although the criteria outlined in \S2.1 require that the host galaxy spectrum shows clear features such as optical emission lines, in practice, no quiescent galaxies were excluded from our sample by this requirement.}. This demonstrates that FRB hosts are forming stars at a similar rate to field galaxies at a given stellar mass. The one star-forming outlier is the host of the apparent non-repeating FRB\,20211127I, which lies well above the main sequence and is the most active galaxy in our sample. We otherwise note that the classification of an FRB as a repeater or non-repeater does not appear to have an effect on the placement of star-forming hosts relative to the SFMS.

In previous works on FRB host galaxies, the majority of the FRB hosts were found to be slightly offset from the SFMS, with smaller star-forming rates than field galaxies of similar stellar mass (e.g., Figure 4 of \citealt{bha+22}). As discussed in Section~\ref{sec: SPPs}, we find slightly larger stellar masses and otherwise similar SFRs. However, previous studies used the PRIMUS catalog \citep{Moustakas_etal_2013} for their field galaxy comparison. When compared to that of COSMOS, the SFMS of PRIMUS is higher by an order of magnitude, effectively resulting in an upward shift in the background galaxy comparison sample and a relative downward shift in the locations of the FRB hosts. The specific interpretation of why the PRIMUS SFMS is significantly higher is beyond the scope of this work, but is likely due to their inclusion of a bursty SFH and lack of IR information. In this work, we use the analysis of the COSMOS dataset from \citet{Leja_etal_2020} derived using identical methodology to our FRB hosts, and are thus free of inter-code systematics. By performing a direct comparison for the first time and employing a quantitative criterion for degree of star formation, we find that only a few are formally ``off'' the SFMS. In other words, the primary difference is the field galaxy modeling, as opposed to any large systematic differences in stellar population properties.

\begin{figure*}
    \centering
    \includegraphics[width=\textwidth]{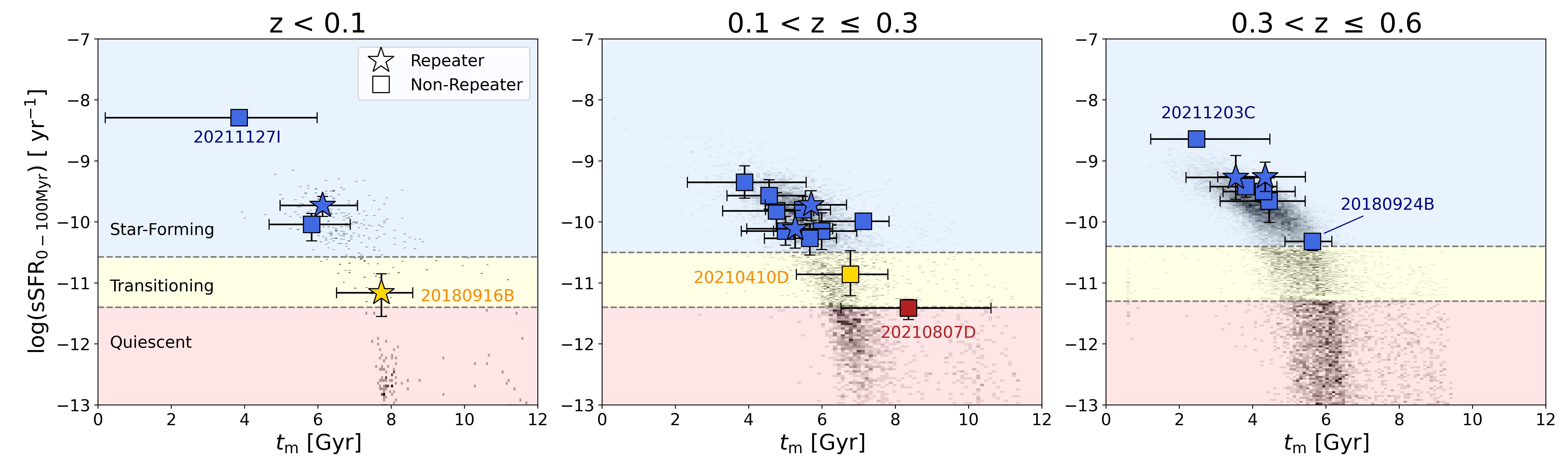}
    \caption{log(${\rm sSFR}_{\rm 0-100 Myr}$) vs. $t_{\rm m}$ of FRB host galaxies compared to COSMOS field galaxies for three redshift bins. The parameter space is divided into Star-Forming, Transitioning, and Quiescent following the \citet{Tacchella+22} classification. Repeaters are denoted by stars and non-repeaters by squares. Both are color-coded by their classification type. Error bars correspond to 68\% confidence. While FRB\,20210807D is on the borderline between transitioning (445 model draws) and quiescent (555 model draws), the mode favors a quiescent classification and is thus classified as such.}
    \label{fig:sSFR-MWA}
\end{figure*}

We next examine the phase space of sSFR versus mass-weighted age (Figure~\ref{fig:sSFR-MWA}). This comparison is meaningful because for non-parametric SFHs, both of these parameters are moments of the SFH. Thus, this serves as a proxy for comparing the SFHs of the FRB hosts to those of field galaxies. We find the star-forming FRB hosts lie in the densest regions occupied by the majority of the COSMOS galaxies, and that there is no apparent distinction between the host galaxies of repeaters and non-repeaters in this phase space. This demonstrates that the SFHs of FRB hosts are also not unique among field galaxies (although we note the presence of outliers such as the host of the very star-forming and relatively young FRB\,20211127I); we discuss further in Section~\ref{sec:discuss-SFMS}.

\subsection{Star Formation Histories} \label{sec:SFH}

\begin{figure*}
    \centering
    \includegraphics[width=\textwidth]{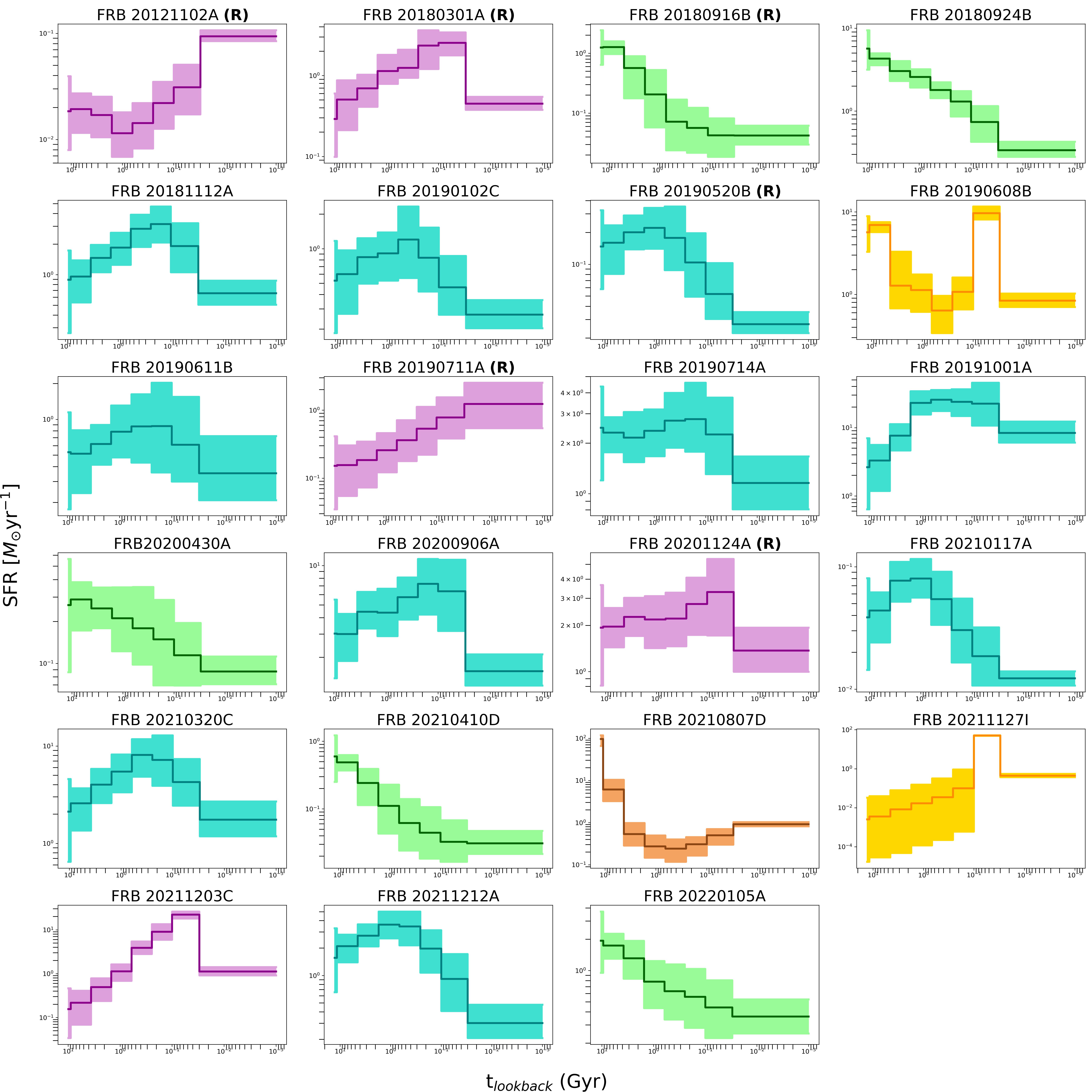}
    \caption{Star formation histories of all FRB host galaxies in the sample. 
    The x-axis is the lookback time, such that the left-hand side is the age of the Universe at the redshift of the galaxy and the right-hand side is the present day. The SFHs are color-coded by SFH type: rising (purple), delayed-$\tau$ exponentially-declining (teal), $\tau$-linear exponentially-declining (green), post-starburst (yellow), and rejuvenating (orange). We denote repeating FRBs with an (R) in the titles.
}
    \label{fig:SFH gallery}
\end{figure*}

Complementary to comparisons involving recent star formation in hosts, we can also leverage our derived SFHs. In particular, given the higher apparent activity of repeating FRBs, it is useful to examine if their progenitors might depend on the level of SF activity over time. In Figure~\ref{fig:SFH gallery}, we show the SFHs of the 23 hosts in our sample over look-back time ($t_{\rm lookback}$), color-coded by SFH type. Specifically, we classify the FRB host SFHs into five types: rising (purple), delayed-$\tau$ exponentially-declining (teal), $\tau$-linear exponentially-declining (green), post-starburst (yellow), and rejuvenating (orange). Rising SFHs, which comprise five galaxies in the sample, are typically associated with dwarf and/or irregular galaxies, and naturally are classified by a consistent rise in SF over time \citep{Papovich+11}. The exponentially-declining SFHs are characteristic of typical L$_{*}$ galaxies. The delayed-$\tau$ (10 hosts) and $\tau$-linear (five hosts) exponentially-declining SFHs are functional forms commonly invoked for parametric SFH modeling (e.g., \citealt{Carnall+19}). Galaxies of both SFH types form most of their stars at early times and decrease in their activities to the present day, with the delayed-$\tau$ model having an important ``delay'' in the onset of the peak of star formation (contributing to a ``rise and fall'' shape in the SFH). Post-starburst SFH galaxies have undergone a recent epoch of high star formation followed by a quenching event (e.g., \citealt{Wild+09,Suess+20,Suess+22}), and comprise two of the hosts. There is a singular possibly rejuvenating host galaxy, FRB\,20210807D. This type of galaxy SFH is fairly rare and is characterized by a recent increase in SF activity after a period of quiescence (e.g., \citealt{Zhang+22}).

Overall, we find that the population of FRB hosts exhibits a diverse range of SFHs, with the majority of FRB hosts (43\%) falling into the delayed-$\tau$ exponentially-declining class in which the peak of SF occurred in the last $\sim 0.1-1$~Gyr; this class includes potentially unexpected hosts like the dwarf-like FRB\,20190520B. Notably, we find that the majority of hosts with clear and prolonged rising SFHs (e.g., increase in SFRs over time) are those of repeating FRBs (FRBs\,20121102A, 20180301A, 20190711A, and 20201124A); however, the non-repeating FRB\,20211203C shows this SFH type as well. We also find repeating FRB hosts with delayed-$\tau$ and $\tau$-linear SFHs. We find evidence of past starburst activity in two host galaxies, both of which originated non-repeating FRBs, corresponding to an occurrence rate of $9\%$ (2/23 hosts). This is in stark contrast to the rate of post-starburst galaxies in SDSS ($0.2\%$), although we note that the selection criteria for the SDSS sample may underestimate the total fraction of galaxies having undergone a starburst event \citep{French2018}. The SDSS criteria is also based on Balmer absorption lines, whereas ours is from the SFHs. Within the final $\sim$Gyr of lookback time, most FRB hosts are rising or at their peak SF activity; some exhibit this behavior even in the final $\sim100$~Myr. Other than these distinctions, we otherwise find no clear patterns or correlations between the host SFHs of repeaters or non-repeaters. 

\subsection{Optical Host Luminosities} \label{sec:lum-z}

\begin{figure}
    \centering
    \includegraphics[width=0.5\textwidth]{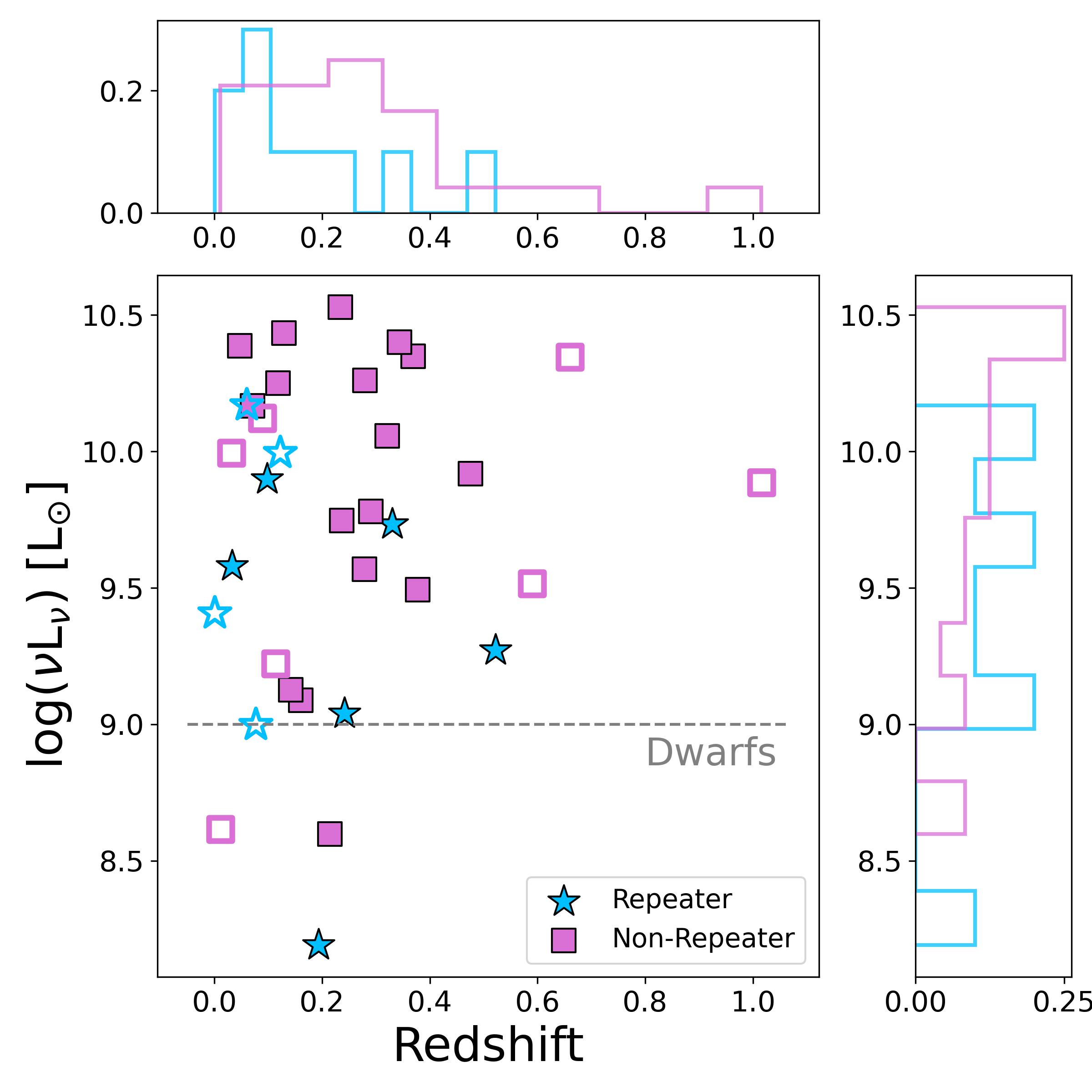}
    \caption{Luminosity--redshift distribution for all FRB hosts to date. We denote repeaters by blue stars and non-repeaters by pink squares. We also include FRB hosts from the literature that did not meet our sample criteria as open symbols.}
    \label{fig:l-z}
\end{figure}

Given the low luminosity of the first identified FRB host galaxy (FRB\,20121102A; \citealt{Tendulkar+17}), it is worthwhile to compare the luminosities of repeating and non-repeating FRBs across redshift. We present the distribution of host optical luminosities ($\nu L_\nu$) and redshift in our sample in Figure~\ref{fig:l-z}, divided into repeaters and non-repeaters. We supplement our sample of 23 FRB hosts with identified hosts of seven~FRBs from the literature that did not meet the criteria for inclusion in our sample (a combination of low PATH probabilities, were detected past the date cutoff of January 2022, or have burst energetics below our cutoff) but nonetheless have claimed host identifications and redshifts. These include the CHIME FRB\,20190425A \citep{Panther23}, DSA FRB\,20190523A \citep{Ravi+19}, \texttt{realfast} FRB\,20190614D \citep{Law+20}, CHIME FRB\,20190110C \citep{Ibik+23}, CHIME FRB\,20200120E \citep{Kirsten21}, CHIME FRB\,20200223B \citep{Ibik+23}, DSA FRB\,20220319D \citep{Ravi23}, DSA FRB\,20220509G \citep{Connor+23,Sharma+23}, ASKAP/CRAFT FRB\,20220610A \citep{Ryder22}, DSA FRB\,20220912A \citep{Ravi22}, and DSA FRB\,20220914A \citep{Connor+23,Sharma+23}. We also plot a demarcation at $\nu L_\nu = 10^{9}L_{\odot}$ below which a host can be classified as a dwarf galaxy (Figure~\ref{fig:l-z}).

First, we find that FRB hosts have a median luminosity of $\approx$ 6$\times 10^{9}\,L_{\odot}$, and span a wide range from the faintest at $\approx 2 \times 10^{8}\,L_{\odot}$ (FRB\,20121102A) to the most luminous at $\approx 3 \times 10^{10}\,L_{\odot}$ (FRB\,20191001A). Next, we find that the hosts of repeaters extend to lower luminosities than those of non-repeaters. Moreover, no repeating FRB hosts in our sample have $\nu L_{\nu} \gtrsim 10^{10}~L_{\odot}$ while 9 non-repeating FRB hosts (or 53\% of the total non-repeating host population in our sample) have luminosities in this range. This is consistent with the finding that non-repeaters also exist in galaxies with larger stellar masses (Section~\ref{sec: SPPs}). We also note that for $z\lesssim 0.6$, repeating and non-repeating FRBs appear to have similar redshift distributions (Figure~\ref{fig:l-z}), although only non-repeating FRBs have been observed at higher redshifts.

Finally, the only repeating FRB whose host falls into the category of dwarf galaxy is FRB\,20120112A,  while two additional hosts of repeaters, FRBs\,20190520B and 20220912A \citep{Ravi22} sit just above the borderline at $\approx 1.1 \times 10^{9}\,L_{\odot}$. FRB\,20190520B has been described as a dwarf galaxy in the literature based on the $J$-band color \citep{Niu+22}, but our modeling reveals it to be slightly more massive than a canonical dwarf galaxy. Notably, the dwarf galaxy luminosity space is not only occupied by repeating FRB hosts; indeed, two non-repeating FRB hosts (FRBs\,20210117A and 20220319D) have low luminosities of $\approx 4 \times 10^{8}\,L_{\odot}$ and can be classified as dwarfs \citep{Bhandari_210117, Ravi23}. This is again consistent with their low stellar masses as discussed in Section~\ref{sec: SPPs}.

\section{Discussion} \label{sec:discussion}

\subsection{The Relationship between FRBs, Star formation, and Implications for the Progenitors} \label{sec:discuss-SFMS}

Numerous lines of evidence support the scenario that at least some fraction of the FRB progenitor population is composed of magnetars: the FRBs' coherent emission \citep{Katz14}, energetics \citep{Margalit20}, durations \citep{Nimmo+22}, rates \citep{CHIME-catalog}, stochastic cycles of strong activity or ``burst storms'' \citep{Lanman22,Marthi22}, evidence for strongly magnetized local environments via their rotation measures \citep{Wang22}, and occasional detection of luminous persistent radio sources co-located with FRBs \citep{Chatterjee+17,Marcote+17,Law21,Niu+22,Ravi22b}. The possibility of magnetars as FRB progenitors was strengthened by the repeating FRB-like emission from a known Galactic magnetar SGR\,1935+2154 \citep{CHIME-magnetar,Bochenek20,Zhang22}. It is therefore natural to consider whether the observed FRB host galaxy population is consistent with all FRBs originating from a magnetar progenitor (although with present data it cannot be ruled out that there are multiple progenitors altogether).

The best-studied pathway to magnetar formation is through the core-collapse supernovae (CCSNe) of recently formed massive stars. Indeed, a young, massive star channel for the origins of Galactic magnetars is supported by observations of core-collapse supernova remnants \citep[see, e.g.,][]{Gaensler04,Vink08,Zhou19,Zhou20}. In addition, isochrone dating of Galactic magnetars shows that they occur in stellar populations with a range of main-sequence turnoff masses from 17--50~$M_{\odot}$ \citep[implying lifetimes of 5--12~Myr;][]{Muno06,Bibby08,Davies09,Tendulkar12}.

However, observations of FRB environments as a whole are difficult to reconcile with a single, young production channel for magnetars that is responsible for all observed FRBs. For instance, FRB\,20200120E was pinpointed to a $\sim 9$~Gyr old globular cluster environment in M81 \citep{Bhardwaj21,Kirsten21}, signifying that at least some FRBs can originate from delayed channels that do not rely on recent massive star formation. It is still viable that magnetars could be responsible for events that originate in older stellar populations, via close binary evolution or the accretion-induced collapse of a white dwarf \citep{Moriya16,Margalit19,Kremer2021}, although progenitors not involving magnetars could also be at play, especially for repeaters (i.e., a neutron star interacting with a companion \citep{Ioka_and_Zhang_2020,Lyutikov+20} or accreting black hole binaries \citep{Sridhar+21}). It has also been predicted that the mergers of two neutron stars (NSs), which can experience potentially long delay times of several Gyr or more, could produce magnetars that may be indefinitely stable to collapse and eventually produce observable FRBs \citep{Totani13,Wang16,Pan22}. The recent claim of an association between the binary NS merger GW190425 and an FRB\,20190425A would be definitive evidence for another such delayed channel \citep{Moroianu+23,Panther23}. These pieces of evidence are consistent with the results of \citet{Li_Zhang_2020} who find that the current sample of FRB host environments are consistent with magnetars formed through multiple formation channels. Multi-wavelength observations of the local environment were also informative for the repeating FRB\,20180916B. Due to its relative proximity (149 Mpc), \citet{Tendulkar+21} found that the FRB was 250~pc from the nearest star-forming region, a location inconsistent with a young magnetar had the progenitor been kicked from this region. \citet{Kaur+22} performed HI mapping of the host and larger-scale environment, finding evidence for a past minor galaxy merger. While they find that the progenitor was likely born from a massive star originating in a burst of star-formation triggered by the merger, they still conclude that the progenitor likely travelled from the nearest star-forming knot identified in \citet{Tendulkar+21}. Studies of the local environments provide unique constraints on progenitor models, but are limited to the closest FRB hosts ($z<0.05$). As the majority of hosts are at redshifts higher than this, we must also rely on the properties gleaned from global host studies.

It is thus instructive to examine the FRB host population and its relation to star formation and star formation history. On sub-galactic scales, high-resolution UV and NIR imaging of a smaller number of FRB hosts have demonstrated that several FRBs occur in or near the spiral arms of their host galaxies, and thus their locations track active star formation within their galaxies \citep{Chittidi21,Mannings+21}. Additional studies based on global properties of FRB hosts have found the majority are near to or slightly below the SFMS, in the band typically occupied by galaxies that are transitioning off the main sequence \citep{Heintz+20,bha+22,Ravi22b}. A number of these also occur outside of the ``blue cloud'' of a color-magnitude diagram, although none to date had formally been categorized as quiescent. 

Here, based on uniformly derived properties and an accurate comparison to like-modeled field galaxies, we find that the majority lie on the SFMS (i.e., they are forming stars at similar rates to field galaxies of the same stellar mass). Furthermore, based on quantitative criteria to classify galaxies by degree of star formation \citep{Tacchella+22}, we find that 87\% (20/23) of FRB hosts are star-forming, two are transitioning, and one is quiescent. We also find a wide range of SFRs (integrated over the past 100~Myr), spanning $2\times10^{-2}$~M$_{\odot}$~yr$^{-1}-36$~M$_{\odot}$~yr$^{-1}$. Two of the galaxies which are transitioning and quiescent host apparently non-repeating FRBs (FRBs\,20210410D and 20210807D) although we note that the repeating FRB\,20180916B host is transitioning as well. Both hosts additionally have older ages of $\approx 7-8$~Gyr. 

The SFH can also inform progenitor models of transients. For example, the declining SFHs of so-called Ca-rich transients indicate that core-collapse SNe progenitors are unlikely \citep{Dong22}, while the post-starburst nature of the hosts of tidal disruption events have been used to exclude O, B, and massive A type stars as likely progenitors \citep{French2017}. In our sample, we find that only a few have monotonically-declining SFHs (which signifies that the large majority of the host stellar mass formed early on in the first few Gyr). Instead, examining the last $\sim 1$~Gyr, most FRB hosts are either rising in SF activity or at their peak, while all four FRB hosts with prolonged rises are repeaters; see Section~\ref{sec:discuss-R/NR} for a further discussion. We also find a few hosts with evidence of past starburst activity $\sim 100$~Myr, commensurate with H\,{\sc i} mapping studies which have shown evidence for minor mergers \citep{Michalowski21,Kaur+22}. While a complete analysis to determine probable progenitor rates would require constructing mass build-up histories and assumptions on delay time distributions, our results indicate that most FRB hosts were fairly active over the last $\lesssim 0.1-1$~Gyr. 

Finally we note that there are three hosts in our sample for which diffuse radio emission has been detected and attributed to star formation \citep{Bhandari+20-191001,Bhandari+20,Fong+21,Ravi22}, and in one case, possibly more localized to the burst site \citep{Piro+21}. While these hosts are not distinct in their SFHs, we find that they all have generally larger SFRs from their SEDs, and two have higher extinction values (in particular, $A_{\rm V, old}$). While many bursts in our sample lack constraining radio observations, it would be useful to compare if bursts with more recent SF activity from their SFHs also have detectable radio emission.

Taken together, FRB hosts thus originate from galaxies of different levels of star formation activity, but have a clear preference for star-forming galaxies representative of those from the field. Our demographics and SFHs also imply that while the majority of FRB progenitor systems are unlikely to come from highly evolved stellar populations \citep[e.g., white dwarfs;][]{Liebert88,Wood92} or other transient events with long delay times \citep[such as merging neutron stars and black holes, e.g.,][]{Zevin22}, the existence of at least a couple of known hosts in less active environments leaves open this possibility. Already, more FRBs are being found in these types of environments. Recent analysis from \citet{Sharma+23} presented the first quiescent, elliptical FRB host galaxy association (FRB\,20220509G), which is additionally part of the galaxy cluster Abell 2311. However, the connection between most FRBs and active star formation supports the young magnetar model theory, where magnetars are formed through the core-collapse supernovae of recently-formed massive stars in the past $100$~Myr, for the majority of FRBs.

\subsection{Comparison of the Host Properties for Repeating and Apparently Non-Repeating FRBs} \label{sec:discuss-R/NR}

As the number of distinct FRB sources now exceeds 600\footnote{\url{https://www.wis-tns.org/}}, various properties have been proposed to distinguish repeating and non-repeating FRBs. For instance, repeaters first exhibited both a downward drift in frequency known as the ``sad trombone'' effect \citep{Hessels+19} and high linear polarization and/or no circular polarization \citep{Nimmo+21}. However, as more apparent non-repeating FRBs were discovered, examples in this class also shared some of these properties (see \citealt{Petroff+22,Zhang22} for a more detailed history). Recently, based on a sample of 18 repeating and 474 non-repeating CHIME FRBs, it was proposed that the populations may be distinct in the duration-bandwidth phase space. Specifically, repeating FRBs exhibit longer intrinsic durations (after de-dispersion) and narrower bandwidths ($\sim 100-200$~MHz) while apparent non-repeaters have shorter durations and wider bandwidths (e.g., Figure 5 of \citealt{Pleunis21}). \citet{Pleunis21} suggest this could be due to a propagation effect, a result of beaming \citep{Connor+20}, different types of bursts from the same source, or factors intrinsic to the populations. \citet{CHIME+23} again performed a similar analysis with a larger sample of 25 repeaters and found a distinction in DM between the two populations (although they additionally find the burst rates between repeaters and non-repeaters are not distinctly bimodal, implying a portion of the non-repeaters may be eventually observed to repeat.) Thus, it is still unclear if the observed distinctions between repeaters and non-repeaters are intrinsic to the objects themselves or not.

Now, we have leveraged new surveys which have enabled a consistent stream of well-localized events (both repeating and non-repeating) and thus robust host associations. From our sample of 23 FRB host galaxies, we find that repeaters and non-repeaters share largely similar distributions of stellar population properties (e.g., mass-weighted age, stellar mass, stellar metallicity, SFH, and ongoing SF), with no statistically significant differences. 

To test the statistical power of our sample, we simulate larger samples of FRBs with one major assumption that the current posterior distributions for repeaters and non-repeaters are representative of the true distributions. We double and triple the current sample while maintaining the same ratio of repeaters to non-repeaters (i.e., 12 repeaters and 34 non-repeaters for the doubled sample). For each stellar population property, we draw random samples from the total posterior distributions of the non-repeating and repeating FRBs. We then perform an AD test in which the null hypothesis is that the classes are from the same parent distribution in that stellar population property. A p-value of $0.05$ indicates the null hypothesis can be rejected. We then repeat this process 5000 times to obtain a distribution of AD test results. We choose 5000 tests because this is the point at which the percentage of statistically significant results are not dependent on the number of tests run. We find that with twice (three times) the current sample size, the null hypothesis is rejected for 84\% (96\%) of the tests in stellar mass, demonstrating that increasing the sample sizes by a modest amount produce statistically distinct distributions in this property. Indeed, even an increase in the sample size by 50\% result in a majority of AD tests with $p<0.05$. Performing this same exercise for all other properties, we find that SFR and stellar metallicity could greatly benefit from sample sizes two and three times the current sample, respectively. Meanwhile, all other properties have distributions in which the repeating and non-repeating hosts remain statistically indistinguishable. We caution that these projections assume that the current posterior distributions of repeaters and non-repeaters are representative of the true parent distributions. Ultimately, continued follow-up and stellar population modeling of new, robustly-associated FRB host galaxies will be required to uncover the true population statistics. 

While our current sample is not large enough to reveal statistically significant distinctions (if they do exist), it still shows a few noteworthy distinctions. For example, the repeating FRB host population as a whole tend to exist in hosts with lower stellar masses than those of non-repeating FRBs. Similarly, only non-repeating FRBs have been found in galaxies with optical luminosities $\nu L_{\nu} \gtrsim 10^{10}\,L_{\odot}$. Additionally, two of the FRBs in the environments with the least star formation activity are apparent non-repeaters (FRBs\,20210410D and 20210807D). Finally, the majority of galaxies with clear and prolonged rises in their star formation histories host repeating FRBs (FRBs\,20121102A, 20180301A, 20190711A and 20201124A), indicating a heightened level of recent SF activity. While expected for low-mass galaxies like the hosts of FRBs\,20121102A and 20190711A \citep{Papovich+11,Leitner2012}, this is particularly surprising for more massive galaxies like the hosts of FRBs\,20201124A and possibly 20180301A. It is plausible that such distinctions in SFH could be enhanced in the context of a larger sample. Overall, with a sample of six repeaters and 17 non-repeaters, our findings are consistent with two possibilities: the progenitor is the same for both populations, or they have distinct progenitors but do not strongly select on any single galaxy property given the still small number of FRB hosts. 

Finally, among the highly secure FRB host sample, we do not find significant differences in the redshift distribution of repeating and non-repeating FRBs. However, when including bursts detected after our date cut-off or with less secure host associations, there are three non-repeating FRB host galaxies identified at $z\gtrsim 0.6$ while no repeating FRB hosts in this regime are known. Actual interpretation is muddied by selection biases based on different discovery experiments with varied observational biases and inconsistent criteria for host galaxy follow-up. However, one might expect that repeating FRBs are easier to identify at low-$z$, as several bursts (presumably some with lower luminosities) must be detected, and we cannot rule out that apparent non-repeaters also have repeat bursts at lower fluences than the detected burst. Moreover, it might be more difficult to discover FRBs in star-forming environments which could potentially harbor foreground columns of ionized gas associated with the host galaxy. The increased dispersion contributions results in larger detected pulse widths which would result in a reduced sensitivity to repeating bursts. If these excess columns contain turbulent gas (as would be expected) scatter-broadening of the emission would also reduce search sensitivity to repeating FRBs. 

With the advantage of a large sample of uniformly modeled hosts, we briefly investigate if repeating FRBs with unique burst properties correlate to unique host properties.  FRB\,20121102A has a PRS \citep{Chatterjee+17,Marcote+17}, large and variable rotation measure \citep{Michilli18}, high burst rate \citep{LiWang21}, and a potential 160 day activity period \citep{Rajwade20}. Additionally, its RM is decreasing whereas its DM is increasing \citet{Hilmarsson+21}. FRB\,20190520B is the second known host to feature a PRS \citep{Niu+22} and has a high and rapidly varying RM \citep{Anna-Thomas22}. Both hosts are notable for having lower stellar masses, optical luminosities, and gas-phase metallicities than the bulk of the population; FRB\,20121102A also has a rise in SFH within the last $\sim 0.1$~Gyr. As a counter-example, \citet{Bhandari_210117} recently found the non-repeating FRB\,20210117A to be in a dwarf galaxy and indeed this is one of the lowest-mass galaxies in our sample. However, this burst lacks a PRS and its environment was not found to be highly magnetized. There may be evidence for the sad-trombone effect commonly associated with repeaters, but it was not detected as strongly as in confirmed repeaters. The repeating FRB\,20180916B has a quasi-periodic $\approx$ 5 day activity window every 16.4 days \citep{CHIME_FRB_2020}, and resides in an old ($\sim 8$~Gyr) transitioning host with a lesser degree of star formation than most repeaters. However, as discussed above, it is possible the merger had some effect on local star formation at the site of the FRB despite its overall low global SFR and declining star formation history. Finally, the highly active repeating FRB\,20201124A undergoes sudden high activity rates \citep{Lanman22}, showing 1863 bursts in 82 hours over 54 days, dramatic RM variations, circular polarization \citep{Kumar+22,Xu22}, and has the widest mean burst width of any repeater \citep{Marthi22}. It also has a rising SFH, but is otherwise unremarkable in terms of its galaxy properties. 

 At face value, there is no strong connection between outliers in FRB burst properties and outliers in host properties. Furthermore, FRB hosts that are outliers in their stellar population properties, such as FRBs\,20210807D and 20211127I in regard to age and SFR, respectively, are unremarkable in their burst properties (Shannon+23 in prep.). The detailed interpretation of the connection between FRB burst properties and host properties may be a fruitful path towards constraining progenitor models. 

\subsection{Comparison of Stellar Population Properties} \label{sec:discuss-SPPs}

To fully contextualize the stellar population properties we derive in this work, it is important to compare to previous studies of FRB host galaxies, as this can inform if there are any systematic biases in stellar population modeling and methodologies. While stellar population properties in some form exist for 16 of the hosts in our sample of 23 (e.g., \citealt{Tendulkar+17, Bannister+19, Prochaska+19, Heintz+20, Marcote_etal_2020, Fong+21, bha+22, Bhandari_210117, Ravi22b, Niu+22}), our study presents uniform modeling, assumes non-parametric SFHs (an assumption which can percolate to systematic offsets in other stellar population properties; e.g., \citealt{Leja2019b}), and derives full posterior distributions in each property allowing for a realistic estimate of uncertainties which we make use of in our population distributions. In addition, while previous papers generally derive SFR and metallicity from nebular emission lines, we use the full SED. 

For the $14$ hosts which are also modeled in these previous compilations, we compare their stellar population properties. For SFR, while those derived from H$\alpha$ and the SED may track different timescales, we do not find that there are any systematic differences. We derive slightly larger stellar masses by $\sim 0.1-0.5$~dex (an expected byproduct of non-parametric versus parametric SFHs; \citealp{Leja2019b}). We also find systematically older mass-weighted ages than \citet{Heintz+20} by a factor of $\approx$ 5 (previously ranging from $0.06-1.6$~Gyr, versus $2.47-8.36$~Gyr here). This is not surprising, as non-parametric SFH modeling is known to result in older ages than the parametric SFH assumption \citep{Leja2019b} as the flexible SFH gives galaxies more time to form mass. Since the non-parametric SFH is a more physically realistic assumption than parametric SFHs (the method employed by the majority of the literature), our ages can be considered more representative of the average age of a star in the galaxy. Furthermore, the placement of the FRB hosts in relation to field galaxies in Figure~\ref{fig:sSFR-MWA} shows that the ages we derive for the FRB hosts are consistent with the field galaxy population given the hosts' sSFRs. In fact, if the hosts have younger ages by an order of magnitude, as found in previous studies, they would occupy a phase space not probed by any galaxies in the COSMOS sample. As FRB hosts appear to track typical galaxies in the Universe in their stellar population properties, significantly younger ages than these galaxies would be highly unusual.

Finally, we emphasize the importance of a direct comparison field galaxy sample for an unambiguous interpretation of results. As discussed in Section~\ref{sec:discuss-SFMS}, our finding that most FRB hosts trace the main sequence of star-forming galaxies (as opposed to offset from it) is primarily due to the background field galaxy catalog used. As each SED modeling code uses different frameworks and libraries for modeling the stellar populations, which introduces their own systematic uncertainties, one must somehow account for the systematic uncertainties in the codes' assumptions which is difficult to quantify. Instead, our work demonstrates the importance of modeling hosts in the same manner as the field galaxy population to perform direct comparisons.

\section{Conclusions and Future Prospects} \label{sec:conclusion}

We have presented the largest collection of highly secure, uniformly modeled FRB host galaxies to date, totaling 23 hosts. We inferred the stellar population properties and star formation histories of the hosts of six repeaters and 17 non-repeaters using the \texttt{Prospector} stellar population synthesis code with non-parametric star formation histories. Our major conclusions are:

\begin{itemize}
    \item FRB hosts have a range of stellar masses of $10^{8.1}-10^{11.0}\,M_{\odot}$ with a median of $\approx 10^{9.9}\,M_{\odot}$. Their mass-weighted ages range from $2.5-8.4$~Gyr with a median of $\approx 5.1$~Gyr. We find SFRs (integrated over the last $100$~Myr) ranging from $0.02-35.8\,M_{\odot}$~yr$^{-1}$ with a median of $\approx 1.3\,M_{\odot}$~yr$^{-1}$.
    \item We find that of the 23 hosts, 87\% (20/23) are actively star-forming. Two hosts (FRB\,20180916B and FRB\,20210410D) are transitioning from star forming to quiescent, and another (FRB\,20210807D) is quiescent. 
    \item Compared to similarly modeled field galaxies in the COSMOS sample at comparable redshifts, star-forming hosts trace the SFMS, demonstrating that they form stars at similar rates compared to field galaxies of the same stellar mass. The one notable exception is FRB\,20211127I which lies well above the SFMS.
    \item We find no statistically significant differences in the stellar population properties of repeating and non-repeating host galaxies. However, the hosts of repeaters tend to extend to lower stellar masses, and the hosts of non-repeaters tend to be more optically luminous. Moreover, the two hosts with the lowest degrees of star formation are both non-repeaters.
    \item FRBs show a diverse range of SFHs. We classify the SFHs into five categories: rising, delayed-$\tau$ exponentially declining, $\tau$-linear exponentially declining, post-starburst, and rejuvenating. The majority peak in star formation in the final Gyr. Repeaters tend to show a clear and prolonged rise in star formation over time indicating a heightened level of more recent activity (although there are also repeaters with delayed-$\tau$ and falling SFHs and one non-repeater with a rising SFH).
    \item FRB hosts are not distinct from the SFH moments (e.g., sSFR and mass-weighted age) of field galaxies in the COSMOS sample. FRB hosts trace the main loci of these phase spaces across redshift evolution.
    \item The large percentage of actively star forming FRB host galaxies, coupled with recent star formation activity, support the young magnetar progenitor model in which the progenitors formed through core-collapse supernovae. However, the presence of transitioning and/or quiescent hosts imply at least a small fraction could originate in more delayed channels.
\end{itemize}

\noindent Our work takes advantage of state-of-the-art galaxy modeling techniques that utilize the full power of photometry and spectroscopy to model the parent stellar populations that host FRBs. While we have outlined several implications for their progenitors, the derived SFHs can be further leveraged to construct probabilistic progenitor rates assuming different delay time distributions. In addition, a full comparison of FRB host SFHs to field galaxy archetypes could further inform if there is anything unusual in their past star formation activity.

Thanks to concerted advancements in radio searches and instrumentation, the era of multiple (sub-)arcsecond host localizations per day is fast approaching, and the number of known FRB host galaxies will rapidly increase.  Moreover, increases in sensitivity will push the FRB detection horizon to higher redshifts of $z\gtrsim 1$ (e.g., \citealt{Ryder22}). Our analysis can soon be extended to hundreds of well-localized FRB host galaxies, which will be crucial for understanding whether subtle differences between the hosts of repeating and non-repeating FRBs are robust, and hence indicative of distinct populations, or a result of small-number statistics. Furthermore, the first large-scale studies of FRB properties and rates as a function of redshift, as well as the FRB delay-time distributions with respect to cosmic star formation, will provide additional constraints on their progenitors and usage as cosmic probes. Spectroscopic redshifts and photometric coverage of FRB hosts to $z\sim 1$ and beyond are the only way to probe their progenitor systems in the cosmic era of peak star formation and uniformly short delay times. In parallel, high-quality stellar population parameter modeling of these galaxies will remain key to understanding the global environments of these events and placing unique constraints on progenitor models. 

\section{Acknowledgements}
The authors thank Kyle Westfall for helpful discussions.
A.C.G. and the Fong Group at Northwestern acknowledges support by the National Science Foundation under grant Nos. AST-1814782, AST-1909358 and CAREER grant No. AST-2047919. W.F. gratefully acknowledges support by the David and Lucile Packard Foundation, the Alfred P. Sloan Foundation, and the Research Corporation for Science Advancement through Cottrell Scholar Award \#28284. C.D.K. acknowledges support from a CIERA postdoctoral fellowship. S.B. is supported by a Dutch Research Council (NWO) Veni Fellowship (VI.Veni.212.058). A.T.D. and K.G. acknowledge support through ARC Discovery Project DP200102243. Y.D. is supported by the National Science Foundation Graduate Research Fellowship under Grant No. DGE-1842165. T.E. is supported by NASA through the NASA Hubble Fellowship grant HST-HF2-51504.001-A awarded by the Space Telescope Science Institute, which is operated by the Association of Universities for Research in Astronomy, Inc., for NASA, under contract NAS5-26555. M.G. is supported by the Australian Government through the Australian Research Council's Discovery Projects funding scheme (DP210102103). L.M. acknowledges the receipt of an MQ-RES scholarship from Macquarie University. R.M.S. acknowledges support through Australian Research Council Future Fellowship FT190100155 and Discovery Project DP220102305. M.C. acknowledges support of an Australian Research Council Discovery Early Career Research Award (project number DE220100819) funded by the Australian Government and the Australian Research Council Centre of Excellence for All Sky Astrophysics in 3 Dimensions (ASTRO 3D), through project number CE170100013. 
J.X.P., A.C.G., Y.D., W.F., T.E., N.T., C.D.K., S.S., and A.G.M. acknowledge support from NSF grants AST-1911140, AST-1910471
and AST-2206490 as members of the Fast and Fortunate for FRB
Follow-up team. MeerTRAP received funding from the European Research Council (ERC) under the European Union's Horizon 2020 research and innovation programme (grant agreement No. 694745).

This research was supported in part through the computational resources and staff contributions provided for the Quest high performance computing facility at Northwestern University which is jointly supported by the Office of the Provost, the Office for Research, and Northwestern University Information Technology.

Based in part on observations obtained at the Southern Astrophysical Research (SOAR) telescope, which is a joint project of the Minist\'{e}rio da Ci\^{e}ncia, Tecnologia e Inova\c{c}\~{o}es (MCTI/LNA) do Brasil, the US National Science Foundation’s NOIRLab, the University of North Carolina at Chapel Hill (UNC), and Michigan State University (MSU).

Based in part on observations obtained at the international Gemini Observatory, a program of NSF’s NOIRLab, which is managed by the Association of Universities for Research in Astronomy (AURA) under a cooperative agreement with the National Science Foundation on behalf of the Gemini Observatory partnership: the National Science Foundation (United States), National Research Council (Canada), Agencia Nacional de Investigaci\'{o}n y Desarrollo (Chile), Ministerio de Ciencia, Tecnolog\'{i}a e Innovaci\'{o}n (Argentina), Minist\'{e}rio da Ci\^{e}ncia, Tecnologia, Inova\c{c}\~{o}es e Comunica\c{c}\~{o}es (Brazil), and Korea Astronomy and Space Science Institute (Republic of Korea).

W. M. Keck Observatory and MMT Observatory access was supported by Northwestern University and the Center for Interdisciplinary Exploration and Research in Astrophysics (CIERA).

Some of the data presented herein were obtained at the W. M. Keck Observatory, which is operated as a scientific partnership among the California Institute of Technology, the University of California and the National Aeronautics and Space Administration. The Observatory was made possible by the generous financial support of the W. M. Keck Foundation.

The authors wish to recognize and acknowledge the very significant cultural role and reverence that the summit of Maunakea has always had within the indigenous Hawaiian community.  We are most fortunate to have the opportunity to conduct observations from this mountain.

Observations reported here were obtained at the MMT Observatory, a joint facility of the Smithsonian Institution and the University of Arizona.

 Based on observations collected at the European Southern Observatory under ESO programmes 2101.A-5005, 0102.A-0450, 0103.A-0101, and 105.204W.003.

This scientific work uses data obtained from Inyarrimanha Ilgari Bundara / the Murchison Radio-astronomy Observatory. We acknowledge the Wajarri Yamaji People as the Traditional Owners and native title holders of the Observatory site. CSIRO’s ASKAP radio telescope is part of the Australia Telescope National Facility (https://ror.org/05qajvd42). Operation of ASKAP is funded by the Australian Government with support from the National Collaborative Research Infrastructure Strategy. ASKAP uses the resources of the Pawsey Supercomputing Research Centre. Establishment of ASKAP, Inyarrimanha Ilgari Bundara, the CSIRO Murchison Radio-astronomy Observatory and the Pawsey Supercomputing Research Centre are initiatives of the Australian Government, with support from the Government of Western Australia and the Science and Industry Endowment Fund.

This research has made use of the Keck Observatory Archive (KOA), which is operated by the W. M. Keck Observatory and the NASA Exoplanet Science Institute (NExScI), under contract with the National Aeronautics and Space Administration.

This research has made use of the NASA/IPAC Infrared Science Archive, which is funded by the National Aeronautics and Space Administration and operated by the California Institute of Technology.

This project used public archival data from the Dark Energy Survey (DES). Funding for the DES Projects has been provided by the U.S. Department of Energy, the U.S. National Science Foundation, the Ministry of Science and Education of Spain, the Science and Technology Facilities Council of the United Kingdom, the Higher Education Funding Council for England, the National Center for Supercomputing Applications at the University of Illinois at Urbana–Champaign, the Kavli Institute of Cosmological Physics at the University of Chicago, the Center for Cosmology and Astro-Particle Physics at the Ohio State University, the Mitchell Institute for Fundamental Physics and Astronomy at Texas A\&M University, Financiadora de Estudos e Projetos, Fundação Carlos Chagas Filho de Amparo à Pesquisa do Estado do Rio de Janeiro, Conselho Nacional de Desenvolvimento Científico e Tecnológico and the Ministério da Ciência, Tecnologia e Inovação, the Deutsche Forschungsgemeinschaft and the Collaborating Institutions in the Dark Energy Survey.

The Collaborating Institutions are Argonne National Laboratory, the University of California at Santa Cruz, the University of Cambridge, Centro de Investigaciones Enérgeticas, Medioambientales y Tecnológicas–Madrid, the University of Chicago, University College London, the DES-Brazil Consortium, the University of Edinburgh, the Eidgenössische Technische Hochschule (ETH) Zürich, Fermi National Accelerator Laboratory, the University of Illinois at Urbana-Champaign, the Institut de Ciències de l’Espai (IEEC/CSIC), the Institut de Física d’Altes Energies, Lawrence Berkeley National Laboratory, the Ludwig-Maximilians Universität München and the associated Excellence Cluster Universe, the University of Michigan, the National Optical Astronomy Observatory, the University of Nottingham, The Ohio State University, the OzDES Membership Consortium, the University of Pennsylvania, the University of Portsmouth, SLAC National Accelerator Laboratory, Stanford University, the University of Sussex, and Texas A\&M University.

Based in part on observations at Cerro Tololo Inter-American Observatory, National Optical Astronomy Observatory, which is operated by the Association of Universities for Research in Astronomy (AURA) under a cooperative agreement with the National Science Foundation.

Funding for SDSS-III has been provided by the Alfred P. Sloan Foundation, the Participating Institutions, the National Science Foundation, and the U.S. Department of Energy Office of Science. The SDSS-III web site is http://www.sdss3.org/.

SDSS-III is managed by the Astrophysical Research Consortium for the Participating Institutions of the SDSS-III Collaboration including the University of Arizona, the Brazilian Participation Group, Brookhaven National Laboratory, Carnegie Mellon University, University of Florida, the French Participation Group, the German Participation Group, Harvard University, the Instituto de Astrofisica de Canarias, the Michigan State/Notre Dame/JINA Participation Group, Johns Hopkins University, Lawrence Berkeley National Laboratory, Max Planck Institute for Astrophysics, Max Planck Institute for Extraterrestrial Physics, New Mexico State University, New York University, Ohio State University, Pennsylvania State University, University of Portsmouth, Princeton University, the Spanish Participation Group, University of Tokyo, University of Utah, Vanderbilt University, University of Virginia, University of Washington, and Yale University.

The Pan-STARRS1 Surveys (PS1) and the PS1 public science archive have been made possible through contributions by the Institute for Astronomy, the University of Hawaii, the Pan-STARRS Project Office, the Max-Planck Society and its participating institutes, the Max Planck Institute for Astronomy, Heidelberg and the Max Planck Institute for Extraterrestrial Physics, Garching, The Johns Hopkins University, Durham University, the University of Edinburgh, the Queen's University Belfast, the Harvard-Smithsonian Center for Astrophysics, the Las Cumbres Observatory Global Telescope Network Incorporated, the National Central University of Taiwan, the Space Telescope Science Institute, the National Aeronautics and Space Administration under Grant No. NNX08AR22G issued through the Planetary Science Division of the NASA Science Mission Directorate, the National Science Foundation Grant No. AST-1238877, the University of Maryland, Eotvos Lorand University (ELTE), the Los Alamos National Laboratory, and the Gordon and Betty Moore Foundation.

This publication makes use of data products from the Two Micron All Sky Survey, which is a joint project of the University of Massachusetts and the Infrared Processing and Analysis Center/California Institute of Technology, funded by the National Aeronautics and Space Administration and the National Science Foundation.

Based on observations obtained as part of the VISTA Hemisphere Survey, ESO Progam, 179.A-2010 (PI: McMahon)

This publication makes use of data products from the Wide-field Infrared Survey Explorer, which is a joint project of the University of California, Los Angeles, and the Jet Propulsion Laboratory/California Institute of Technology, and NEOWISE, which is a project of the Jet Propulsion Laboratory/California Institute of Technology. WISE and NEOWISE are funded by the National Aeronautics and Space Administration.

This research is based in part on data collected at the Subaru Telescope, which is operated by the National Astronomical Observatory of Japan. We are honored and grateful for the opportunity of observing the Universe from Maunakea, which has the cultural, historical, and natural significance in Hawaii.

This work has made use of data from the European Space Agency (ESA) mission
{\it Gaia} (\url{https://www.cosmos.esa.int/gaia}), processed by the {\it Gaia}
Data Processing and Analysis Consortium (DPAC,
\url{https://www.cosmos.esa.int/web/gaia/dpac/consortium}). Funding for the DPAC
has been provided by national institutions, in particular the institutions
participating in the {\it Gaia} Multilateral Agreement.

%

\vspace{5mm}
\facilities{
Blanco (DECam),
CTIO:2MASS,
Gemini:Gillett (GMOS),
Gemini:South (GMOS),
\textit{HST} (WFC3),
Keck:I (LRIS),
Keck:II (DEIMOS),
MMT (MMIRS, Binospec),
NOT (ALFOSC),
PS1,
Sloan (SDSS),
SOAR (Goodman),
\textit{Spitzer} (IRAC),
Subaru (MOIRCS),
ESO:VISTA (VIRCAM),
VLT:Antu (FORS2),
VLT:Yepun (HAWK-I),
WISE}


\software{
\texttt{astropy} \citep{astropy}, \texttt{DoPhot} \citep{Schechter93}, \texttt{dynesty} \citep{Speagle2020}, \texttt{FRBs/FRB} \citep{F4_repo}, \texttt{linetools} \citep{linetools}, \texttt{matplotlib} \citep{matplotlib},
\texttt{numpy} \citep{numpy}, 
\texttt{pandas} \citep{pandas}, 
\texttt{photpipe} \citep{Rest+05},
\texttt{photutils} \citep{photutils}, 
\texttt{POTPyRI}\footnote{https://github.com/CIERA-Transients/POTPyRI},
\texttt{Prospector}\footnote{https://ui.adsabs.harvard.edu/abs/2022zndo...6192136J/abstract} \citep{Johnson+21},
\texttt{PypeIt} \citep{pypeit:joss_pub, pypeit:zenodo},
\texttt{python-fsps} \citep{Conroy2009, Conroy2010},
\texttt{SAOImageDS9} \citep{DS9},
\texttt{scipy} \citep{scipy},
\texttt{sedpy} \citep{sedpy},
\texttt{SWarp} \citep{swarp}}


\clearpage

\appendix

\section{Prospector Modeling Details}
\label{app:prospector}

Here we present additional details on our \texttt{Prospector} modeling. For hosts with photometric coverage $\geq$ 2 microns in the rest-frame (i.e. WISE or \textit{Spitzer} coverage), we include additional parameters on IR dust emission and/or the presence of an AGN. To model IR dust emission, we use the three-component \citet{DraineandLi07} dust emission model included in \texttt{FSPS}. As the WISE and \textit{Spitzer} data available for these hosts do not extend into the FIR, we choose to only set \texttt{duste\char`_qpah}, the polycyclic aromatic hydrocarbon mass fraction, as a free parameter and not include the other two components. This choice balances the dimensionality of the model with the available data, ensuring the model is not underconstrained by our dataset. For the AGN prior, we use the two-component AGN model from \citet{Nenkova08} with both components set free. We alternate turning the dust emission and AGN models on and off in various linear combinations with the other parameters in order to determine which model fits the data best. The final model is then chosen through a combination of visual inspection of the agreement between the model and data and the evidence value of the model, a statistical measure of the ``goodness" of the fit.

For the fitting of nebular emission lines, \texttt{Prospector}'s nebular marginalization template fits a Gaussian to each emission line. The Gaussians have the same widths in velocity-space. As the marginalization procedure is purely mathematical (see Appendix E of \citet{Johnson+21} for further details), there are no physics in place to prevent the prediction of negative emission lines; this more likely to happen for spectra with low S/N. In order to determine when to include this template in the final model, we use a combination of visual inspection of the model spectrum and the convergence of the \texttt{eline\char`_sigma} parameter (which describes the emission line widths). 14 FRB hosts had spectra with high enough S/N to include nebular marginalization in the final model. For the remaining eight hosts with usable spectra, we instead use the \texttt{nebemlineinspec} prior, which adds the emission lines to the model spectrum following a pre-built CLOUDY grid \citep{Byler+17}. 

In this work, we use a non-parametric SFH, characterized by the \texttt{continuity} prior in \texttt{Prospector}. This prior prefers a flat SFH: in other words, any observed deviation from a constant SFH is driven by the data. By definition, a non-parametric SFH does not impose an {\it a priori} functional form onto the galaxy's SFH. Instead, the galaxy is allowed to form mass as it sees fit within each age bin, leading to a data-driven SFH that is more physically realistic than parametric models.

As we use eight age bins in this work, the \texttt{continuity} prior outputs seven parameters describing the logarithm of the ratio of the star formation rates between adjacent age bins. To convert this to star formation rates in each age bin, we use the convenience function \texttt{logsfr\char`_ratios\char`_to\char`_sfrs} in the \texttt{prospect.models.transforms} module of \texttt{Prospector}\footnote{https://ui.adsabs.harvard.edu/abs/2022zndo...6192136J/abstract}. We then construct the star formation history from these values. We calculate mass-weighted age by summing the product of the SFR for each age bin and the square of the length of the age bin, then dividing by the mass formed. This value is calculated for each model iteration, weighted by their likelihood weights. We then construct a distribution of these values and report the median, 16th, and 84th quantiles. Additionally, we report the 100 Myr integrated star formation rate ($\log({\rm SFR}_{\rm 0-100 Myr})$) -- the average of the two most recent age bins, spanning 0--30 Myr and 30--100 Myr, weighted by the width of the age bin. This calculation is done using 1000 representative samples of the model, weighted by their likelihood weights. This metric describes the current day star formation rate and is sensitive to both the older (30--100 Myr) and younger (0--30 Myr) recently formed stars.

To calculate the stellar mass formed, we retrieve the mass fraction by calling \texttt{model.predict} for a random sample of 1000 model iterations. We multiply by the total mass formed for the associated model iteration. The resulting 1000 values are thus a representative distribution of stellar masses, from which we report the median and 68\% confidence intervals.

\section{SEDs} \label{sec:SEDs}

Here we present the SEDs of all hosts modeled in this work. See Section~\ref{sec:prospector} for more details on the \texttt{Prospector} modeling.

\begin{figure*}
    \includegraphics[width=\textwidth]{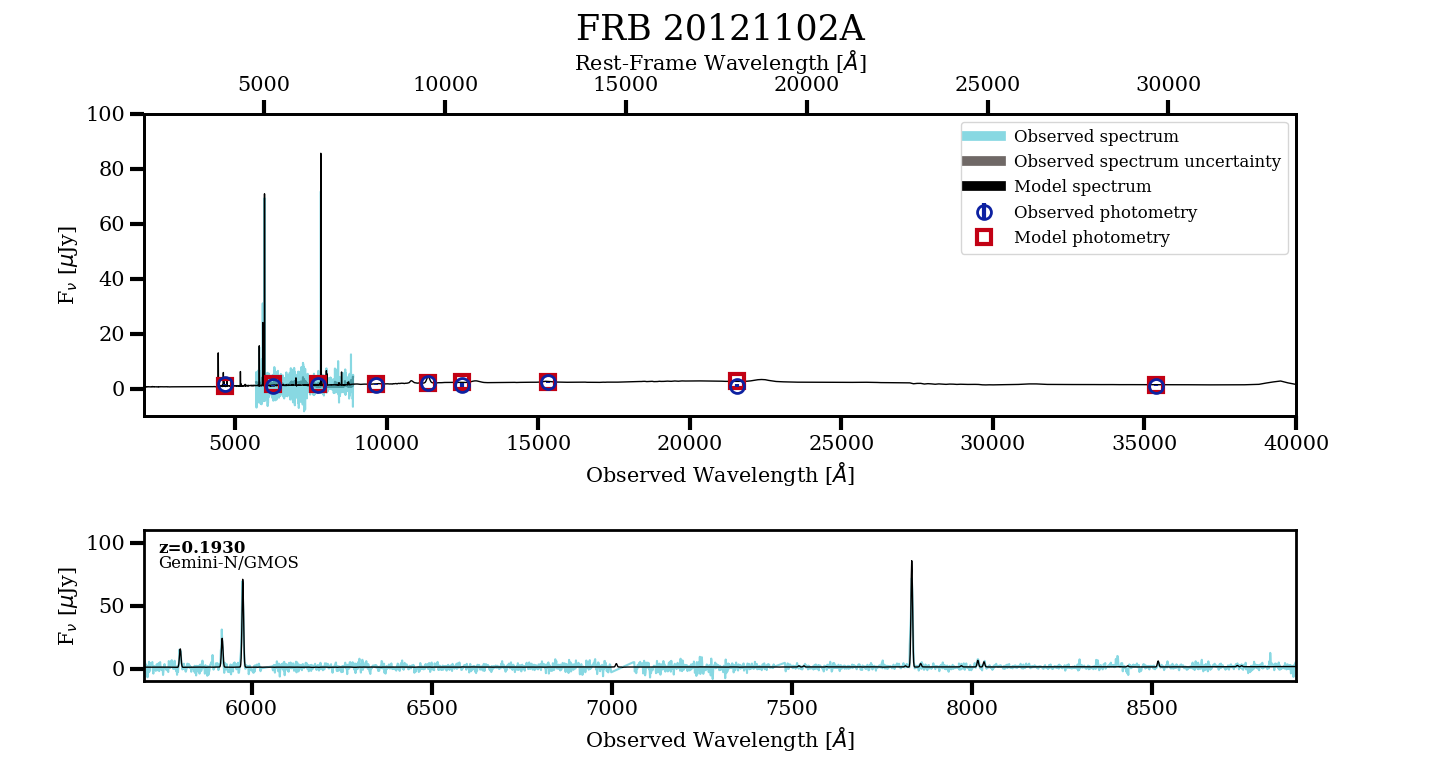}
    \label{fig:121102_SED}
    \caption{SED of FRB\,20121102A.}
\end{figure*}

\begin{figure*}
     \includegraphics[width=\textwidth]{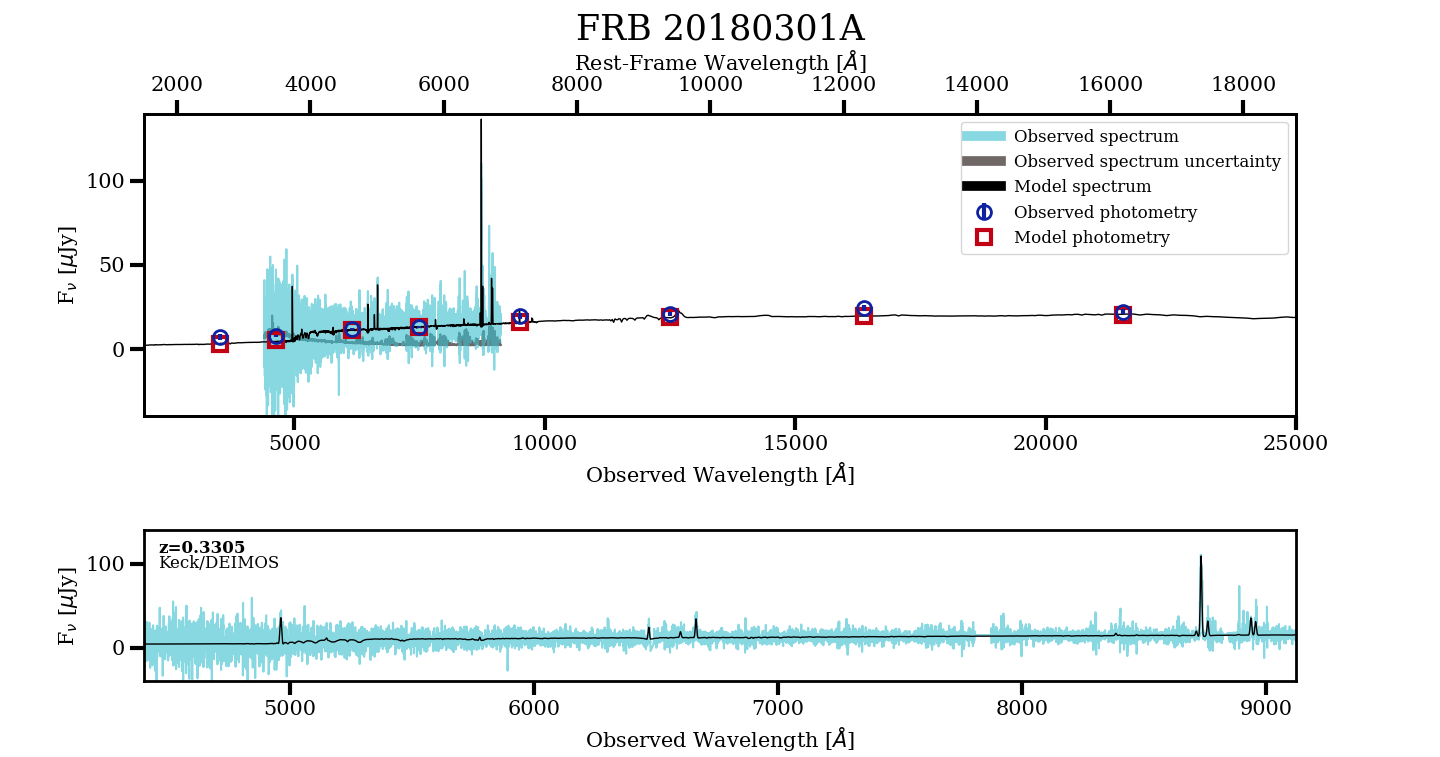}
     \label{fig:180301_SED}
     \caption{SED of FRB\,20180301A.}
\end{figure*}

\begin{figure*}
     \includegraphics[width=\textwidth]{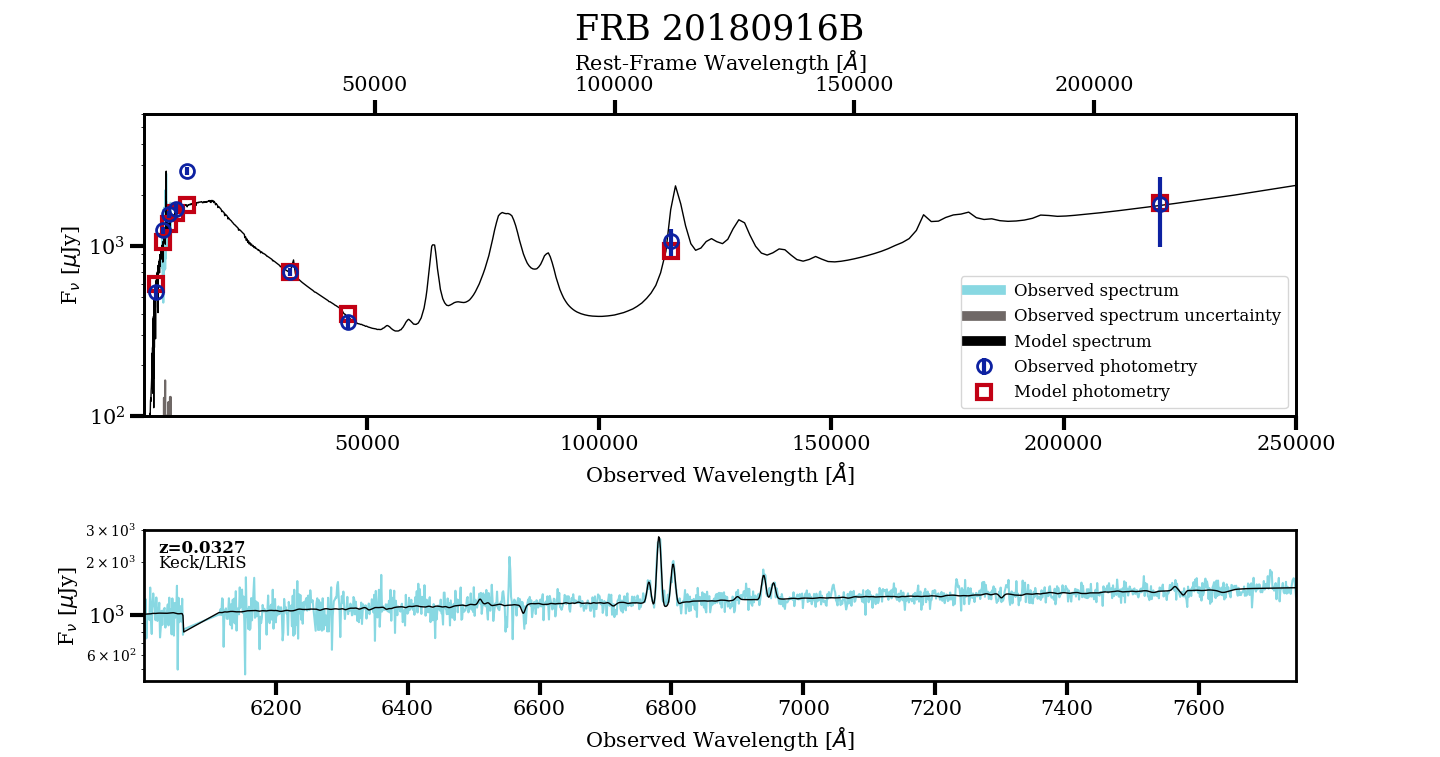}
     \label{fig:180916_SED}
     \caption{SED of FRB\,20180916B.}
\end{figure*}

\begin{figure*}
     \includegraphics[width=\textwidth]{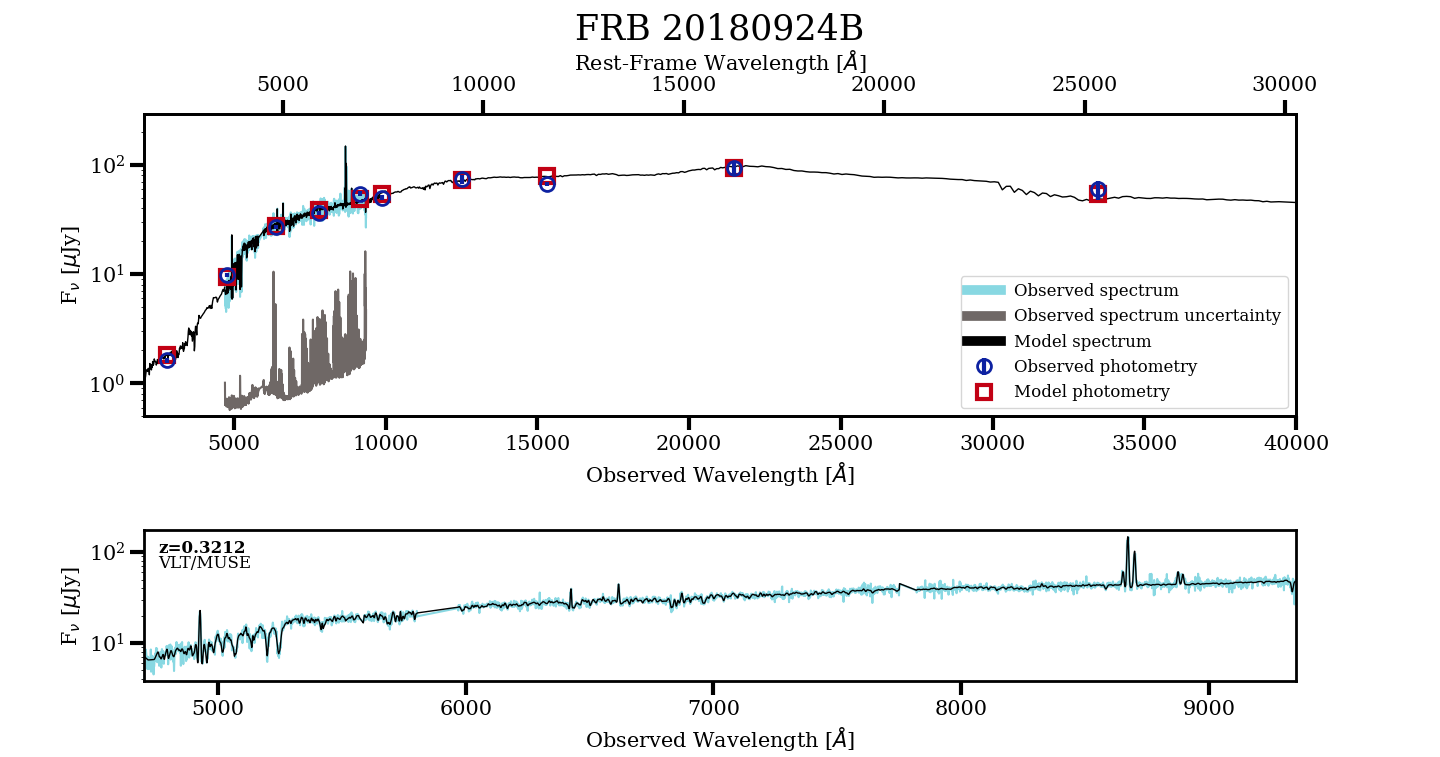}
     \label{fig:180924_SED}
     \caption{SED of FRB\,20180924B.}
\end{figure*}

\begin{figure*}
     \includegraphics[width=\textwidth]{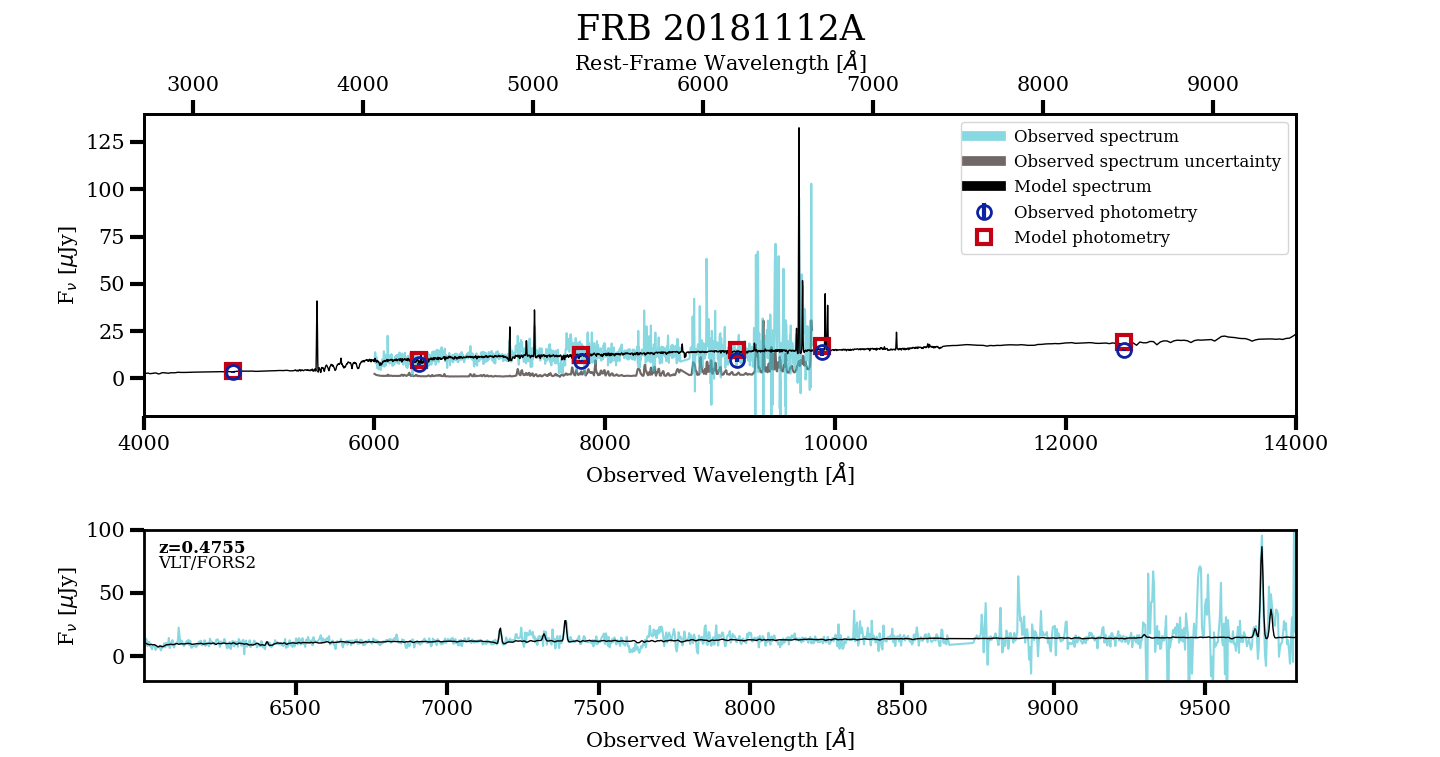}
     \label{fig:181112_SED}
     \caption{SED of FRB\,20181112A.}
\end{figure*}

\begin{figure*}
     \includegraphics[width=\textwidth]{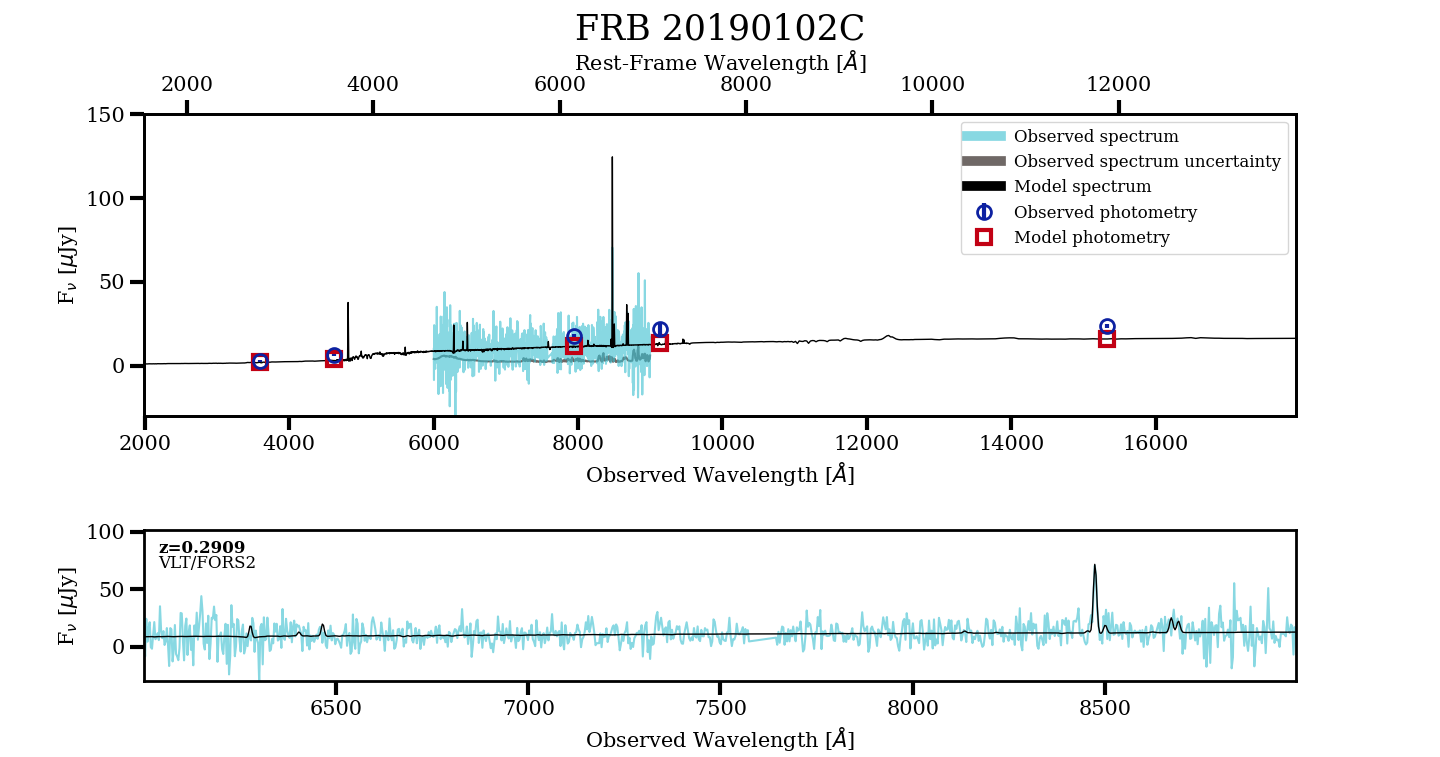}
     \label{fig:190102_SED}
     \caption{SED of FRB\,20190102C.}
\end{figure*}

\begin{figure*}
     \includegraphics[width=\textwidth]{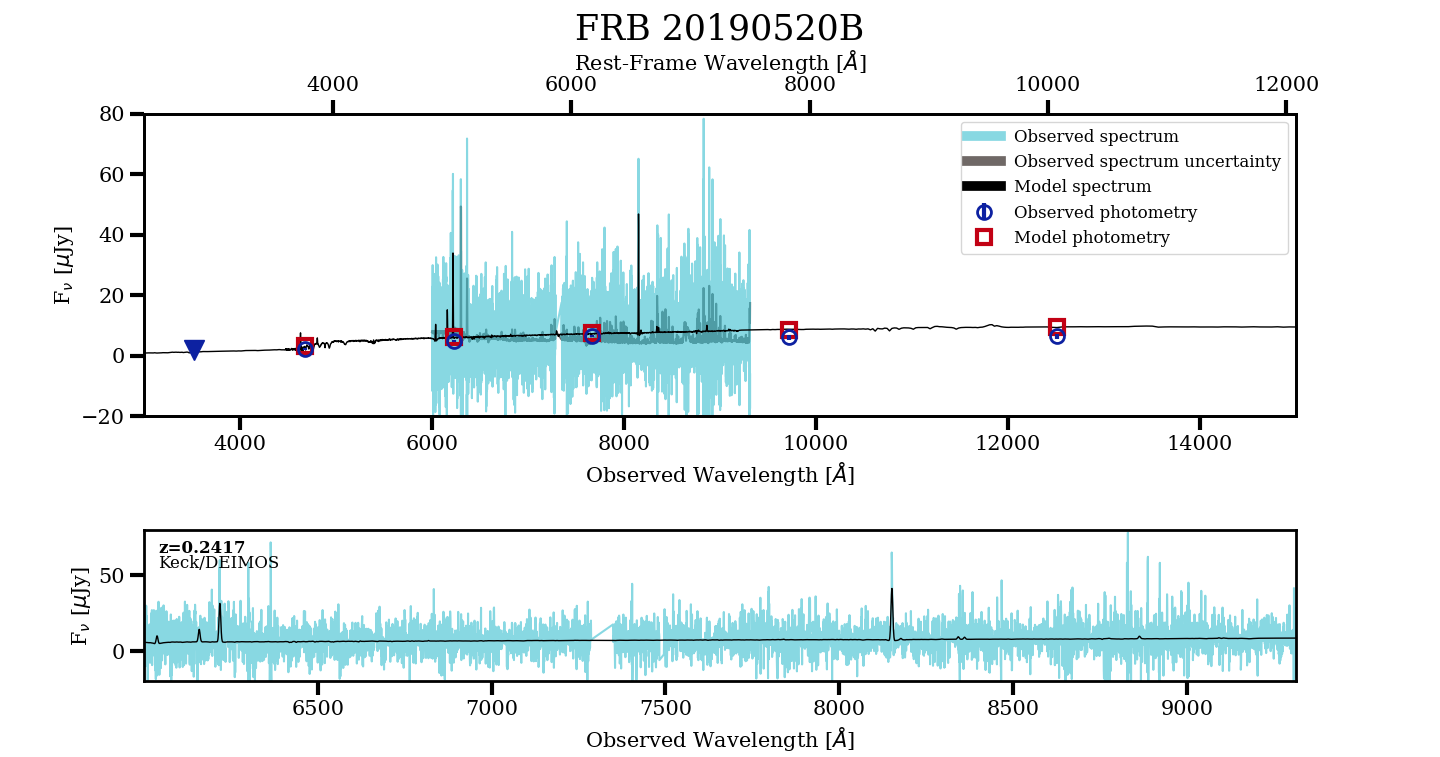}
     \label{fig:190520_SED}
     \caption{SED of FRB\,20190520B.}
\end{figure*}

\begin{figure*}
     \includegraphics[width=\textwidth]{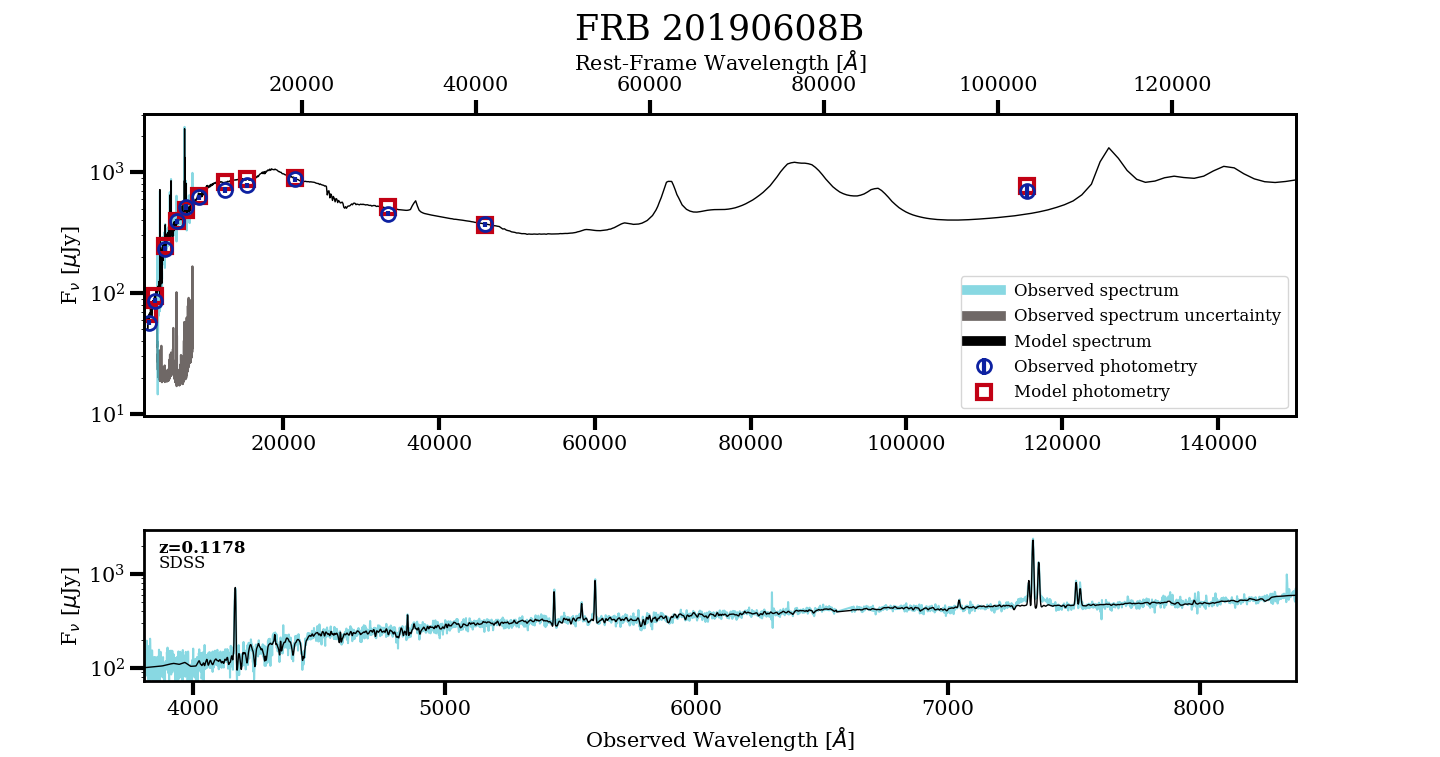}
     \label{fig:190608_SED}
     \caption{SED of FRB\,20190608B.}
\end{figure*}

\begin{figure*}
     \includegraphics[width=\textwidth]{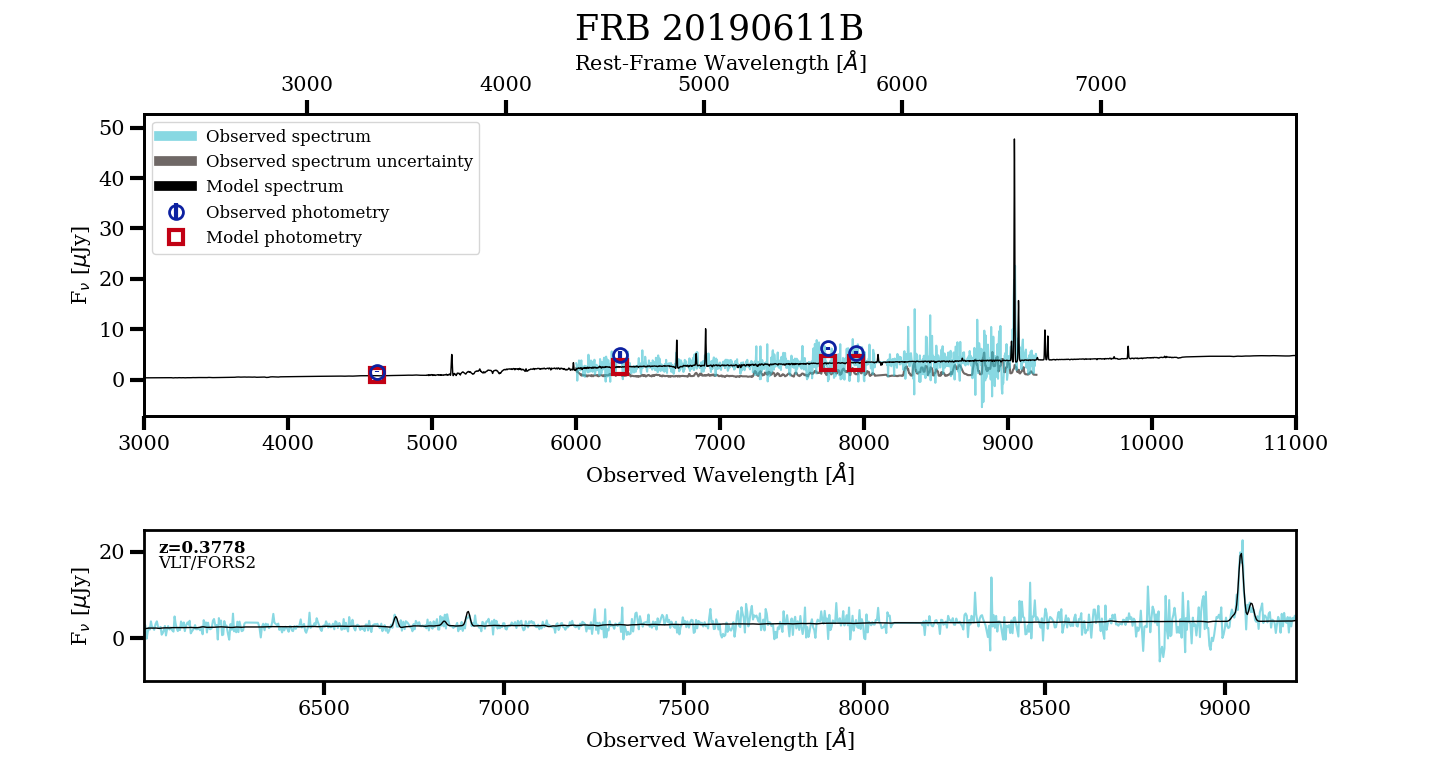}
     \label{fig:190611_SED}
     \caption{SED of FRB\,20190611B.}
\end{figure*}

\begin{figure*}
     \includegraphics[width=\textwidth]{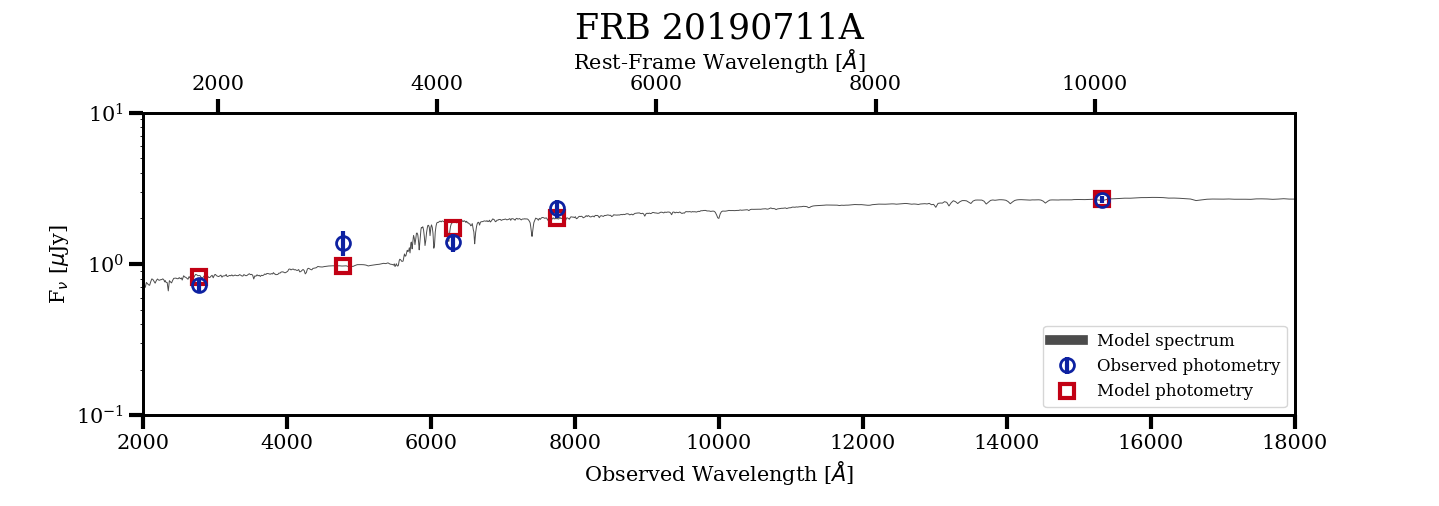}
     \label{fig:190711_SED}
     \caption{SED of FRB\,20190711A.}
\end{figure*}

\begin{figure*}
     \includegraphics[width=\textwidth]{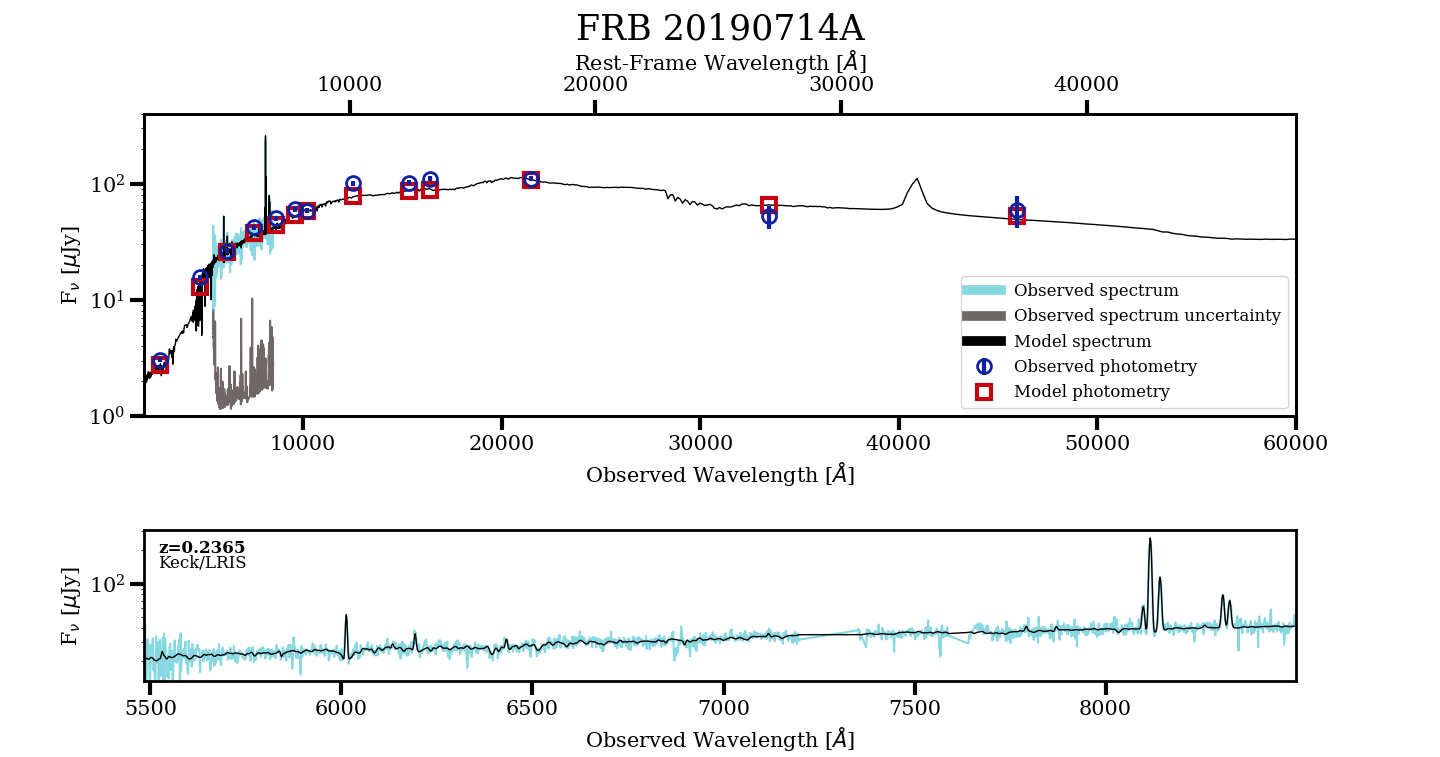}
     \label{fig:190714_SED}
     \caption{SED of FRB\,20190714A.}
\end{figure*}

\begin{figure*}
     \includegraphics[width=\textwidth]{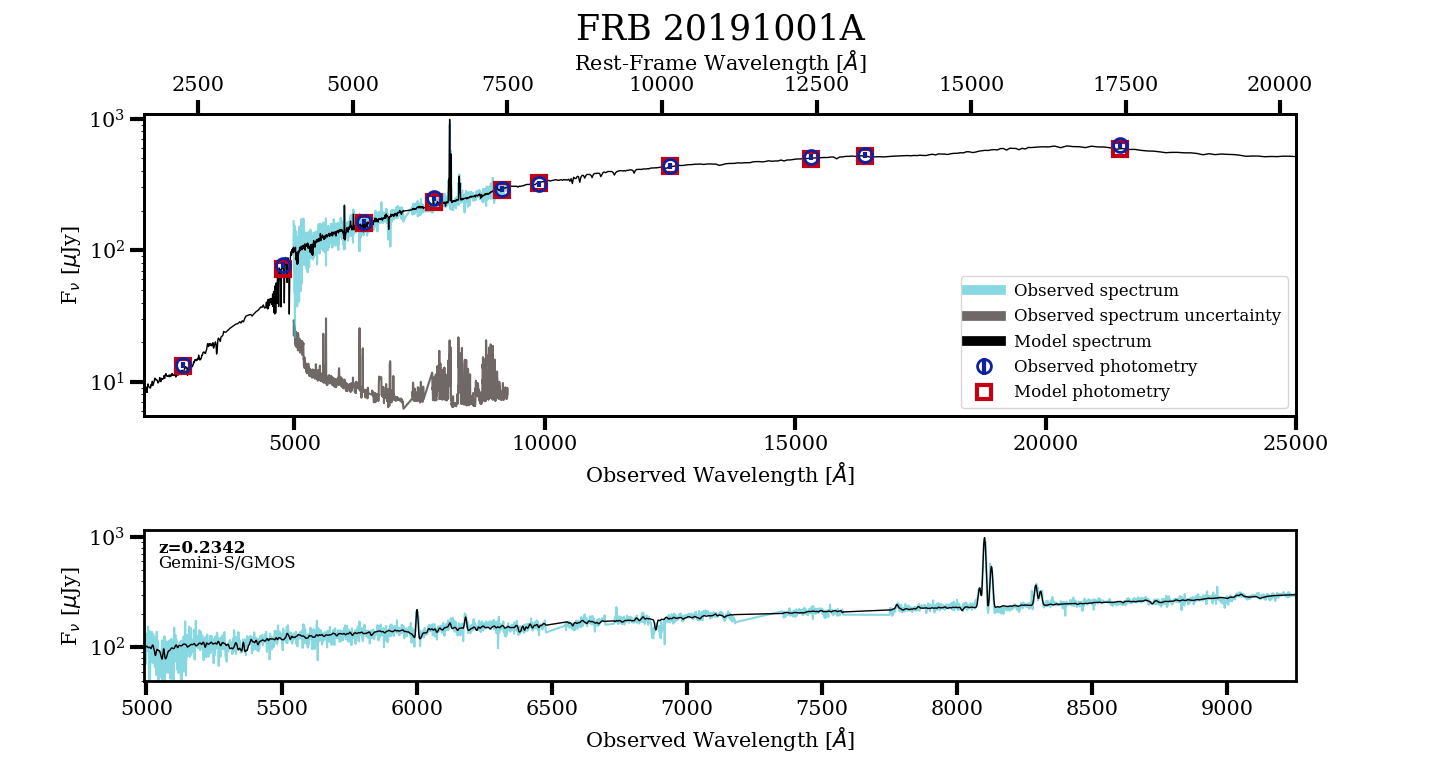}
     \label{fig:191001_SED}
     \caption{SED of FRB\,20191001A.}
\end{figure*}

\begin{figure*}
     \includegraphics[width=\textwidth]{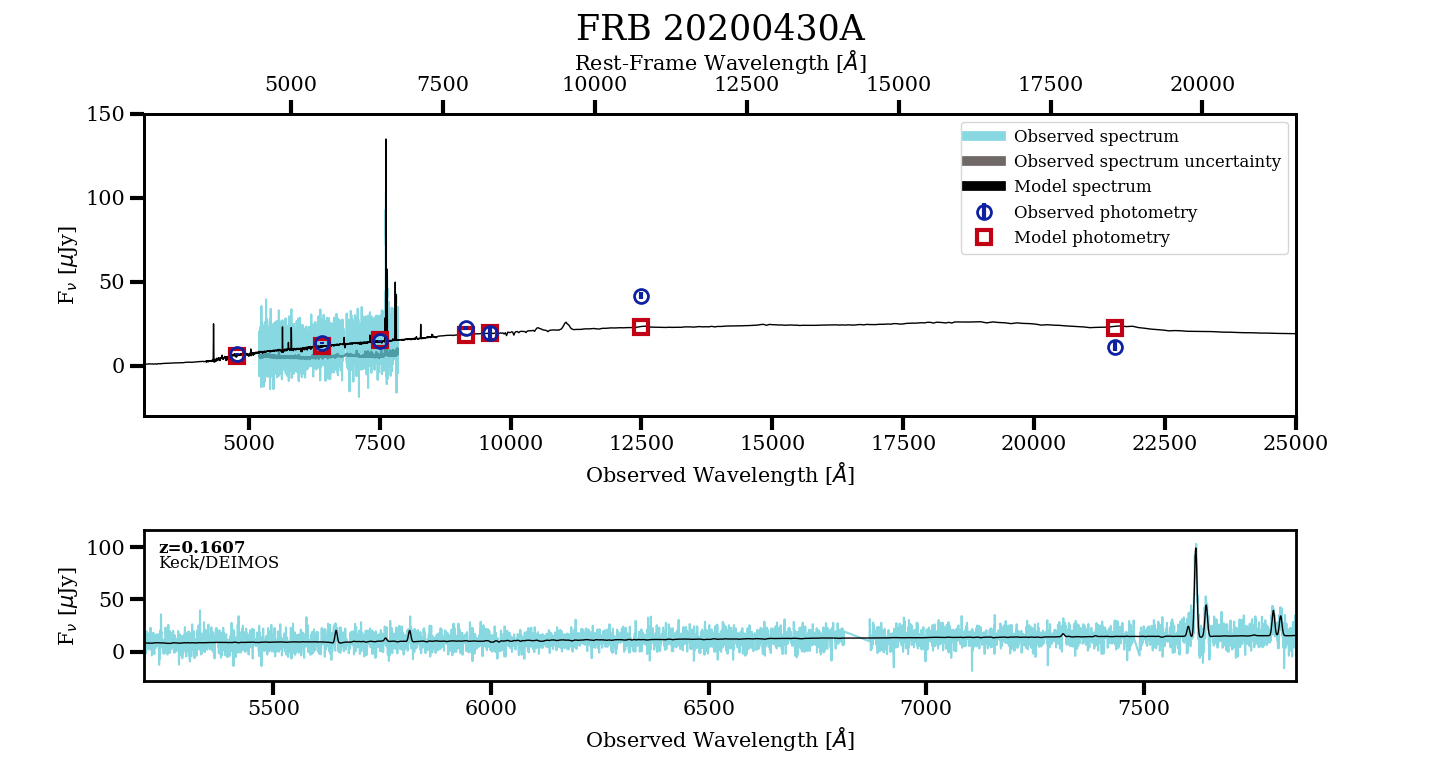}
     \label{fig:200430_SED}
     \caption{SED of FRB\,20200430A.}
\end{figure*}

\begin{figure*}
     \includegraphics[width=\textwidth]{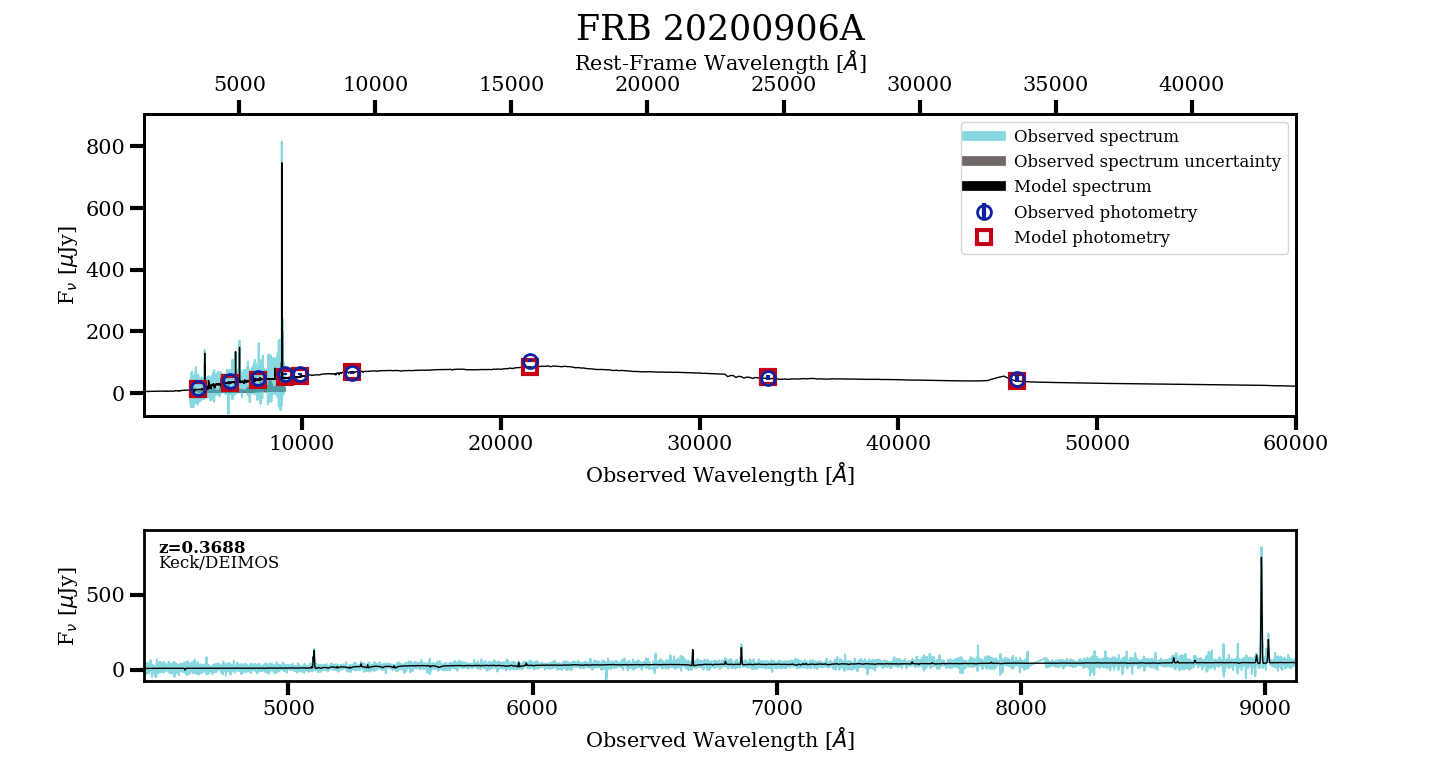}
     \label{fig:200906_SED}
     \caption{SED of FRB\,20200906A.}
\end{figure*}

\begin{figure*}
     \includegraphics[width=\textwidth]{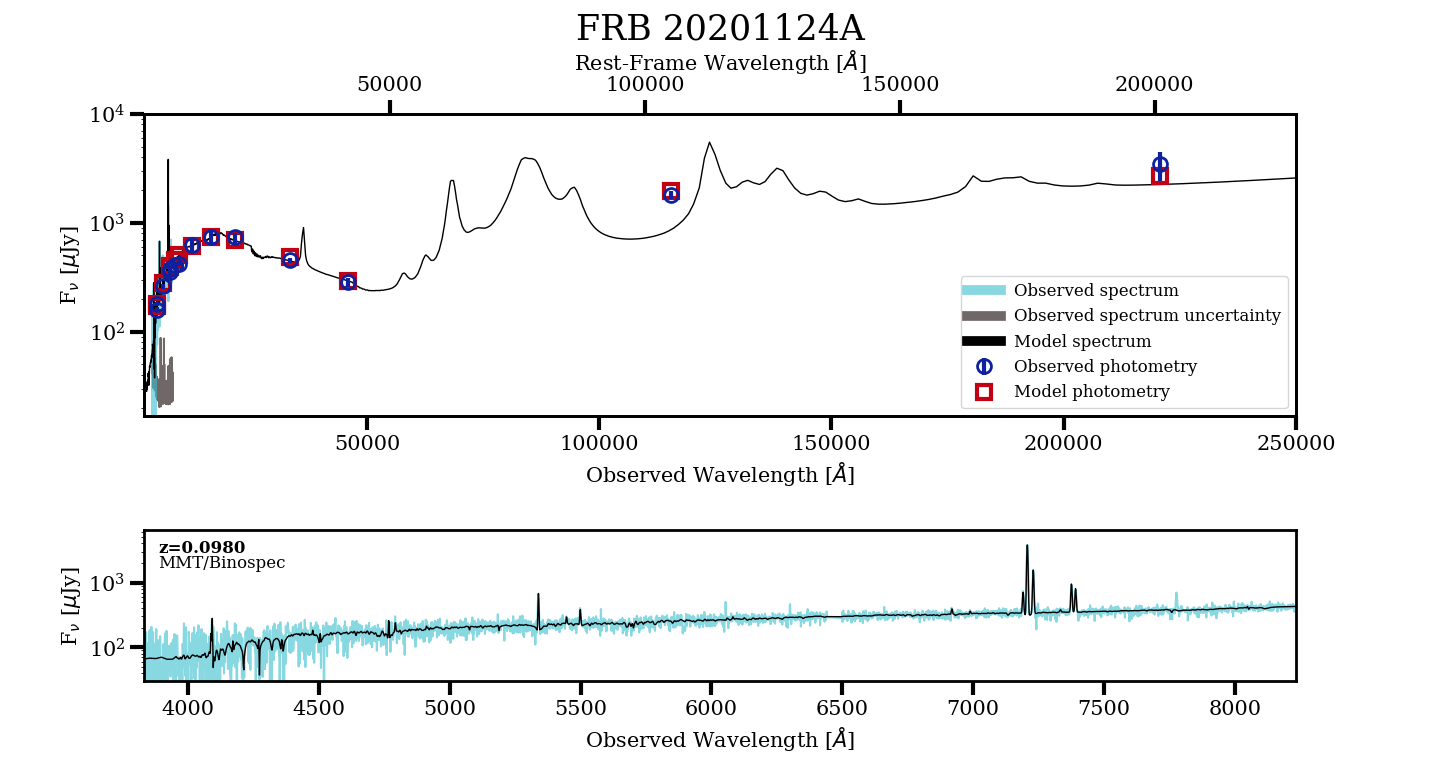}
     \label{fig:201124_SED}
     \caption{SED of FRB\,20201124A.}
\end{figure*}

\begin{figure*}
     \includegraphics[width=\textwidth]{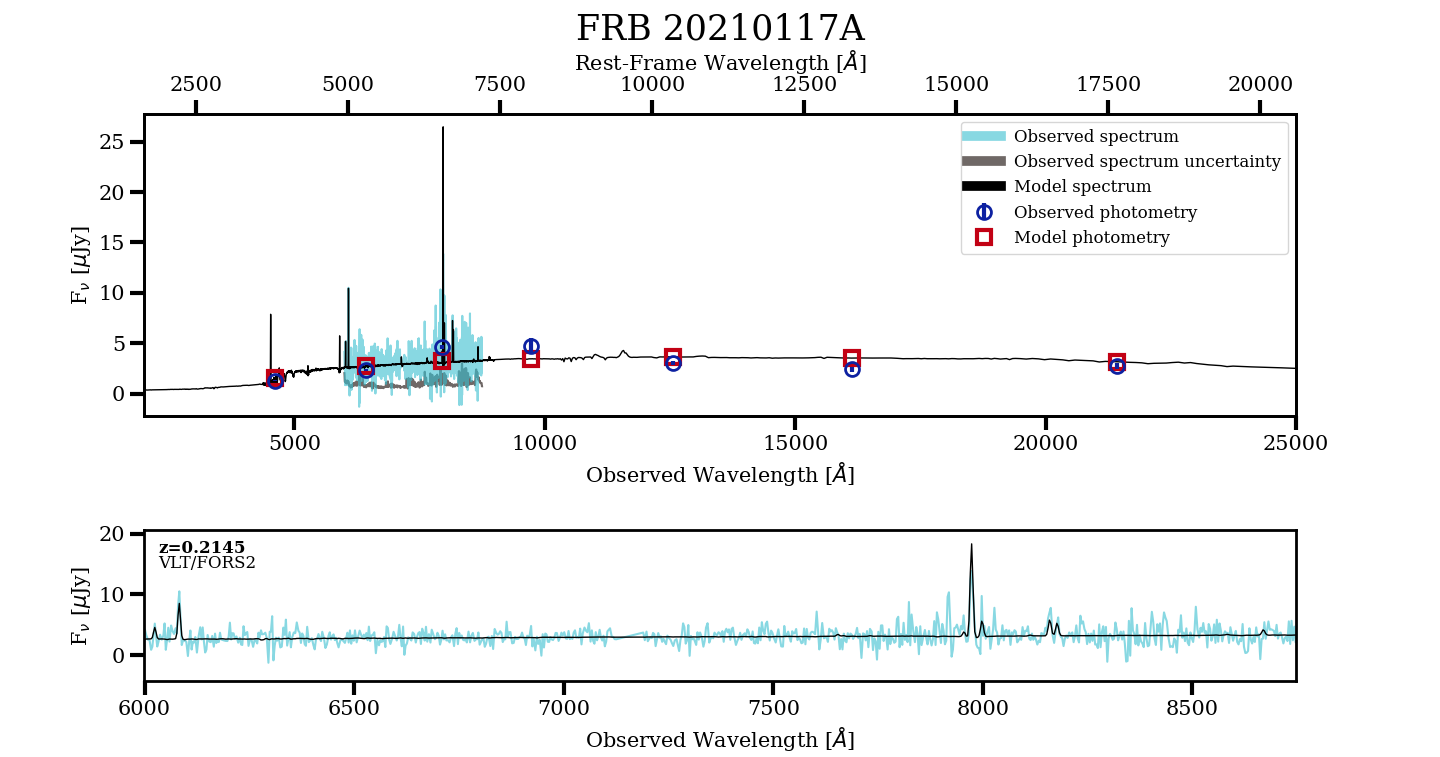}
     \label{fig:210117_SED}
     \caption{SED of FRB\,20210117A.}
\end{figure*}

\begin{figure*}
     \includegraphics[width=\textwidth]{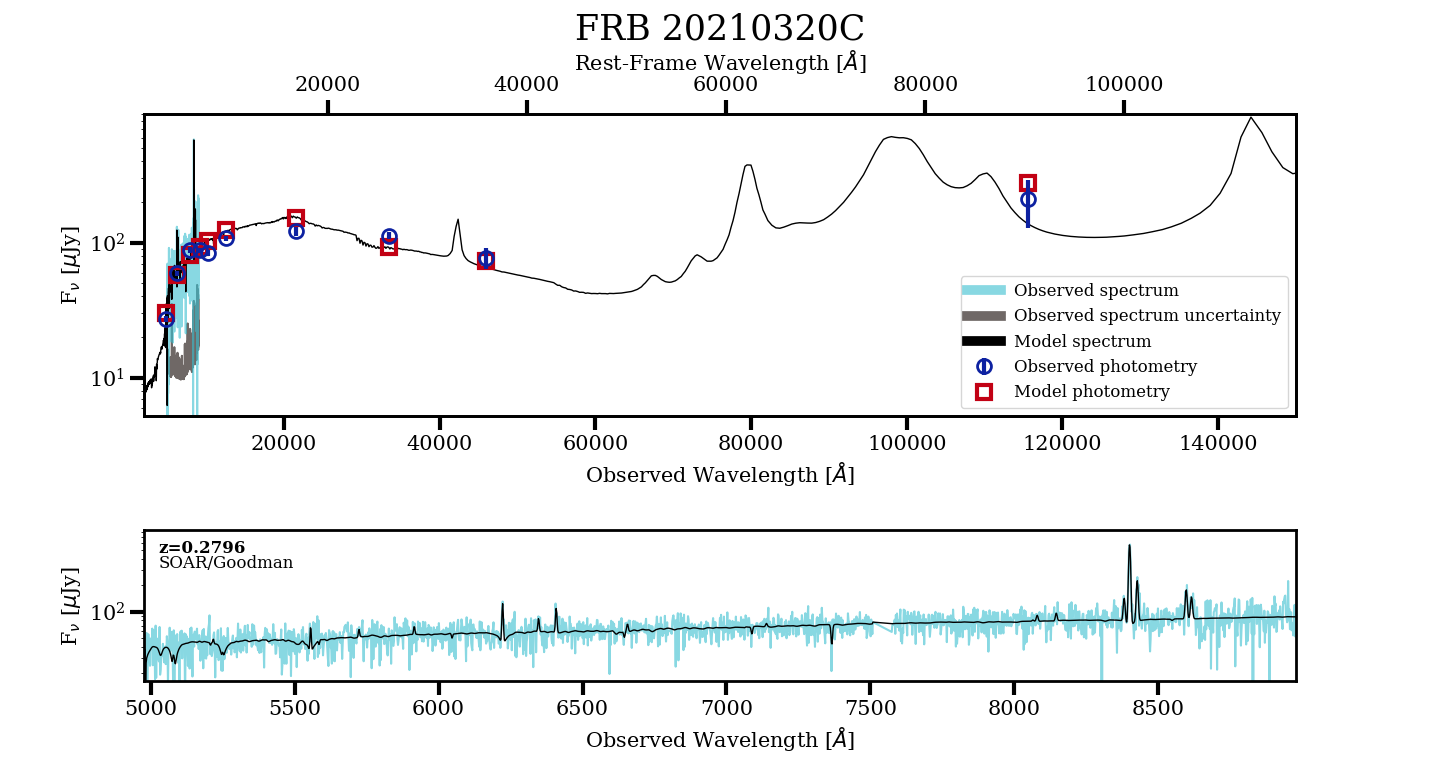}
     \label{fig:210320_SED}
     \caption{SED of FRB\,20210320C.}
\end{figure*}

\begin{figure*}
     \includegraphics[width=\textwidth]{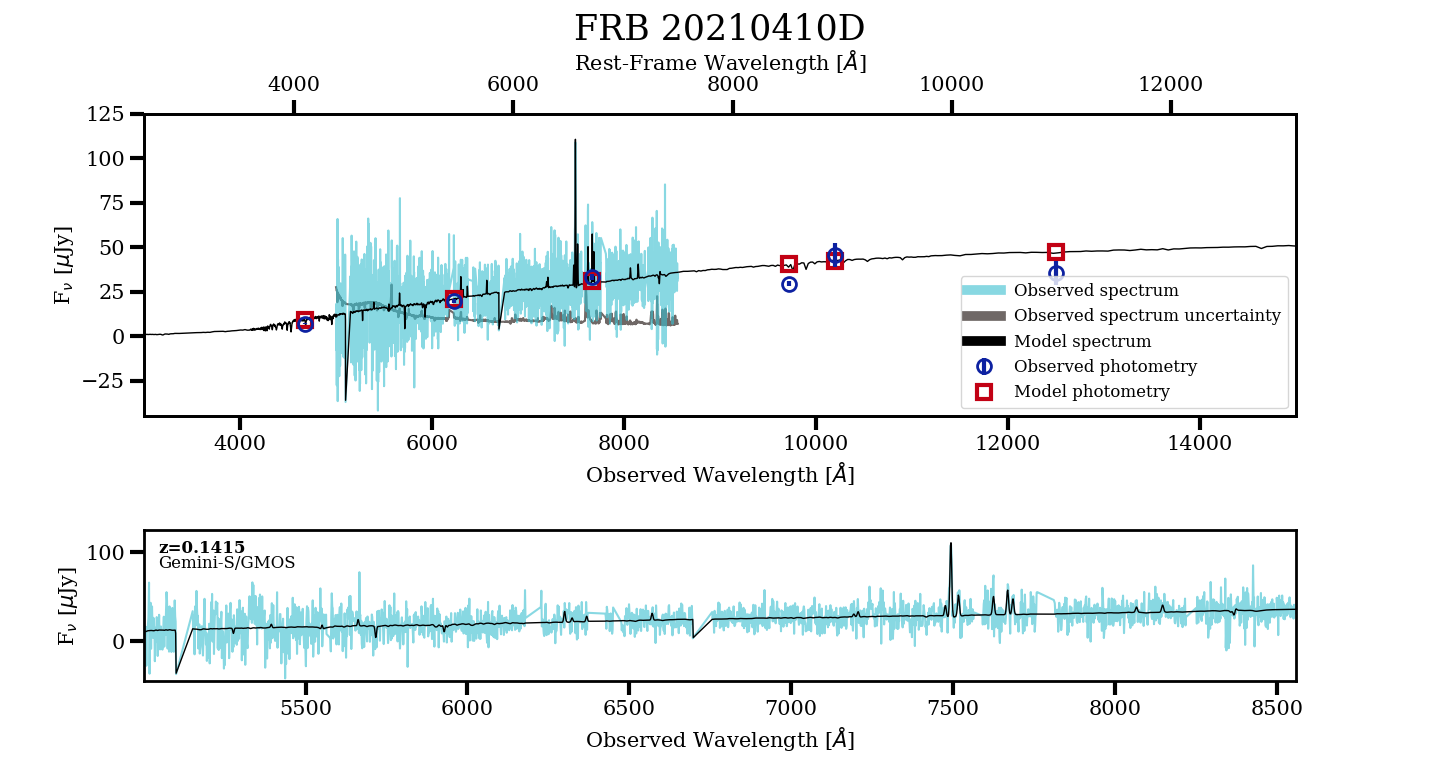}
     \label{fig:210410_SED}
     \caption{SED of FRB\,20210410D.}
\end{figure*}

\begin{figure*}
     \includegraphics[width=\textwidth]{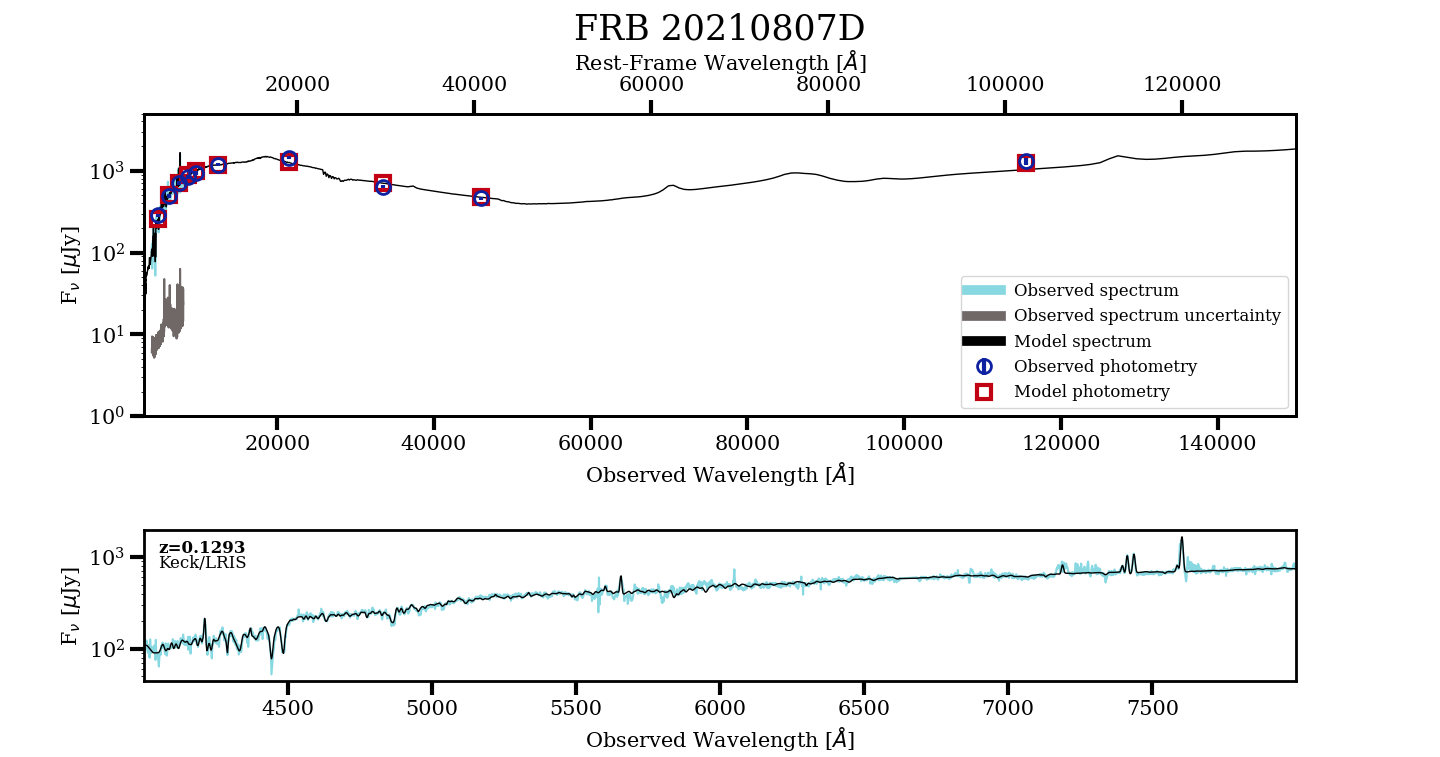}
     \label{fig:210807_SED}
     \caption{SED of FRB\,20210807D.}
\end{figure*}

\begin{figure*}
     \includegraphics[width=\textwidth]{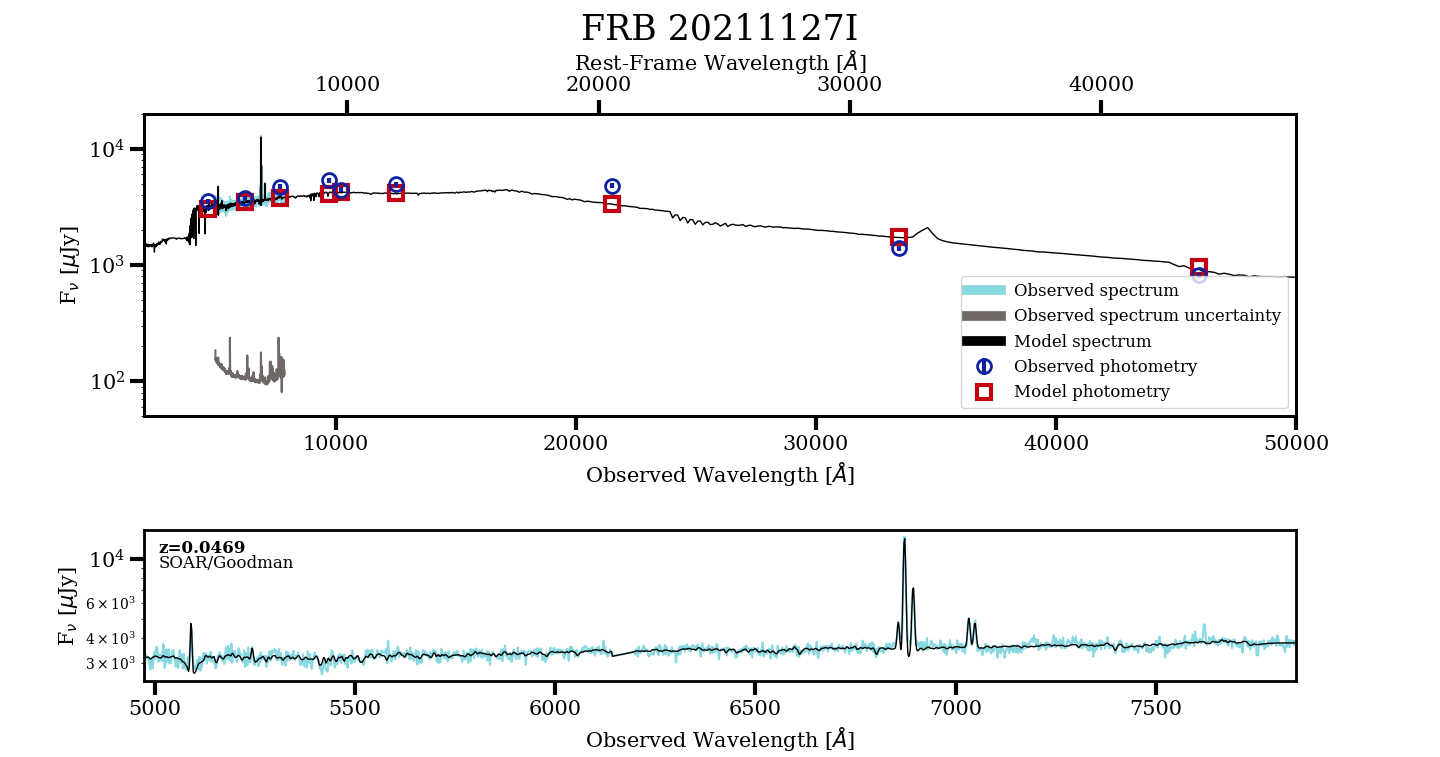}
     \label{fig:211127_SED}
     \caption{SED of FRB\,20211127I.}
\end{figure*}

\begin{figure*}
     \includegraphics[width=\textwidth]{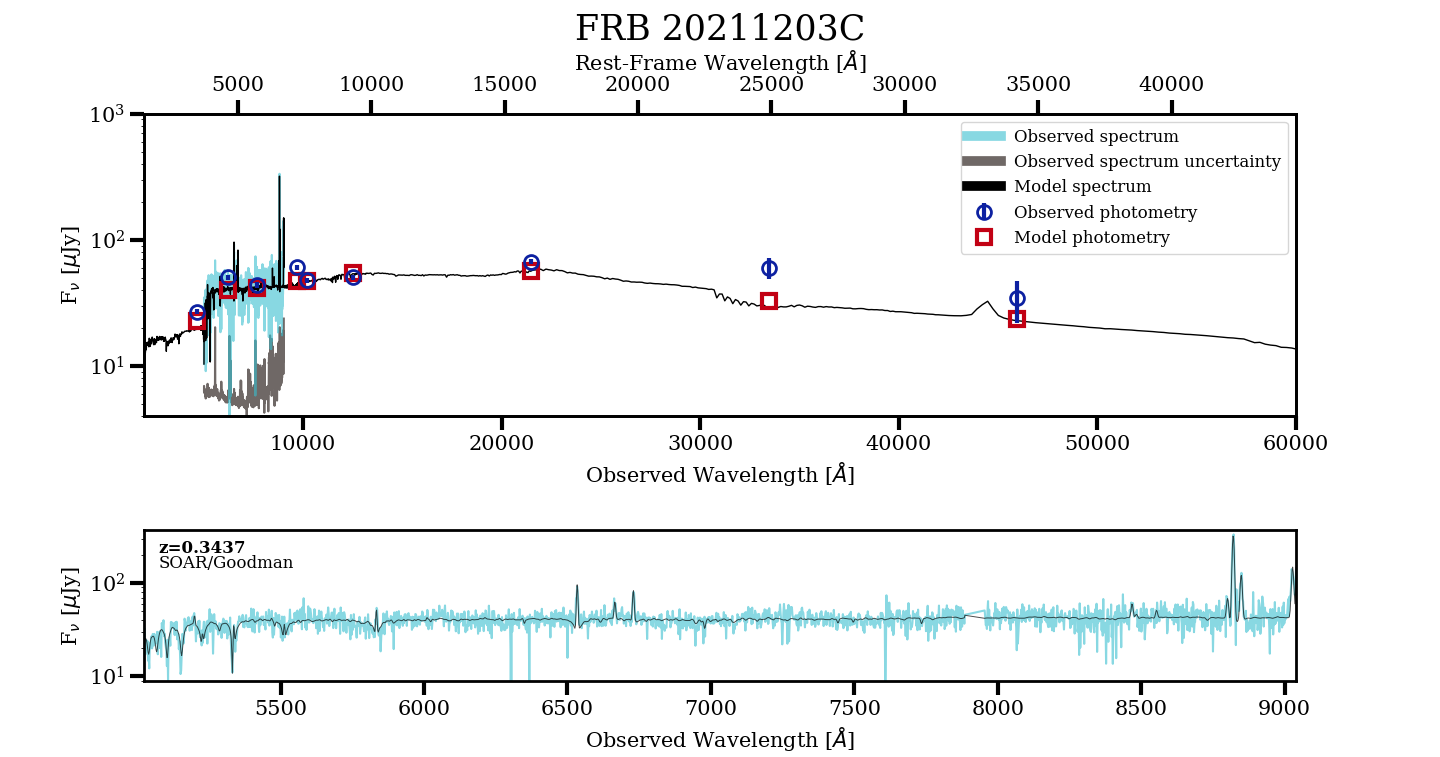}
     \label{fig:211203_SED}
     \caption{SED of FRB\,20211203C.}
\end{figure*}

\begin{figure*}
     \includegraphics[width=\textwidth]{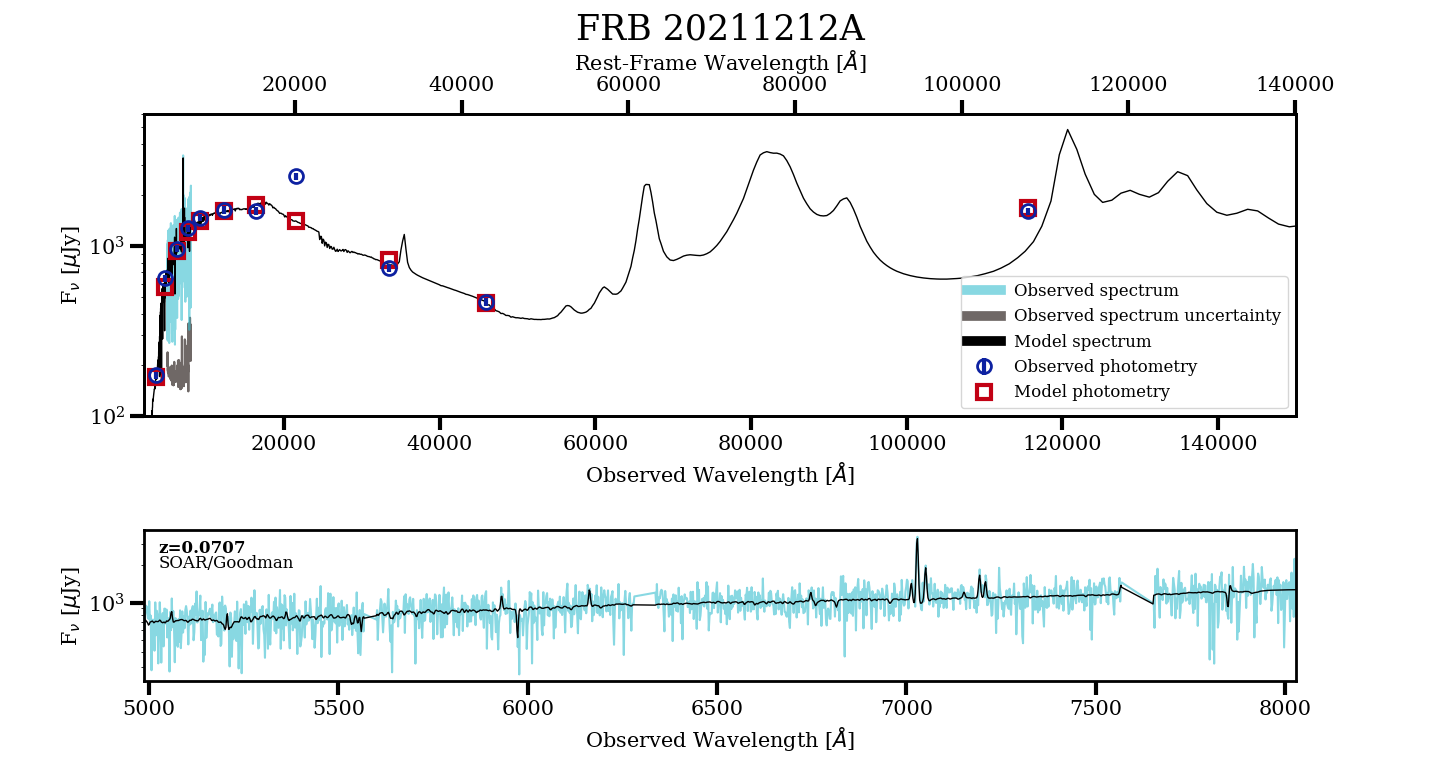}
     \label{fig:211212_SED}
     \caption{SED of FRB\,20211212A.}
\end{figure*}

\begin{figure*}
     \includegraphics[width=\textwidth]{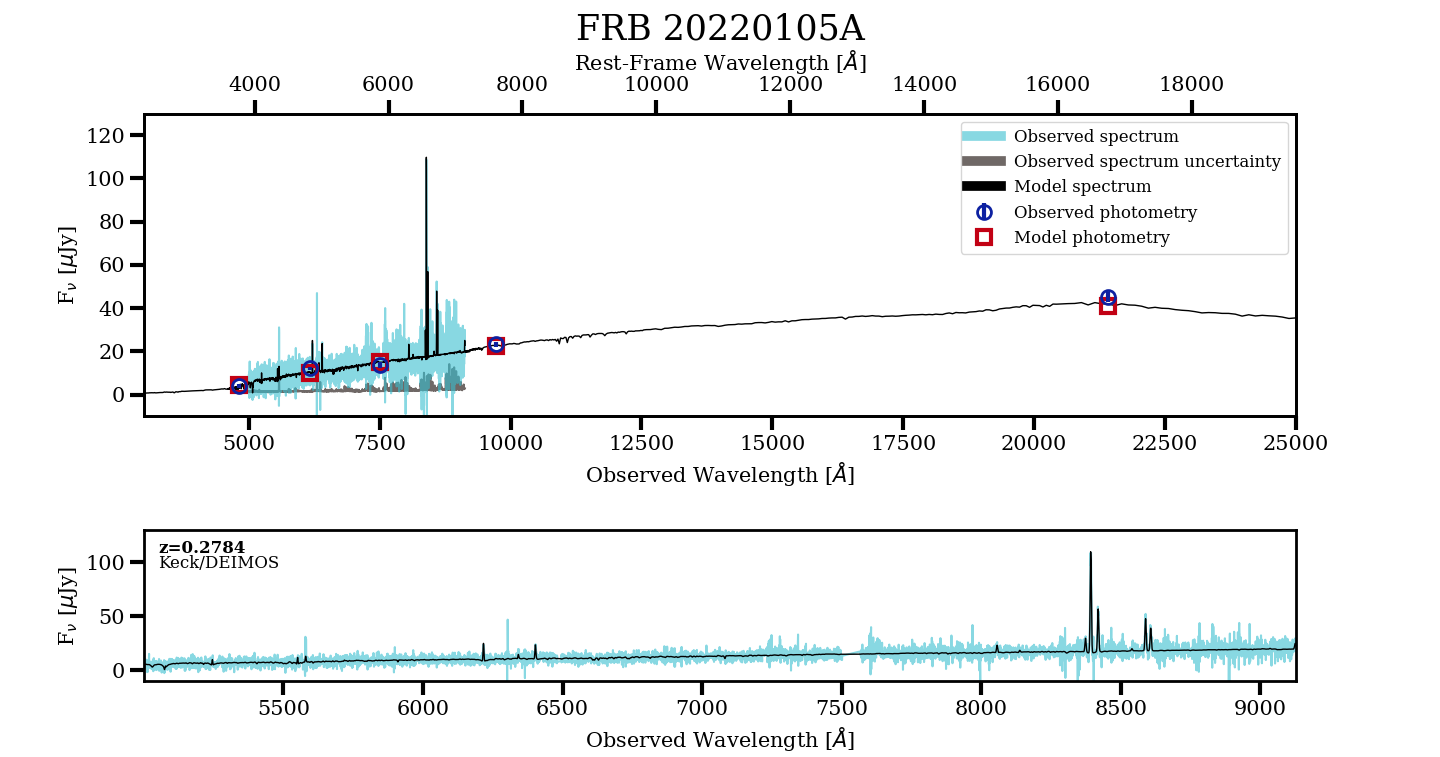}
     \label{fig:220105_SED}
     \caption{SED of FRB\,20220105A.}
\end{figure*}
\clearpage
\section{Photometry} \label{sec: phot}

Here we present all FRB host galaxy photometry used in our modeling. We include details on the facility, instrument, observation date, filter, photometry, and references. See Section~\ref{sec: imaging} for additional details about the observation and data reduction procedures.

\startlongtable
\begin{deluxetable*}{l|ccccccc}
\tabletypesize{\footnotesize}
\tablewidth{0pc}
\tablecaption{Log of FRB Host Galaxy Imaging
\label{tab:imaging}}
\tablehead{
\colhead{FRB}	 &
\colhead{Facility} &
\colhead{Instrument} &
\colhead{Observation Date(s)} &
\colhead{Filter} &
\colhead{Magnitude} &
\colhead{Program ID} &
\colhead{Reference} \\
\colhead{} &  
\colhead{} &  
\colhead{} &  
\colhead{} &  
\colhead{} & 
\colhead{[AB]}  &
\colhead{} &  
\colhead{} 
 }
\startdata
20121102A & Gemini North & GMOS & 2016 Dec 29 UT & g & 23.33$\pm$0.12 & GN-2016B-DD-2 & 1 \\
 & Gemini North & GMOS & 2016 Dec 29 UT & r & 23.73$\pm$0.14 &  & 1 \\
 & Gemini North & GMOS & 2016 Nov 02 UT & i & 23.54$\pm$0.09 &  & 2 \\
 & Gemini North & GMOS & 2016 Nov 02 UT & z & 23.49$\pm$0.13 &  & 2 \\
 & MMT & MMIRS & 2021 Dec 21, 22 UT & J & 23.51$\pm$0.051 & UAO-G177-21B & This Work \\
 & MMT & MMIRS & 2021 Dec 17 UT & Ks & 23.73$\pm$0.59 &  & This Work \\
 & HST & WFC3 & 2017 Feb 23 UT & F110W & 23.08$\pm$0.01 & GO-14890 & 1, 3 \\
 & HST & WFC3 & 2017 Feb 23 UT & F160W & 22.96$\pm$0.03 &  & 1, 3 \\
 & Spitzer & IRAC & 2017 Jan 04 UT & CH1 & 23.79$\pm$0.20 & Obs ID 62322432 & 1 \\
\hline
20180301A & NOT & ALFOSC & 2021 Oct 26-Dec 14 UT & u & 21.77$\pm$0.30 & 62-503 & 4 \\
 & NOT & ALFOSC & 2021 Oct 26-Dec 14 UT & g & 21.64$\pm$0.09 &  & 4 \\
 & NOT & ALFOSC & 2021 Oct 26-Dec 14 UT & r & 21.21$\pm$0.06 &  & 4 \\
 & NOT & ALFOSC & 2021 Oct 26-Dec 14 UT & i & 21.11$\pm$0.06 &  & 4 \\
 & NOT & ALFOSC & 2021 Oct 26-Dec 14 UT & z & 20.66$\pm$0.11 &  & 4 \\
 & MMT & MMIRS & 2021 Feb 27 UT & J & 20.61$\pm$0.07 & UAO-G195-21A & 4 \\
 & MMT & MMIRS & 2021 Feb 27 UT & H & 20.43$\pm$0.08 &  & 4 \\
 & MMT & MMIRS & 2021 Feb 28 UT & Ks & 20.53$\pm$0.08 &  & 4 \\
\hline
20180916B & SDSS &  &  & g & 17.08$\pm$0.08 &  & 5 \\
 & SDSS &  &  & r & 16.17$\pm$0.03 &  & 5 \\
 & SDSS &  &  & i & 15.93$\pm$0.02 &  & 5 \\
 & SDSS &  &  & z & 15.85$\pm$0.06 &  & 5 \\
 & HST & WFC3 & 2020 July 17 UT & F110W & 15.30$\pm$0.01 & 16072 & 3,6 \\
 & WISE &  &  & W1 & 17.04$\pm$0.03 &  & 7 \\
 & WISE &  &  & W2 & 17.73$\pm$0.05 &  & 7 \\
 & WISE &  &  & W3 & 15.71$\pm$0.08 &  & 7 \\
 & WISE &  &  & W4 & 15.67$\pm$0.52 &  & 7 \\
\hline
20180924B & DECaLS & DECam &  & g & 21.42$\pm$0.02 &  & 8 \\
 & DECaLS & DECam &  & r & 20.33$\pm$0.01 &  & 8 \\
 & DES & DECam &  & i & 20.01$\pm$0.01 &  & 9 \\
 & DECaLS & DECam &  & z & 19.56$\pm$0.01 &  & 8 \\
 & DES & DECam &  & Y & 19.65$\pm$0.05 &  & 9 \\
 & HST & WFC3 & 2019 Nov 26 UT & F300X & 23.37$\pm$0.06 & 15878 & 3 \\
 & HST & WFC3 & 2019 Nov 27 UT & F160W & 19.34$\pm$0.002 &  & 3 \\
 & VISTA & VIRCAM &  & J & 19.24$\pm$0.11 &  & 10, This Work \\
 & VISTA & VIRCAM &  & Ks & 18.97$\pm$0.17 &  & 10, This Work \\
 & WISE &  &  & W1 & 19.46$\pm$0.22 &  & 7 \\
\hline
20181112A & DES & DECam &  & g & 22.64$\pm$0.09 &  & 9 \\
 & DES & DECam &  & r & 21.68$\pm$0.05 &  & 9 \\
 & DES & DECam &  & i & 21.46$\pm$0.06 &  & 9 \\
 & DES & DECam &  & z & 21.42$\pm$0.11 &  & 9 \\
 & DES & DECam &  & Y & 21.05$\pm$0.17 &  & 9 \\
 & VISTA & VIRCAM &  & J & 20.96$\pm$0.02 &  & 10, This Work \\
\hline
20190102C & VLT & FORS2 & 2019 Jun 17 UT & u & 22.77$\pm$0.20 & 0103.A-0101(A) & 4 \\
 & VLT & FORS2 & 2019 Jan 12 UT & g & 21.87$\pm$0.10 &  & 4 \\
 & VLT & FORS2 & 2019 Jan 12 UT & I & 20.77$\pm$0.05 &  & 4 \\
 & VLT & FORS2 & 2019 Jun 17 UT & z & 20.54$\pm$0.20 &  & 4 \\
 & HST & WFC3 & 2020 Jan 14 UT & F160W & 20.45$\pm$0.01 & 15878 & 3 \\
\hline
20190520B & SOAR & Goodman & 2022 Sept 01 UT & u & $<$23.23 & SOAR2022B-007 & This Work \\
 & SOAR & Goodman & 2022 Sept 14 UT & g & 23.03$\pm$0.14 &  & This Work \\
 & SOAR & Goodman & 2022 Aug 26 UT & r & 22.16$\pm$0.06 &  & This Work \\
 & SOAR & Goodman & 2022 Aug 20 UT & i & 21.85$\pm$0.07 &  & This Work \\
 & SOAR & Goodman & 2022 Sept 14,29 UT & z & 21.95$\pm$0.16 &  & This Work \\
 & Subaru & MOIRCS & 2020 Aug 05 UT & J & 21.88$\pm$0.14 &  & 11 \\
\hline
20190608B & SDSS &  &  & u & 19.06$\pm$0.04 &  & 5 \\
 & DECaLS & DECam &  & g & 17.98$\pm$0.001 &  & 8 \\
 & DECaLS & DECam &  & r & 17.41$\pm$0.002 &  & 8 \\
 & SDSS &  &  & i & 17.12$\pm$0.01 &  & 5 \\
 & DECaLS & DECam &  & z & 16.92$\pm$0.001 &  & 8 \\
 & HST & WFC3 & 2019 Oct 11 UT & F300X & 19.51$\pm$0.01 & 15878 & 3 \\
 & HST & WFC3 & 2019 Dec 01 UT & F160W & 16.67$\pm$0.001 &  & 3 \\
 & VISTA & VIRCAM &  & J & 16.76$\pm$0.02 &  & 10, This Work \\
 & VISTA & VIRCAM &  & Ks & 16.55$\pm$0.04 &  & 10, This Work \\
 & WISE &  &  & W1 & 16.97$\pm$0.03 &  & 7 \\
 & WISE &  &  & W2 & 17.13$\pm$0.06 &  & 7 \\
 & WISE &  &  & W3 & 15.93$\pm$0.15 &  & 7 \\
\hline
20190611B & VLT & FORS2 & 2020 Sept 19 UT & g & 23.36$\pm$0.09 & 105.204W.001 & 4 \\
 & Gemini South & GMOS & 2019 Sept 26 UT & r & 22.15$\pm$0.15 & GS-2019B-Q-132 & 4 \\
 & Gemini South & GMOS & 2019 Dec 27 UT & i & 21.90$\pm$0.02 &  & 4 \\
 & VLT & FORS2 & 2019 July 12 UT & I & 22.07$\pm$0.07 & 0103.A-0101(A) & 4 \\
\hline
20190711A & Gemini South & GMOS & 2019 Nov 28 UT & g & 23.55$\pm$0.20 & GS-2019B-Q-132 & 12 \\
 & Gemini South & GMOS & 2019 Nov 23, 27 UT & r & 23.54$\pm$0.15 &  & 12 \\
 & Gemini South & GMOS & 2019 Nov 28 UT & i & 22.98$\pm$0.15 &  & 12 \\
 & HST & WFC3 & 2020 May 09 UT & F300X & 24.25$\pm$0.12 & 16080 & 3 \\
 & HST & WFC3 & 2020 May 11 UT & F160W & 22.84$\pm$0.01 &  & 3 \\
\hline
20190714A & Pan-STARRS &  &  & g & 20.91$\pm$0.04 &  & 13 \\
 & Pan-STARRS &  &  & r & 20.34$\pm$0.03 &  & 13 \\
 & Pan-STARRS &  &  & i & 19.84$\pm$0.02 &  & 13 \\
 & Pan-STARRS &  &  & z & 19.64$\pm$0.03 &  & 13 \\
 & Pan-STARRS &  &  & y & 19.44$\pm$0.06 &  & 13 \\
 & HST & WFC3 & 2020 May 19 UT & F300X & 22.68$\pm$0.05 & 16080 & 3 \\
 & HST & WFC3 & 2020 Apr 30 UT & F160W & 18.88$\pm$0.002 & & 3 \\
 & VISTA & VIRCAM &  & Y & 19.48$\pm$0.02 &  & 10, This Work \\
 & VISTA & VIRCAM &  & J & 18.90$\pm$0.01 &  & 10, This Work \\
 & VISTA & VIRCAM &  & H & 18.80$\pm$0.01 &  & 10, This Work \\
 & VISTA & VIRCAM &  & Ks & 18.79$\pm$0.01 &  & 10, This Work \\
 & WISE &  &  & W1 & 19.31$\pm$0.24 &  & 7 \\
 & WISE &  &  & W2 & 19.11$\pm$0.33 &  & 7 \\
\hline
20191001A & DECaLS & DECam &  & g & 19.18$\pm$0.01 &  & 8 \\
 & DECaLS & DECam &  & r & 18.36$\pm$0.003 & & 8 \\
 & DES & DECam &  & i & 17.92$\pm$0.002 &  & 9 \\
 & DECaLS & DECam &  & z & 17.73$\pm$0.004 &  & 8 \\
 & DES & DECam &  & Y & 17.64$\pm$0.01 &  & 9 \\
 & HST & WFC3 & 2020 Apr 25 UT & F300X & 21.07$\pm$0.02 & 16080 & 3 \\
 & HST & WFC3 & 2020 Apr 28 UT & F160W & 17.12$\pm$0.001 &  & 3 \\
 & VISTA & VIRCAM &  & J & 17.30$\pm$0.01 &  & 10, This Work \\
 & VISTA & VIRCAM &  & H & 17.09$\pm$0.01 &  & 10, This Work \\
 & VISTA & VIRCAM &  & Ks & 16.90$\pm$0.01 &  & 10, This Work \\
\hline
20200430A & DECaLS & DECam &  & g & 21.78$\pm$0.03 &  & 8 \\
 & DECaLS & DECam &  & r & 21.05$\pm$0.02 &  & 8 \\
 & Pan-STARRS &  &  & i & 20.98$\pm$0.05 &  & 13 \\
 & DECaLS & DECam &  & z & 20.52$\pm$0.03 &  & 8 \\
 & Pan-STARRS &  &  & y & 20.68$\pm$0.18 &  & 13 \\
 & MMT & MMIRS & 2022 June 12 UT & J & 19.85$\pm$0.05 & UAO-G193-22A & This Work \\
 & MMT & MMIRS & 2022 June 15 UT & Ks & 21.26$\pm$0.25 &  & This Work \\
\hline
20200906A & DES & DECam &  & g & 20.84$\pm$0.01 &  & 9 \\
 & DES & DECam &  & r & 19.95$\pm$0.01 &  & 9 \\
 & DES & DECam &  & i & 19.69$\pm$0.01 &  & 9 \\
 & DES & DECam &  & z & 19.43$\pm$0.02 &  & 9 \\
 & DES & DECam &  & Y & 19.40$\pm$0.06 &  & 9 \\
 & VISTA & VIRCAM &  & J & 19.36$\pm$0.01 &  & 10, This Work \\
 & VISTA & VIRCAM &  & Ks & 18.84$\pm$0.01 &  & 10, This Work \\
 & WISE &  &  & W1 & 19.35$\pm$0.15 &  & 7 \\
 & WISE &  &  & W2 & 19.36$\pm$0.23 &  & 7 \\
\hline
20201124A & Pan-STARRS &  &  & g & 18.40$\pm$0.04 &  & 13 \\
 & Keck & LRIS & 2022 Oct 29 UT & G & 18.26$\pm$0.02 & O287 & This Work \\
 & Pan-STARRS &  &  & r & 17.86$\pm$0.03 &  & 13 \\
 & Pan-STARRS &  &  & i & 17.53$\pm$0.03 &  & 13 \\
 & Keck & LRIS & 2022 Oct 29 UT & I & 17.46$\pm$0.02 & O287 & This Work \\
 & Pan-STARRS &  &  & z & 17.36$\pm$0.03 &  & 13 \\
 & Pan-STARRS &  &  & y & 17.34$\pm$0.06 &  & 13 \\
 & 2MASS &  &  & J & 16.92$\pm$0.12 &  & 14 \\
 & 2MASS &  &  & H & 16.74$\pm$0.12 &  & 14 \\
 & 2MASS &  &  & Ks & 16.74$\pm$0.12 &  & 14 \\
 & WISE &  &  & W1 & 17.01$\pm$0.04 &  & 7 \\
 & WISE &  &  & W2 & 17.34$\pm$0.05 &  & 7 \\
 & WISE &  &  & W3 & 15.01$\pm$0.06 &  & 7 \\
 & WISE &  &  & W4 & 14.59$\pm$0.26 &  & 7 \\
\hline
20210117A & VLT & FORS2 & 2021 June 12 UT & g & 23.06$\pm$0.02 & 105.204W.001 & 15 \\
 & Keck & DEIMOS & 2021 June 10, 11 UT & R & 22.97$\pm$0.04 & O316 & 15 \\
 & VLT & FORS2 & 2021 June 12 UT & I & 22.23$\pm$0.05 & 105.204W.001 & 15 \\
 & SOAR & Goodman & 2022 Nov 14 UT & z & 22.20$\pm$0.16 & SOAR2022B-007 & This Work \\
 & VLT & HAWK-I & 2022 June 10 UT & J & 22.69$\pm$0.08 & 108.21ZF.005 & 15 \\
 & VLT & HAWK-I & 2022 June 10 UT & H & 22.94$\pm$0.10 &  & 15 \\
 & VLT & HAWK-I & 2022 June 10 UT & Ks & 22.80$\pm$0.10 &  & 15 \\
\hline
20210320C & Pan-STARRS &  &  & g & 20.31$\pm$0.04 &  & 13, This Work \\
 & SOAR & Goodman & 2022 Feb 02 UT & r & 19.47$\pm$0.02 & SOAR2021B-002 & This Work \\
 & VLT & FORS2 & 2021 Apr 15 UT & I & 19.04$\pm$0.005 & 105.204W.003 & This Work \\
 & Gemini South & GMOS & 2022 Jan 19 UT & z & 19.04$\pm$0.03 & GS-2021B-Q-138 & This Work \\
 & VISTA & VIRCAM &  & J & 19.09$\pm$0.04 &  & 10, This Work \\
 & VISTA & VIRCAM &  & H & 18.82$\pm$0.05 &  & 10, This Work \\
 & VISTA & VIRCAM &  & Ks & 18.68$\pm$0.08 &  & 10, This Work \\
 & WISE &  &  & W1 & 18.77$\pm$0.07 &  & 7 \\
 & WISE &  &  & W2 & 19.18$\pm$0.19 &  & 7 \\
 & WISE &  &  & W3 & 18.10$\pm$0.42 &  & 7 \\
\hline
20210410D & SOAR & Goodman & 2022 Aug 10, 20 UT & g & 21.77$\pm$0.05 & SOAR2022B-007 & 16 \\
 & SOAR & Goodman & 2021 July 19 UT & r & 20.65$\pm$0.03 & SOAR2021A-010 & 16 \\
 & SOAR & Goodman & 2022 Aug 20 UT & i & 20.10$\pm$0.02 & SOAR2022B-007 & 16 \\
 & SOAR & Goodman & 2022 Sept 03 UT & z & 20.23$\pm$0.04 &  & 16 \\
 & VISTA & VIRCAM &  & Y & 19.76$\pm$0.16 &  & 10, 16 \\
 & VISTA & VIRCAM &  & J & 20.02$\pm$0.21 &  & 10, 16 \\
\hline
20210807D & Pan-STARRS &  &  & g & 17.76$\pm$0.01 &  & 13, This Work \\
 & Pan-STARRS &  &  & r & 17.17$\pm$0.01 &  & 13, This Work \\
 & Pan-STARRS &  &  & i & 16.77$\pm$0.01 &  & 13, This Work \\
 & Pan-STARRS &  &  & z & 16.58$\pm$0.01 &  & 13, This Work \\
 & Pan-STARRS &  &  & y & 16.46$\pm$0.02 &  & 13, This Work \\
 & VISTA & VIRCAM &  & J & 16.21$\pm$0.01 &  & 10, This Work \\
 & VISTA & VIRCAM &  & Ks & 16.00$\pm$0.02 &  & 10, This Work \\
 & WISE &  &  & W1 & 16.88$\pm$0.04 &  & 7 \\
 & WISE &  &  & W2 & 17.24$\pm$0.04 &  & 7 \\
 & WISE &  &  & W3 & 16.11$\pm$0.10 &  & 7 \\
\hline
20211127I & SOAR & Goodman & 2022 Jan 07, 27 UT & g & 15.03$\pm$0.01 & SOAR2021B-002 & This Work \\
 & SOAR & Goodman & 2022 Jan 07, 27 UT & r & 14.96$\pm$0.01 &  & This Work \\
 & SOAR & Goodman & 2022 Jan 07 UT & i & 14.72$\pm$0.01 &  & This Work \\
 & SOAR & Goodman & 2022 Feb 02 UT & z & 14.57$\pm$0.01 &  & This Work \\
 & VISTA & VIRCAM &  & Y & 14.78$\pm$0.01 &  & 10, This Work \\
 & VISTA & VIRCAM &  & J & 14.66$\pm$0.01 &  & 10, This Work \\
 & VISTA & VIRCAM &  & Ks & 14.70$\pm$0.01 &  & 10, This Work \\
 & WISE &  &  & W1 & 16.05$\pm$0.01 &  & 7 \\
 & WISE &  &  & W2 & 16.62$\pm$0.02 &  & 7 \\
 & WISE &  &  & W3 & 14.99$\pm$0.03 &  & 7 \\
 & WISE &  &  & W4 & 14.17$\pm$0.11 &  & 7 \\
\hline
20211203C & SOAR & Goodman & 2022 Feb 01 UT & g & 20.32$\pm$0.02 & SOAR2021B-002 & This Work \\
 & SOAR & Goodman & 2022 Jan 28 UT & r & 19.64$\pm$0.03 &  & This Work \\
 & SOAR & Goodman & 2022 Jan 27 UT & i & 19.79$\pm$0.01 &  & This Work \\
 & SOAR & Goodman & 2022 Jan 27 UT & z & 19.44$\pm$0.03 &  & This Work \\
 & VISTA & VIRCAM &  & Y & 19.70$\pm$0.02 &  & 10, This Work \\
 & VISTA & VIRCAM &  & J & 19.64$\pm$0.02 &  & 10, This Work \\
 & VISTA & VIRCAM &  & Ks & 19.33$\pm$0.01 &  & 10, This Work \\
 & WISE &  &  & W1 & 19.45$\pm$0.20 &  & 7 \\
 & WISE &  &  & W2 & 20.05$\pm$0.40 &  & 7 \\
\hline
20211212A & SDSS &  &  & u & 18.30$\pm$0.03 &  & 5 \\
 & SOAR & Goodman & 2022 Jan 25 UT & g & 16.88$\pm$0.01 & SOAR2021B-002 & This Work \\
 & SOAR & Goodman & 2022 Jan 25 UT & r & 16.44$\pm$0.01 &  & This Work \\
 & SOAR & Goodman & 2022 Jan 25, 27 UT & i & 16.14$\pm$0.01 &  & This Work \\
 & Gemini South & GMOS & 2022 Jan 18 UT & z & 15.98$\pm$0.03 & GS-2021B-Q-138 & This Work \\
 & 2MASS &  &  & J & 15.87$\pm$0.03 &  & 14 \\
 & 2MASS &  &  & H & 15.89$\pm$0.03 &  & 14 \\
 & 2MASS &  &  & Ks & 15.38$\pm$0.01 &  & 14 \\
 & WISE &  &  & W1 & 16.72$\pm$0.02 &  & 7 \\
 & WISE &  &  & W2 & 17.22$\pm$0.03 &  & 7 \\
 & WISE &  &  & W3 & 15.89$\pm$0.05 &  & 7 \\
\hline
20220105A & Pan-STARRS &  &  & g & 22.36$\pm$0.21 &  & 13, This Work \\
 & Pan-STARRS &  &  & r & 21.19$\pm$0.08 &  & 13, This Work \\
 & Pan-STARRS &  &  & i & 21.05$\pm$0.11 &  & 13, This Work \\
 & SOAR & Goodman & 2022 Jan 25 UT & z & 20.48$\pm$0.01 & SOAR2021B-002 & This Work \\
 & VLT & HAWK-I & 2022 Mar 24 UT & Ks & 19.76$\pm$0.02 & 108.21ZF.005 & This Work
\enddata
\tablecomments{All imaging observations included in this work. All photometry is corrected for Galactic extinction using the \cite{Fitzpatrick:2007} extinction law. \\
References:
1. \citet{Bassa+17},
2. \citet{Tendulkar+17},
3. \citet{Mannings+21},
4. \citet{bha+22},
5. \citet{SDSS}, 
6. \citet{Tendulkar+21}, 
7. \citet{WISE}, 
8. \citet{DECALS},
9. \citet{DES},
10. \citet{VISTA},
11. \citet{Niu+22},
12. \citet{Heintz+20},
13. \citet{PS1},
14. \citet{2MASS},
15. \citet{Bhandari_210117},
16. \citet{Caleb+23}}
\end{deluxetable*}

\section{Prospector Priors} \label{sec: priors}

Here we list the full details of the priors used in the \texttt{Prospector} modeling. We define each prior, list its range or fixed value, and note any additional relevant information.

\startlongtable
\begin{deluxetable*}{l|p{60mm}p{45mm}p{40mm}}
\tablewidth{0pc}
\tablecaption{Priors
\label{tab:priors}}
\tablehead{
\colhead{Parameter}	 &
\colhead{Definition} &
\colhead{Prior Range or Value} &
\colhead{Notes}
}
\startdata
z & Spectroscopic host redshift & $\mathcal{U}(z-0.01,z+0.01)$ &  \\
dust2 & Dust attenuation of stellar light & $C\mathcal{N}$($\mu$=0.3,$\sigma$=1,\text{min}=0.0,\text{max}=4.0) &  \\
dust1\underline{{ }}fraction & Optical depth of dust attenuating young stars and nebular emission & $C\mathcal{N}$($\mu$=0.3,$\sigma$=1,\text{min}=0.0,\text{max}=2.0) &  \\
dust1 & Dust attenuation of young stellar light & dust2$\times$dust1\underline{{ }}fraction &  \\
dust\underline{{ }}index & Power-law modification of slope from the \citet{Calzetti00} attenuation curve & $\mathcal{U}$(-1.0,4.0) &  \\
logsfr\underline{{ }}ratios & Ratio of star formation rate between adjacent agebins & $\mathcal{T}$($\mu$=0,$\sigma$=0.3,$\nu$=2) &  \\
log(M/M$\odot$) & Total mass formed & (8.0, 12.0) & Follows \citet{Gallazzi2005} mass-metallicity relation \\
log(Z/Z$\odot$) & Metallicity & (-2.0, 0.19) & Follows \citet{Gallazzi2005} mass-metallicity relation \\
duste\underline{{ }}qpah & Grain size distribution of polycyclic aromatic hydrocarbons & $\mathcal{U}$(0.5,7.0) & Only included if data $\geq 2\mu$m are available \\
fagn & Fraction of total AGN luminosity relative to the bolometric stellar luminosity & $L\mathcal{U}$(1$e-$5,3.0) & Only included if data $\geq 2\mu$m are available \\
agn\underline{{ }}tau & Optical depth of the AGN dust torus & $L\mathcal{U}$(5.0,150.0) & Only included if data $\geq 2\mu$m are available \\
gas\underline{{ }}logz & Gas-phase metallicity & $\mathcal{U}$(-2.0,0.5) & Only included if spectrum is used \\
gas\underline{{ }}logu & Gas ionization parameter & $\mathcal{U}$(-4,-1) & Only included if spectrum is used \\
eline\underline{{ }}sigma & Emission line width & $\mathcal{U}$(30,300) & Only included if spectrum and nebular marginalization used \\
sigma\underline{{ }}smooth & Velocity dispersion in km~s$^{-1}$ & $\mathcal{U}$(40.0,400.0) & Only included if spectrum is used \\
f\underline{{ }}outlier\underline{{ }}spec & Fraction of spectral pixels that are considered outliers & $\mathcal{U}$(1$e-$5,0.5) & Only if spectrum used \\
spec\underline{{ }}jitter & Multiplicative noise inflation term in spectroscopic pixels & $\mathcal{U}$(1.0,3.0) & Only if spectrum used \\
\hline
SFH & Continuity SFH & 3 &  \\
imf\underline{{ }}type & \citet{Kroupa01} IMF & 2 &  \\
dust\underline{{ }}type & \citet{KriekandConroy13} dust attenuation curve & 4 & \\
smoothtype & Type of spectral smoothing & vel & Smoothing in velocity dispersion \\
fftsmooth & Use Fast Fourier Transform to perform spectral smoothing & True &  \\
add\underline{{ }}neb\underline{{ }}emission & Turn on nebular emission & True &  \\
add\underline{{ }}neb\underline{{ }}continuum & Turn off nebular continuum & True &  \\
nebemlineinspec & Add nebular emission lines to the model spectrum & False & Turned off if nebular marginalization is on \\
marginalize\underline{{ }}elines & Fit and marginalize over observed emission lines & True & Only included if spectrum and nebular marginalization used \\
use\underline{{ }}eline\underline{{ }}prior & Use prior on width of nebular emission lines & True & Only included if spectrum and nebular marginalization used \\
eline\underline{{ }}prior\underline{{ }}width & Width of the prior on line luminosity in units of true luminosity divided by FSPS prediction & 1.0 & Only included if spectrum and nebular marginalization used \\
lines\underline{{ }}to\underline{{ }}fit & Specify which lines to marginalize over & All & Only included if spectrum and nebular marginalization used \\
polyorder & Chebyshev polynomial order for fitting the observed spectrum & 12 & Only if spectrum used \\
poly\underline{{ }}regularization & Regularization of polynomial coefficients for the spectral calibration vector & 0 & Only if spectrum used \\
spec\underline{{ }}norm & Normalization of the spectrum in units of true flux divided by observed flux & 1.0 & Only if spectrum used. This parameter is fixed if the spectral polynomial calibration vector is set to be optimized \\
nsigma\underline{{ }}outlier\underline{{ }}spec & Factor of inflation for errors determined by f\underline{{ }}outlier\underline{{ }}spec & 50 & Only if spectrum used \\
f\underline{{ }}outlier\underline{{ }}phot & Fraction of photometric bands considered outliers & 0 &  \\
nsigma\underline{{ }}outlier\underline{{ }}phot & Factor of inflation for errors determined by f\underline{{ }}outlier\underline{{ }}phot & 50 &  \\
\enddata
\tablecomments{Details of the free and fixed \texttt{Prospector} priors used in this work. $\mathcal{U}$ denotes a Uniform distribution. $C\mathcal{N}$ denotes a Clipped Normal distribution. $\mathcal{T}$ denotes a Student T-distribution. $L\mathcal{U}$ denotes a Log Uniform distribution.}
\end{deluxetable*}

\bibliography{main}{}
\bibliographystyle{aasjournal}


\end{CJK*}
\end{document}